\documentclass[sort&compress]{elsarticle}

\makeatletter
\def\ps@pprintTitle{%
\let\@oddhead\@empty
\let\@evenhead\@empty
\def\@oddfoot{}%
\let\@evenfoot\@oddfoot}
\makeatother

\usepackage{amsfonts}
\usepackage{amssymb}
\usepackage{amsmath}
\usepackage{amsthm}
\usepackage{mathrsfs}
\usepackage{bm}
\usepackage{graphicx}
\usepackage{enumerate}
\usepackage{multirow}
\usepackage{hyperref}
\usepackage[hmargin=1.1in,vmargin=1.1in]{geometry}
\usepackage{float}
\usepackage{footnote}
\usepackage{framed}
\usepackage{natbib}
\usepackage{adjustbox}
\usepackage{appendix}
\usepackage{color}

\hypersetup{colorlinks=true}

\newtheoremstyle{definitionstyle}
{\baselineskip\@plus.2\baselineskip\@minus.2\baselineskip}
{\baselineskip\@plus.2\baselineskip\@minus.2\baselineskip}
{}
{}
{\bfseries}
{.}
{ }
{\thmname{#1}\thmnumber{~#2}\thmnote{~(#3)}}
\newtheoremstyle{nameddefinitionstyle}
{\baselineskip\@plus.2\baselineskip\@minus.2\baselineskip}
{\baselineskip\@plus.2\baselineskip\@minus.2\baselineskip}
{}
{}
{\bfseries}
{.}
{ }
{\thmnote{#3}}
\newtheoremstyle{framednameddefinitionstyle}
{0}
{0}
{}
{}
{\bfseries}
{.}
{ }
{\thmnote{#3}}
\newtheoremstyle{theoremstyle}
{\baselineskip\@plus.2\baselineskip\@minus.2\baselineskip}
{\baselineskip\@plus.2\baselineskip\@minus.2\baselineskip}
{}
{}
{\bfseries}
{.}
{ }
{\thmname{#1}\thmnumber{~#2}\thmnote{~(#3)}}
\newtheoremstyle{framedtheoremstyle}
{0}
{0}
{}
{}
{\bfseries}
{.}
{ }
{\thmname{#1}\thmnumber{~#2}\thmnote{~(#3)}}
\newtheoremstyle{proofstyle}
{\baselineskip\@plus.2\baselineskip\@minus.2\baselineskip}
{\baselineskip\@plus.2\baselineskip\@minus.2\baselineskip}
{}
{}
{}
{.}
{ }
{\textsc{\thmname{#1}\thmnote{~#3}}}

\theoremstyle{theoremstyle}
\newtheorem{nul}{}[section]
\newtheorem{lem}[nul]{Lemma}
\newtheorem{prp}[nul]{Proposition}
\newtheorem{thm}[nul]{Theorem}
\newtheorem{cor}[nul]{Corollary}

\theoremstyle{framedtheoremstyle}
\newtheorem{rslt}{Physical Result}

\theoremstyle{definitionstyle}
\newtheorem{dfn}[nul]{Definition}

\theoremstyle{definitionstyle}

\theoremstyle{definitionstyle}

\theoremstyle{definitionstyle}
\newtheorem{exm}[nul]{Example}
\newtheorem{cnstr}[nul]{Construction}

\theoremstyle{nameddefinitionstyle}
\newtheorem{nameddef}{Definition}

\theoremstyle{framednameddefinitionstyle}
\newtheorem*{framednameddef}{}

\theoremstyle{proofstyle}
\newtheorem{pf}{Proof}
\newtheorem{ag}{Arguments}

\theoremstyle{definitionstyle}
\newtheorem*{ack}{Acknowledgments}

\newcommand{\fromto}{\rightarrow}
\newcommand{\oneone}{\hookrightarrow}
\newcommand{\onto}{\twoheadrightarrow}
\newcommand{\xfromto}[1]{\xrightarrow{#1}}

\newcommand{\ZZZ}{\mathbb{Z}}
\newcommand{\NNN}{\mathbb{N}}

\newcommand{\RRR}{\mathbb{R}}
\newcommand{\CCC}{\mathbb{C}}

\newcommand{\DD}{\mathbf{D}}

\renewcommand{\SS}{\mathbf{S}}

\newcommand{\C}{\mathcal{C}}
\newcommand{\D}{\mathcal{D}}

\newcommand{\F}{\mathcal{F}}
\newcommand{\G}{\mathcal{G}}

\newcommand{\K}{\mathcal{K}}
\newcommand{\M}{\mathcal{M}}

\newcommand{\T}{\mathcal{T}}
\newcommand{\U}{\mathcal{U}}

\DeclareMathOperator{\Map}{Map}
\DeclareMathOperator{\identity}{id}

\newcommand{\Set}{\operatorname{\mathbf{Set}}}

\newcommand{\Top}{\operatorname{\mathbf{Top}}}

\newcommand{\Toph}{\operatorname{\mathbf{Toph}}}

\newcommand{\Grp}{\operatorname{\mathbf{Grp}}}
\newcommand{\Ab}{\operatorname{\mathbf{Ab}}}

\DeclareMathOperator{\Obj}{Obj}
\DeclareMathOperator{\Mor}{Mor}
\DeclareMathOperator{\Hom}{Hom}

\DeclareMathOperator{\dom}{dom}
\DeclareMathOperator{\cod}{cod}

\newcommand{\For}{\mathcal For}

\newcommand{\colim}{\underset{\longrightarrow}{\lim}}

\DeclareMathOperator{\kernel}{ker}

\newcommand{\coloneq}{\mathrel{\mathop:}=}
\newcommand{\eqcolon}{=\mathrel{\mathop:}}
\newcommand{\homeomorphic}{\approx}
\newcommand{\homotopic}{\simeq}

\newcommand{\isomorphic}{\cong}

\newcommand{\bars}[1]{\left| #1 \right|}
\newcommand{\paren}[1]{\left( #1 \right)}
\newcommand{\angles}[1]{\left\langle #1 \right\rangle}
\newcommand{\brackets}[1]{\left[ #1 \right]}
\newcommand{\braces}[1]{\left\{ #1 \right\}}

\newcommand{\bra}[1]{\left\langle #1\right|}
\newcommand{\ket}[1]{\left|#1\right\rangle}
\newcommand{\braket}[2]{\left\langle#1\middle|#2\right\rangle}
\newcommand{\ketbra}[2]{\left| #1 \right\rangle\left\langle #2\right|}

\def\ind\hspace{0.2in}

\DeclareMathOperator{\pt}{pt}

\DeclareMathOperator{\Ext}{Ext}
\newcommand{\SPT}{\operatorname{\mathcal{SPT}}}

\begin{document}

\title{Minimalist approach to the classification of symmetry protected topological phases}
\author{Charles Zhaoxi Xiong\corref{cor1}}
\ead{zxiong@g.harvard.edu}

\cortext[cor1]{Corresponding author}

\address{Department of Physics, Harvard University, Cambridge, MA 02138, USA\tnoteref{add1}}
\address{Department of Physics, University of California, Berkeley, CA 94720, USA\tnoteref{add2}}

\tnotetext[add1]{Since September 2016}
\tnotetext[add2]{Until August 2016}


\begin{abstract}
A number of proposals with differing predictions (e.g. Borel group cohomology, oriented cobordism, group supercohomology, spin cobordism, etc.) have been made for the classification of symmetry protected topological (SPT) phases. Here we treat various proposals on an equal footing and present rigorous, general results that are independent of which proposal is correct. We do so by formulating a minimalist Generalized Cohomology Hypothesis, which is satisfied by existing proposals and captures essential aspects of SPT classification. From this Hypothesis alone, formulas relating classifications in different dimensions and/or protected by different symmetry groups are derived. Our formalism is expected to work for fermionic as well as bosonic phases, Floquet as well as stationary phases, and spatial as well as on-site symmetries. As an application, we predict that the complete classification of 3-dimensional bosonic SPT phases with space group symmetry $G$ is $H^4_{\rm Borel}\paren{G;U(1)} \oplus H^1_{\rm group}\paren{G;\ZZZ}$, where the $H^1$ term classifies phases beyond the Borel group cohomology proposal.
\end{abstract}

\begin{keyword}
topological phases of matter \sep symmetry protected topological phases \sep generalized cohomology theories
\end{keyword}

\maketitle

\setcounter{tocdepth}{2}

\tableofcontents


\section{Introduction\label{sec:introduction}}

The quest for a complete understanding of phases of matter has been a driving force in condensed matter physics. From the Landau-Ginzburg-Wilson paradigm \cite{Landau_Lifshitz} to topological insulators and superconductors \cite{kane2005A, kane2005B, Qi_Hughes_Zhang, Schnyder, Kitaev_TI, Hasan_Kane, Qi_Zhang, Chiu_Teo_Schnyder_Ryu} to topological orders \cite{Wen_TO, WenNiu, Wen_book} to symmetry protected topological (SPT) phases \cite{SPT_origin} to symmetry enriched topological phases \cite{Wen_Definition}, we have witnessed an infusion of ideas from topology into this century-old field.
SPT phases are a relatively simple class of non-symmetry-breaking, gapped quantum phases and have been a subject of intense investigation in recent years \cite{Wen_review_2016}. As an interacting generalization of topological insulators and superconductors and intimate partner of topological orders \cite{1410.4540}, they exhibit such exotic properties as the existence of gapless edge modes, and harbor broad applications. They have also been increasingly integrated into other novel concepts such as many-body localization and Floquet phases \cite{Keyserlingk_Floquet, Else_Floquet, Potter_Floquet, Roy_Floquet, PhysRevB.93.245146, PhysRevB.94.085112, 1607.05277, 1512.04190, 1610.07611, 1610.06899, PhysRevLett.117.090402, 1610.03485, 1609.00006, PhysRevB.93.134207, 1605.03601}.

Despite tremendous progress \cite{Wen_1d, Cirac, Wen_2d, Wen_Boson, Wen_Fermion, Kapustin_Boson, Kapustin_Fermion, Kapustin_equivariant, Freed_SRE_iTQFT, Freed_ReflectionPositivity, Kitaev_Stony_Brook_2011_SRE_1, Kitaev_Stony_Brook_2011_SRE_2, Kitaev_Stony_Brook_2013_SRE, Kitaev_IPAM, Husain, 2dChiralBosonicSPT, 2dChiralBosonicSPT_erratum, 3dBTScVishwanathSenthil, 3dBTScWangSenthil, 3dBTScBurnell, 3dFTScWangSenthil_1, 3dFTScWangSenthil_2, 3dFTScWangSenthil_2_erratum, 2dFermionGExtension, Lan_Kong_Wen_1, Lan_Kong_Wen_2, SOinfty, Else_edge, Jiang_sgSPT, Thorngren_sgSPT, Wang_Levin_invariants, Wang_intrinsic_fermionic, Huang_dimensional_reduction}, a complete classification of SPT phases for arbitrary symmetries in arbitrary dimensions remains elusive.
A number of classification proposals have been made in the general case:
the Borel group cohomology proposal \cite{Wen_Boson}, the oriented cobordism proposal \cite{Kapustin_Boson}, the Freed-Hopkins proposal \cite{Freed_SRE_iTQFT, Freed_ReflectionPositivity}, and Kitaev's proposal \cite{Kitaev_Stony_Brook_2011_SRE_1, Kitaev_Stony_Brook_2013_SRE} in the bosonic case; and the group supercohomology proposal \cite{Wen_Fermion}, the spin cobordism proposal \cite{Kapustin_Fermion}, the Freed-Hopkins proposal \cite{Freed_SRE_iTQFT, Freed_ReflectionPositivity}, and Kitaev's proposal \cite{Kitaev_Stony_Brook_2013_SRE, Kitaev_IPAM} in the fermionic case. These proposals give differing predictions in certain dimensions for certain symmetry groups that cannot be attributed to differences in definitions or conventions. While more careful analysis \cite{2dChiralBosonicSPT, 2dChiralBosonicSPT_erratum, 3dBTScVishwanathSenthil, 3dBTScWangSenthil, 3dBTScBurnell, 3dFTScWangSenthil_1, 3dFTScWangSenthil_2, 3dFTScWangSenthil_2_erratum, 2dFermionGExtension, Wang_intrinsic_fermionic} has uncovered previously overlooked phases and input from topological field theories \cite{Kapustin_Boson, Kapustin_Fermion, Freed_SRE_iTQFT, Freed_ReflectionPositivity} has brought us closer than ever to our destination, we believe that we can do much more.

In this paper, we will take a novel, minimalist approach to the classification problem of SPT phases, by appealing to the following principle of Mark Twain's \cite{MarkTwain}:
\begin{quote}
Distance lends enchantment to the view.
\end{quote}
In this spirit, we will not
commit ourselves to any particular construction of SPT phases,
specialize to specific dimensions or symmetry groups,
or investigate the completeness of any of the proposals above.
Instead, we will put various proposals under one umbrella and present results that are independent of which proposal is correct. This will begin with the formulation of a hypothesis, we dub the Generalized Cohomology Hypothesis, that encapsulates essential attributes of SPT classification. These attributes will be shown to be possessed by various existing proposals and argued, on physical grounds, to be possessed by the unknown complete classification should it differ from existing ones. The results we present will be rigorously derived from this Hypothesis alone. Because we are taking a ``meta" approach, we will not be able to produce the exact classification in a given dimension protected by a given symmetry group. We will be able, however, to \emph{relate} classifications in different dimensions and/or protected by different symmetry groups. Such relations will be interpreted physically -- this may require additional physical input, which we will keep to a minimum and state explicitly. A major advantage of this formalism is the universality of our results, which, as we said, are not specific to any particular construction.

What will enable us to relate different dimensions and symmetry groups is ultimately the fact that the Hypothesis is a statement about all dimensions and all symmetry groups at once. Furthermore, due to a certain ``symmetry" the Hypothesis carries, the relations we derive will hold in arbitrarily high dimensions. Finally, the Hypothesis is supposed to apply to fermionic phases as well as bosonic phases. Thus our formalism is not only independent of construction, but also independent of physical dimension and particle content, that is, bosons vs.\,fermions.

More specifically, the Hypothesis will be based Kitaev's proposal that the classification of SPT phases should carry the structure of a generalized cohomology theory \cite{Kitaev_Stony_Brook_2011_SRE_1, Kitaev_Stony_Brook_2013_SRE, Kitaev_IPAM}. We will clarify the needed ingredients and formulate the ideas in a language amenable to rigorous treatment. While the Hypothesis is informed by Refs.\,\cite{Kitaev_Stony_Brook_2011_SRE_1, Kitaev_Stony_Brook_2013_SRE, Kitaev_IPAM}, our philosophy is fundamentally different. The goal of Refs.\,\cite{Kitaev_Stony_Brook_2011_SRE_1, Kitaev_Stony_Brook_2013_SRE, Kitaev_IPAM} was to classify SPT phases in physical ($\leq 3$) dimensions by building a specific generalized cohomology theory using one's current understanding of invertible topological orders. The goal of this paper is to make maximally general statements about the classification of SPT phases that would apply to all generalized cohomology theories, by refraining from incorporating any additional data. The approach of Refs.\,\cite{Kitaev_Stony_Brook_2011_SRE_1, Kitaev_Stony_Brook_2013_SRE, Kitaev_IPAM} was concrete, whereas ours is minimalist.

This paper is best regarded as a proof of concept aimed at fleshing out the basic structure of the minimalist approach to SPT phases, and we will restrict to the case of unitary symmetries for simplicity (see Sec.\,\ref{subsec:particle_content_dimensionality_symmetry_action}) -- the generalization to antiunitary symmetries will be covered in Ref.\,\cite{Xiong_Alexandradinata}. We will derive some simple physical consequences of the Hypothesis and compare them to the literature. We will also apply the Hypothesis to crystalline SPT phases and address the open question of what the complete classification of 3-dimensional bosonic SPT phases with space group symmetries might be. Other applications, including a proof that the recently discovered ``hourglass fermions" \cite{Nonsymm_Shiozaki, Hourglass, Cohomological, Ma_discoverhourglass} -- a topological insulator featuring an hourglass-shaped surface band structure -- are robust to interactions \cite{Xiong_Alexandradinata}, will be the subject of future works.

This paper is organized as follows. First, we will summarize the main physical results of the paper in Sec.\,\ref{sec:main_results}. Next, in Sec.\,\ref{sec:generalities}, we will establish conventions, clarify our definition of SPT phases, and comment on two elementary properties of SPT phases -- additivity and functoriality -- that will play a role in the Hypothesis. Then, in Sec.\,\ref{sec:generalized_cohomology_hypothesis}, we will introduce necessary mathematical concepts and formulate the Generalized Cohomology Hypothesis. In the succeeding Sec.\,\ref{sec:justification_hypothesis}, we will justify the Hypothesis from several different perspectives. In particular, we will present the physical interpretations of $\Omega$-spectrum (Sec.\,\ref{subsec:physical_interpretations_Omega_spectrum}), justify that SPT phases described by generalized cohomology theories are realizable by local Hamiltonians (Sec.\,\ref{subsec:realizability}), and elaborate upon the applicability of the Hypothesis to spatiotemporal symmetries (Sec.\,\ref{subsec:spatiotemporal}). In Sec.\,\ref{sec:consequences_hypothesis_physical_implications}, we will present some simple physical consequences of the Hypothesis and compare them with the literature. Finally, in Sec.\,\ref{sec:an_application}, we will apply the Hypothesis to crystalline SPT phases and make a prediction for the complete classification of 3-dimensional bosonic SPT phases with space group symmetries. We end the paper in Sec.\,\ref{sec:summary_outlook} with a summary and suggestions for future directions.

\begin{ack}
I am grateful to my advisor, Ashvin Vishwanath, for his guidance and support.
I also want to thank Ammar Husain, Ryan Thorngren, Benjamin Gammage, and Richard Bamler for introducing me to the subject of generalized cohomology theories;
Hoi-Chun Po, Alexei Kitaev, Christian Schmid, Yen-Ta Huang, Yingfei Gu, Dominic Else, Shengjie Huang, Shenghan Jiang, Drew Potter, and Chong Wang for numerous inspiring discussions;
and Judith H\"oller, Alex Takeda, and Byungmin Kang for their invaluable comments on an early draft of the paper.
This work was supported in part by the 2016 Boulder Summer School for Condensed Matter and Materials Physics through NSF grant DMR-13001648.
\end{ack}

\section{Main Results\label{sec:main_results}}

In this section, we summarize the main physical results of the paper.

\begin{enumerate}[(i)]
\item We will be able to relate the original definition of SPT phases \cite{Wen_Definition, Cirac} to the one currently being developed by Refs.\,\cite{Kitaev_Stony_Brook_2011_SRE_2, Kitaev_Stony_Brook_2013_SRE, Kapustin_Boson, Freed_SRE_iTQFT, Freed_ReflectionPositivity, McGreevy_sSourcery}, which is in terms of invertibility \cite{FreedMoore2006} of phases and uniqueness of ground state on arbitrary spatial slices. According to the latter definition, the classification of SPT phases can be nontrivial even without symmetry. (For instance, the integer quantum Hall state represents an SPT phase in that sense.) Assuming $G$ is unitary, we will show that SPT phases in the old sense and new sense are related as follows:\footnote{The direct sum is with respect to an abelian group structure of classification that we will describe later. Note that we could have used the direct product notation $\times$ for groups, but the direct sum notation $\oplus$ is more common for abelian groups in the mathematical literature.}
\begin{eqnarray}
\braces{\parbox{3.4cm}{$d$-dimensional $G$-protected SPT phases in the new sense}}
\isomorphic
\braces{\parbox{3.4cm}{$d$-dimensional $G$-protected SPT phases in the old sense}}
\oplus
\braces{\parbox{2.4cm}{$d$-dimensional invertible topological orders}},
\end{eqnarray}
where \emph{invertible topological orders} is a different name for SPT phases without symmetry (from now on, SPT phases will always mean SPT phases in the new sense), and $d$ and $G$ are arbitrary; see Sec.\,\ref{subsec:unification_old_new_definitions_SPT_phases}.
This result is consistent with Ref.\,\cite{Kapustin_Fermion}.

\item We will be able to relate the classification of translationally invariant SPT phases to the classification of SPT phases without translational symmetry.
The former are protected by a discrete spatial translational symmetry $\ZZZ$ as well as an internal symmetry $G$, whereas the latter are protected by $G$ alone.
More precisely, we will prove that there is a decomposition
\begin{eqnarray}
\braces{\parbox{2.8cm}{$d$-dimensional $(\ZZZ \times G)$-protected SPT phases}}
\isomorphic
\braces{\parbox{2.9cm}{$(d-1)$-dimensional $G$-protected SPT phases}}
\oplus
\braces{\parbox{2.2cm}{$d$-dimensional $G$-protected SPT phases}};
\end{eqnarray}
see Sec.\,\ref{subsec:strong_weak_topological_indices_interacting_world}.
This result is consistent with Refs.\,\cite{Wen_1d, Cirac, Wen_sgSPT_1d}.


\item We will generalize the relation above to $d$-dimensional SPT phases protected by discrete translation in $n$ directions, so that the full symmetry is $\ZZZ^n \times G$. We will see a hierarchy of lower-dimensional classifications enter the decomposition, with $n \choose k$ terms in dimension $d-k$, which correspond to layering $(d-k)$-dimensional SPT phases in $n \choose k$ different ways; see Sec.\,\ref{subsec:hierarchy_strong_weak_topological_indices}. This result is consistent with one's physical intuition.

\item We will reinterpret the $\ZZZ$ above as a discrete temporal translational symmetry. Accordingly, there will be a decomposition
\begin{eqnarray}
\braces{\parbox{2.8cm}{$d$-dimensional $G$-protected Floquet SPT phases}}
\isomorphic
\braces{\parbox{3.4cm}{$(d-1)$-dimensional $G$-\par protected (stationary) SPT phases}}
\oplus
\braces{\parbox{2.9cm}{$d$-dimensional $G$-\par protected (stationary) SPT phases}};
\end{eqnarray}
see Sec.\,\ref{subsec:pumping_Floquet_eigenstates_classification_Floquet_SPT_phases}. This result is consistent with Refs.\,\cite{Keyserlingk_Floquet, Else_Floquet, Potter_Floquet}.

\item We will show that a similar decomposition exists for semidirect products $\ZZZ \rtimes G$, and more generally $G_1 \rtimes G_2$, which have implications for the classification of space group-protected SPT phases; see Sec.\,\ref{subsec:applications_space_group_protected_SPT_phases}.

\item 
In Sec.\,\ref{subsec:obstruction_free_enlargement_symmetry_group}, we will address the question as to when one can extend a symmetry $G'$ of an SPT phase to a bigger symmetry $G \supset G'$. 
Specifically, if there exists a subgroup $G''\subset G$ such that $G'' \rtimes G' = G$,
then we will show that every $G'$-protected SPT phase is representable by some $G$-protected SPT phase.

\item As the main application of the Hypothesis in this paper, in Sec.\,\ref{sec:an_application}, we will make a prediction for the complete classification of 3D bosonic SPT phases with space group symmetry $G$:
\begin{equation}
    H^4_{\rm Borel}\paren{G;U(1)} \oplus H^1_{\rm group}\paren{G;\ZZZ},
\end{equation}
where the coefficients $U(1)$ and $\ZZZ$ are twisted and an element $g$ of $G$ acts nontrivially on them iff it reverses the spatial orientation. Here, the $H^4$ term represents the crystalline analogue \cite{Huang_dimensional_reduction, Thorngren_sgSPT} of internal SPT phases described by the Borel group cohomology proposal \cite{Wen_Boson}. The $H^1$ term is previously unknown and classifies 3D bosonic crystalline SPT phases beyond group cohomology.

\end{enumerate}

\section{Generalities\label{sec:generalities}}

\subsection{Particle content, dimensionality, and symmetry action\label{subsec:particle_content_dimensionality_symmetry_action}}

Locality is defined differently for fermionic systems than for bosonic (i.e.\ spin) systems \cite{Wen_Fermion_vs_Boson}.
For this reason, classifications of bosonic phases and fermionic phases are traditionally done separately. While we will follow that tradition, our formalism works identically in the two cases. Therefore, we can omit the qualifiers ``fermionic" and ``bosonic" and simply speak of ``SPT phases."

By the dimension of a physical system, we always mean the spatial dimension. When it comes to mathematical construction, it is convenient to allow dimensions to be negative. If a purely mathematical result in this paper appears to contain a free variable $d$, then it should be understood that this result is valid for all $d\in \ZZZ$. If a physical result appears to contain a free variable $d$, then it should be understood that this result is valid for all $d\in \ZZZ$ for which all dimensions involved are non-negative.

For simplicity, we assume all symmetry actions to be linear unitary. A generalization to antilinear antiunitary actions is possible (see Sec.\,\ref{sec:summary_outlook}) but beyond the purview of this paper.

We allow all topological groups satisfying the basic technical conditions in App.\,\ref{subapp:technical_conventions} to be symmetry groups. Thus, a symmetry group can be finite or infinite, and discrete or non-discrete (also called ``continuous").
In the non-discrete case, one must define what it means for a symmetry group $G$ to act on a Hilbert space $\mathscr H$, that is whether we want a representation $\rho: G \fromto U(\mathscr H)$ to be continuous, measurable\footnote{The measurability of $(d+1)$-cochains as postulated in Ref.\,\cite{Wen_Boson} reduces to the measurability of $\rho$ when $d=0$.}, or something else, where $U(\mathscr H)$ denotes the space of unitary operators on $\mathscr H$ \cite{Wen_Boson}.
Conceivably, the Hypothesis can hold for one definition but fail for another,
so some care is needed.

\subsection{Definition of SPT phases\label{subsec:definition_SPT_phases}}

\subsubsection{Old definition of SPT phases\label{subsubsec:old_definition_SPT_phases}}

Traditionally, the definition of SPT phases goes as follows \cite{Wen_Definition, Cirac}. First, one defines a trivial system to be a local, gapped system whose unique ground state is a product state. Then, one defines a short-range entangled (SRE) system to be a local, gapped system that can be deformed to a trivial one via local, gapped systems.\footnote{We do not consider a system with accidental degeneracy in the thermodynamic limit to be gapped.} Finally, one defines a $G$-protected SPT phase to be an equivalence class of $G$-symmetric, non-symmetry-breaking\footnote{Note that ``$G$-symmetric" is an adjective qualifying Hamiltonians while ``non-symmetry-breaking" is an adjective qualifying ground states.} SRE systems with respect to the following equivalence relation: two such systems are equivalent if they can be deformed into each other via $G$-symmetric, non-symmetry-breaking SRE systems.

\subsubsection{New definition of SPT phases\label{subsubsec:new_definition_SPT_phases}}

Explicit as the definition above is, we shall adopt a different definition that will turn out to be 
convenient for our formalism, at the expense of including more phases. The set of SPT phases in the old sense will be shown to sit elegantly inside the set of SPT phases in the new sense, undisturbed, and they can be readily recovered. The definition spelled out below is based on the ideas in Refs.\,\cite{FreedMoore2006, Kitaev_Stony_Brook_2011_SRE_2, Kitaev_Stony_Brook_2013_SRE, Kapustin_Boson, Freed_SRE_iTQFT, Freed_ReflectionPositivity, McGreevy_sSourcery}.

To begin, let us assume that the terms ``system," ``local," ``gapped," ``$G$-symmetric," ``non-symmetry-breaking," and ``deformation" have been defined. Given two arbitrary systems $a$ and $b$ of the same dimension, we write $a+b$ (no commutativity implied; this is just a notation) for the composite system formed by stacking $b$ on top of $a$. However the aforementioned terms may be defined, it seems reasonable to demand the following:
\begin{enumerate}[(i)]
\item $a+b$ is well-defined.

\item If both $a$ and $b$ are local, gapped, $G$-symmetric, or non-symmetry-breaking, then $a+b$ is also local, gapped, $G$-symmetric, or non-symmetry-breaking, respectively.

\item A deformation of either $a$ or $b$ also constitutes a legitimate deformation of $a+b$.
\end{enumerate}
We will speak of deformation class, which, as usual, is an equivalence class of systems with respect to the equivalence relation defined by deformation (possibly subject to constraints, as discussed in the next paragraph)\footnote{If a deformation is defined to be a path in a space of systems that comes with a topology, then a deformation class is nothing but a path component of the space.}.

\begin{figure}[t]
\centering
\includegraphics[height=2.9in]{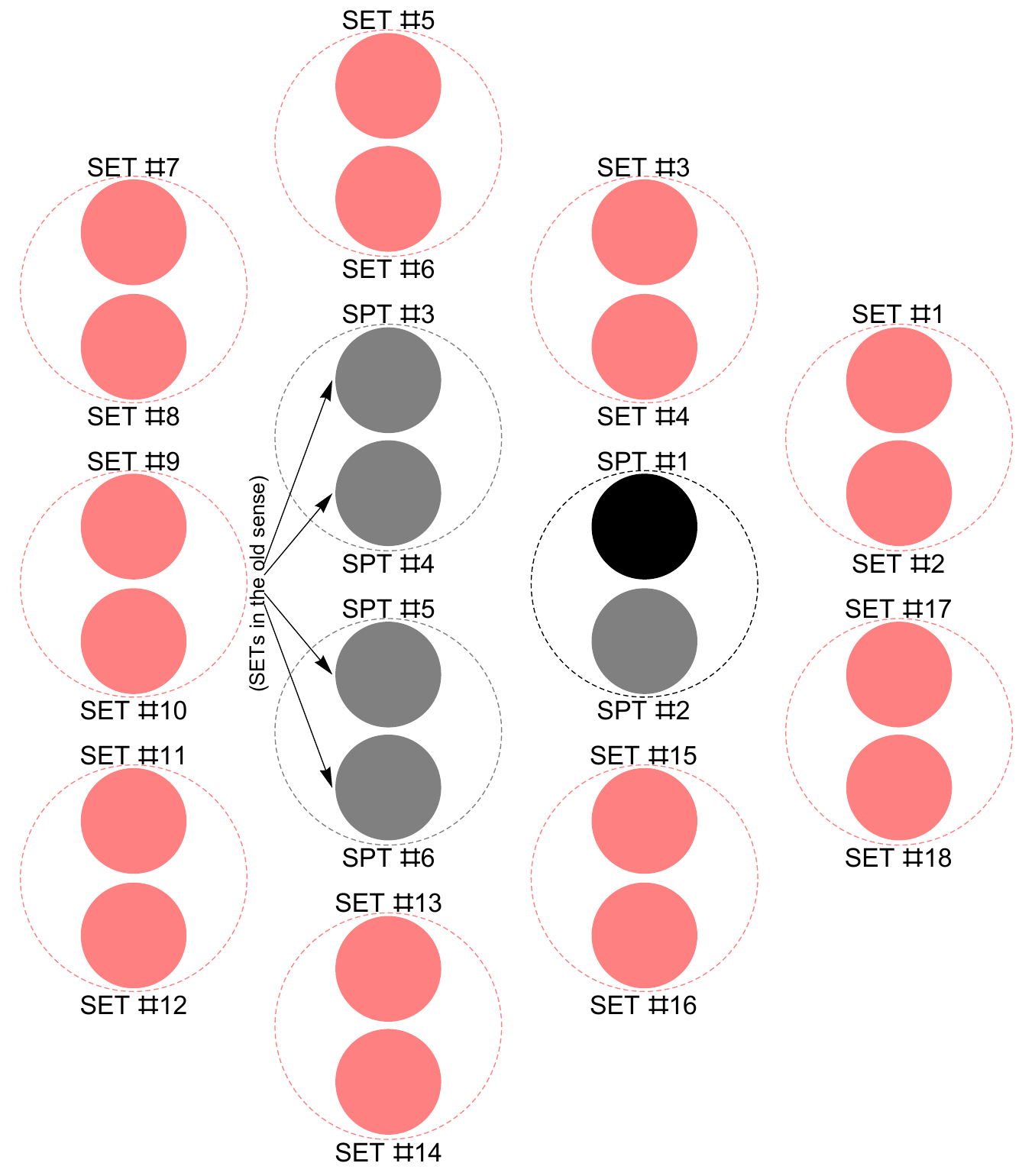}
\caption{(color online). Schematic illustration of the structure of the space of $d$-dimensional, $G$-symmetric, non-symmetry-breaking, local, gapped systems. Each deformation class, shown as a patch here, is called a $G$-protected topological phase. Each invertible (respectively non-invertible) class, shown as a gray or black (respectively pink) patch, is called an SPT (respectively SET) phase. The identity class, shown as a black patch, is called the trivial SPT phase.
Dashed circles are meant to indicate, by forgetting the symmetry, that more systems will be allowed and that distinct phases can become one.}
\label{fig:SPT_SET_old_new}
\end{figure}

Now, let $G$ be a symmetry group and $d$ be a non-negative integer. Consider the set $\M^d(G)$ of deformation classes of $d$-dimensional, local, gapped, $G$-symmetric, non-symmetry-breaking systems. We have seen that there is a binary operation on the set of such systems, given by stacking, which descends to a binary operation on $\M^d(G)$, owing to property (iii). We define the \emph{trivial $d$-dimensional $G$-protected SPT phase} to be the identity of $\M^d(G)$ with respect to the said binary operation. We define a \emph{$d$-dimensional $G$-protected SPT phase} to be an invertible element of $\M^d(G)$. We define a \emph{$d$-dimensional $G$-protected symmetry enriched topological (SET) phase} to be a non-invertible element of $\M^d(G)$. In general, we call an element of $\M^d(G)$ a \emph{$d$-dimensional $G$-protected topological phase}. An illustration of these concepts appears in Fig.\,\ref{fig:SPT_SET_old_new}.

In mathematical jargon, SPT phases are thus the group of invertible elements of the monoid $\M^d(G)$ of $d$-dimensional $G$-protected topological phases. We will see later that $\M^d(G)$ is commutative. This means that the $d$-dimensional $G$-protected SPT phases form not just a group, but an abelian group. This is elaborated upon in Sec.\,\ref{subsubsec:additivity}.

Note that we have made no mention of SRE systems so far. Instead, SPT and SET phases naturally fall out of the binary operation given by stacking. The uniqueness of identity and inverses and the abelian group structure of SPT phases come about for free. This is in line with the minimalism we are after.

\begin{figure}[t]
\centering
\includegraphics[height=2.9in]{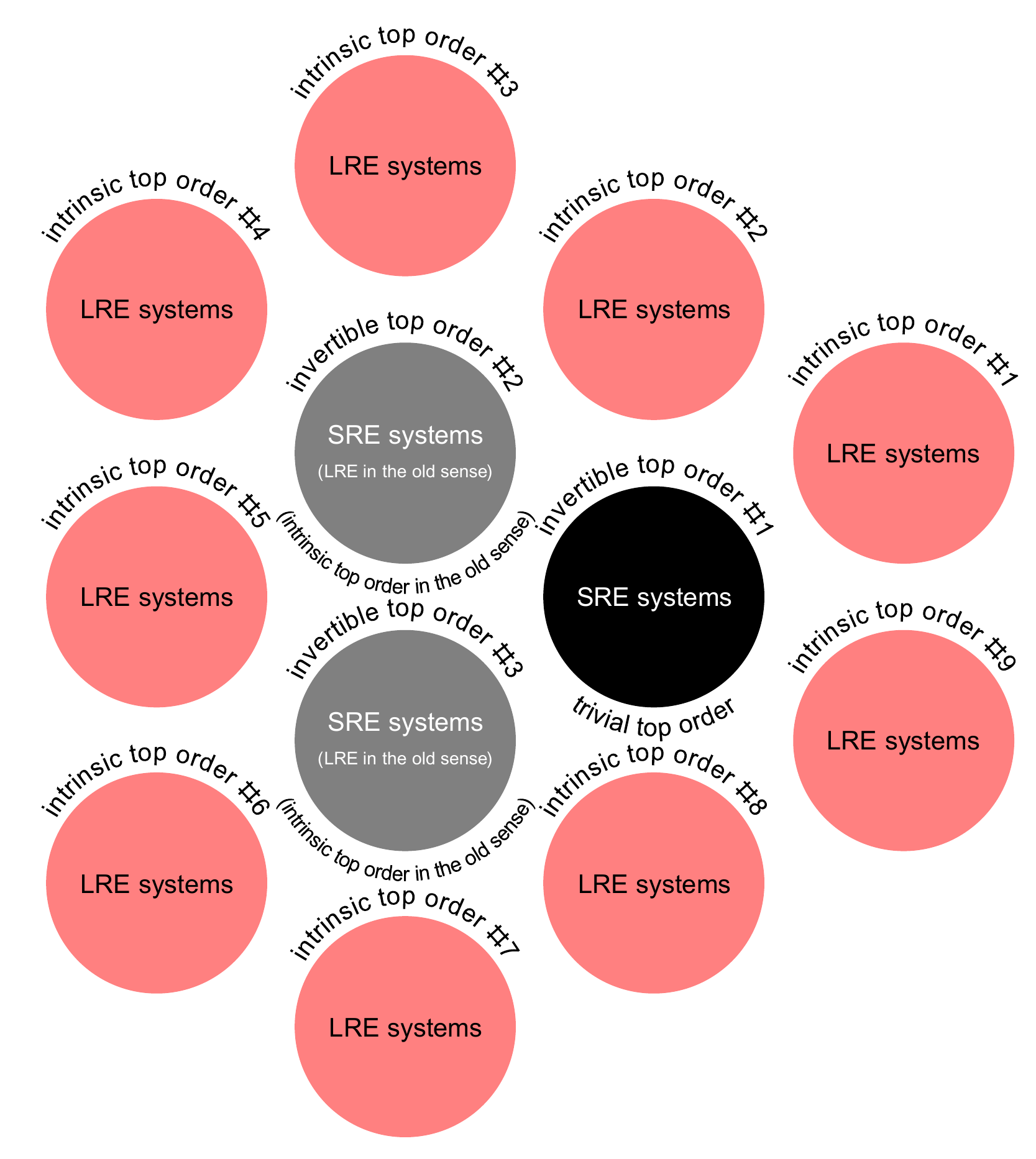}
\caption{(color online). Schematic illustration of the structure of the space of $d$-dimensional local, gapped systems. Each deformation class, shown as a patch here, is called a topological order. Each invertible (respectively non-invertible) class, shown as a gray or black (respectively pink) patch, is called an invertible (respectively intrinsic) topological order. The identity class, shown as a black patch, is called the trivial topological order, which is in particular invertible. A system is called SRE (respectively LRE) if it belongs to an invertible (respectively intrinsic) topological order.}
\label{fig:SRE_LRE_old_new}
\end{figure}

Let us introduce special names for the special case of trivial symmetry group $G=0$. The trivial SPT phase in this case can be called the \emph{trivial topological order}; an SPT phase, an \emph{invertible topological order}; an SET phase, an \emph{intrinsic topological order}; and any element of $\M^d(0)$, a \emph{topological order}. We may call a system \emph{short-range entangled (SRE)} if it represents an invertible topological order, and \emph{long-range entangled (LRE)} otherwise. An illustration of these concepts appears in Fig.\,\ref{fig:SRE_LRE_old_new}.

\subsubsection{Comparison between old and new definitions of SPT phases\label{subsubsec:comparison_definition_SPT_phases}}

To make contact with the old definition of SPT phases \cite{Wen_Definition, Cirac}, we note that all trivial systems in the old sense represent the identity element of $\M^d(0)$, where $0$ denotes the trivial group. Hence, SRE systems in the old sense are precisely those SRE systems in our sense that happen to lie in this identity class. Similarly, SPT phases in the old sense are precisely those SPT phases in our sense that, by forgetting the symmetry, represent the said identity class. This shows that the SPT phases in the old sense are a subset of the SPT phases in our sense. One of our results in this paper is that the former form a subgroup, in fact a direct summand, of the latter. These are illustrated in Figs.\,\ref{fig:SPT_SET_old_new} and \ref{fig:SRE_LRE_old_new}.

What is also clear is that the classification of SPT phases (according to our definition; same below) can be nontrivial even for the trivial symmetry group. This amounts to saying that there can exist nontrivial invertible topological orders, or that the set of SRE systems are partitioned into more than one deformation classes in the absence of symmetry. Examples of systems that represent nontrivial invertible topological orders are given in Table \ref{table:SPT_examples}. While this may seem to contradict the original idea \cite{Wen_Definition} of symmetry protection, it is the new notion of short-range entanglement not the old one that is closely related and potentially equivalent to the condition of unique ground state on spatial slices of arbitrary topology, and in two dimensions, the condition of no nontrivial anyonic excitations \cite{Kitaev_Stony_Brook_2011_SRE_2, Kitaev_Stony_Brook_2013_SRE, Kapustin_Boson, Freed_SRE_iTQFT, Freed_ReflectionPositivity, McGreevy_sSourcery, Wen_review_2016}, both of which are more readily verifiable, numerically and experimentally, than the deformability to product states.

\begin{table}[t]
\caption{Examples of systems that represent nontrivial invertible topological orders \cite{Kitaev_Stony_Brook_2013_SRE}. They are legitimate representatives of SPT phases according to our definition but fall outside the realm of Refs.\,\cite{Wen_Definition,Cirac}.\label{table:SPT_examples}}
\begin{tabular}{lll}
\hline
\hline
Particle content & Dimension & System \\
\hline
Fermion & 0 & An odd number of fermions \\
Fermion & 1 & The Majorana chain \cite{Majorana_chain} \\
Fermion & 2 & $\paren{p+ip}$-superconductors \cite{Volovik_p+ip, Read_p+ip, Ivanov_p+ip} \\
Boson & 2 & The $E_8$-model \cite{Kitaev_honeycomb, 2dChiralBosonicSPT, Kitaev_KITP} \\
\hline
\hline
\end{tabular}
\end{table}

\subsection{Elementary properties of SPT phases\label{subsec:elementary_properties_SPT}}

In this subsection, we discuss two elementary properties of the classification of SPT phases that will play a role in the Hypothesis. These follow essentially from the definition and should be features of any classification proposal.

\subsubsection{Additivity\label{subsubsec:additivity}}

\begin{figure}[t]
\centering
\includegraphics[width=3.3in]{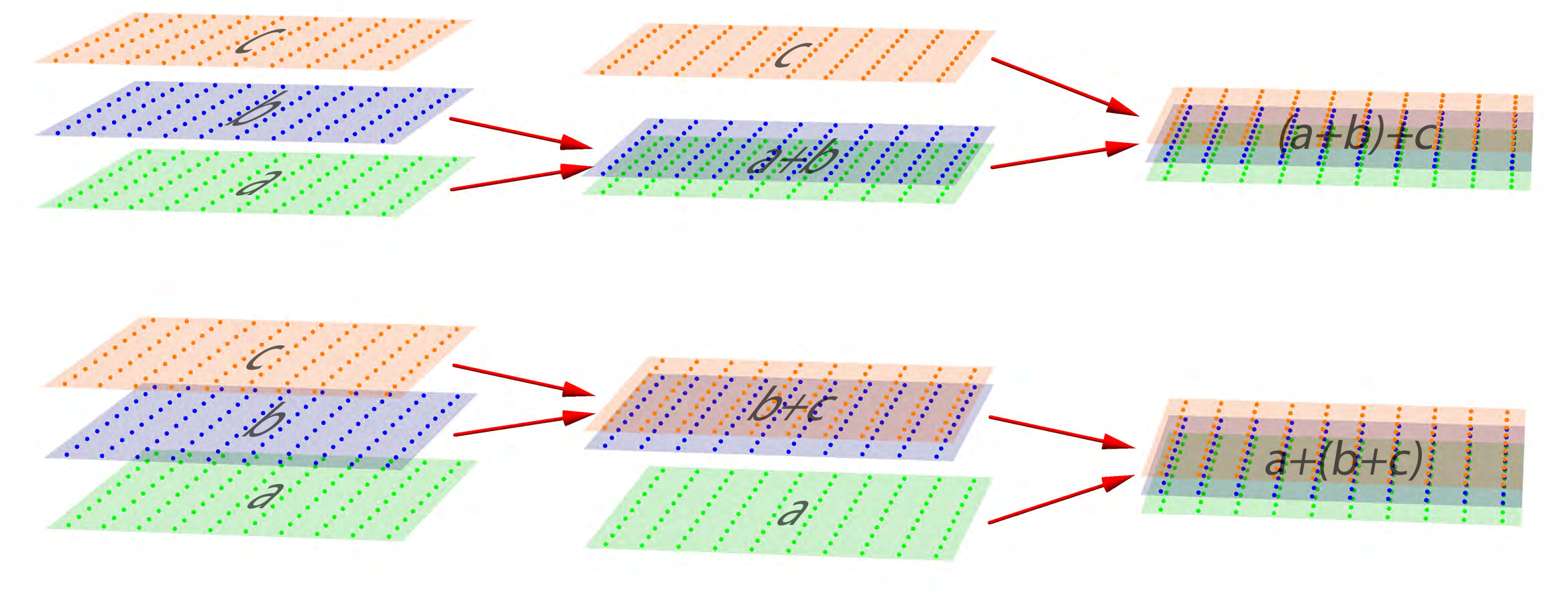}
\caption{(color online). Stacking is associative. Given three systems, $a$ (green), $b$ (blue), and $c$ (orange), combining $a$ and $b$ first and then $c$ (upper panel) produces the same system as combining $b$ and $c$ first and then $a$ (lower panel) does.}
\label{fig:associativity}
\end{figure}

Additivity says that the $d$-dimensional $G$-protected SPT phases form a discrete\footnote{Recall that ``group" in this paper means ``topological group." This is why we need the adjective ``discrete" here, as the abelian group of SPT phases is not endowed with a topology.} abelian group with respect to stacking. To see this, we first note that stacking of $d$-dimensional $G$-protected topological phases is tautologically associative (Fig.\,\ref{fig:associativity}). We then note that any $G$-symmetric system with a product state as the unique gapped ground state, which always exists, represents an identity with respect to stacking. Since SPT phases are invertible by definition, a discrete group structure is defined.

This leaves commutativity. We recall, in order to compare systems defined on different Hilbert spaces, that one would usually allow for ``embedding" of smaller Hilbert spaces into larger Hilbert spaces\footnote{More precisely, we want to ``embed" representations of the symmetry group rather than Hilbert spaces.}. This is known as an isometry \cite{Vidal_1, Vidal_2, Wen_1d, Cirac}. Given two Hilbert spaces $\mathscr H_1$ and $\mathscr H_2$ -- these are supposed to be associated to individual sites of two different systems -- the Hilbert spaces $\mathscr H_1 \otimes \mathscr H_2$ and $\mathscr H_2 \otimes \mathscr H_1$ are isomorphic. Embedding them into $\paren{\mathscr H_1 \otimes \mathscr H_2} \oplus \paren{\mathscr H_2\otimes \mathscr H_1} \isomorphic \CCC^2 \otimes \mathscr H_1 \otimes \mathscr H_2$, we can then interpolate between the two in a canonical, symmetry-preserving fashion. Therefore, the resulting phase is independent of the order of stacking.

Note that the above also shows that the $d$-dimensional $G$-protected topological phases form a discrete commutative monoid $\M^d(G)$. 

(Some definitions of SPT phases admit the coexistence of multiple trivial phases \cite{SPt,Hermele_torsor}, but this can always be salvaged by declaring the identity under stacking to be the true trivial phase, which is unique by elementary group theory.)

\subsubsection{Functoriality \label{subsubsec:functoriality}}

Functoriality says that every homomorphism $\varphi: G' \fromto G$ between any symmetry groups $G'$ and $G$ induces a homomorphism $\varphi^\ast$ from the discrete abelian group of $d$-dimensional $G$-protected SPT phases to the discrete abelian group of $d$-dimensional $G'$-protected SPT phases. Note that the direction of mapping is reversed. Implicit here is the assumption that the coherence relation $(\varphi \circ \phi)^\ast = \phi^\ast \circ \varphi^\ast$ be satisfied for all composable homomorphisms $\varphi$ and $\phi$.

\begin{figure}[t]
\centering
\includegraphics[width=5in]{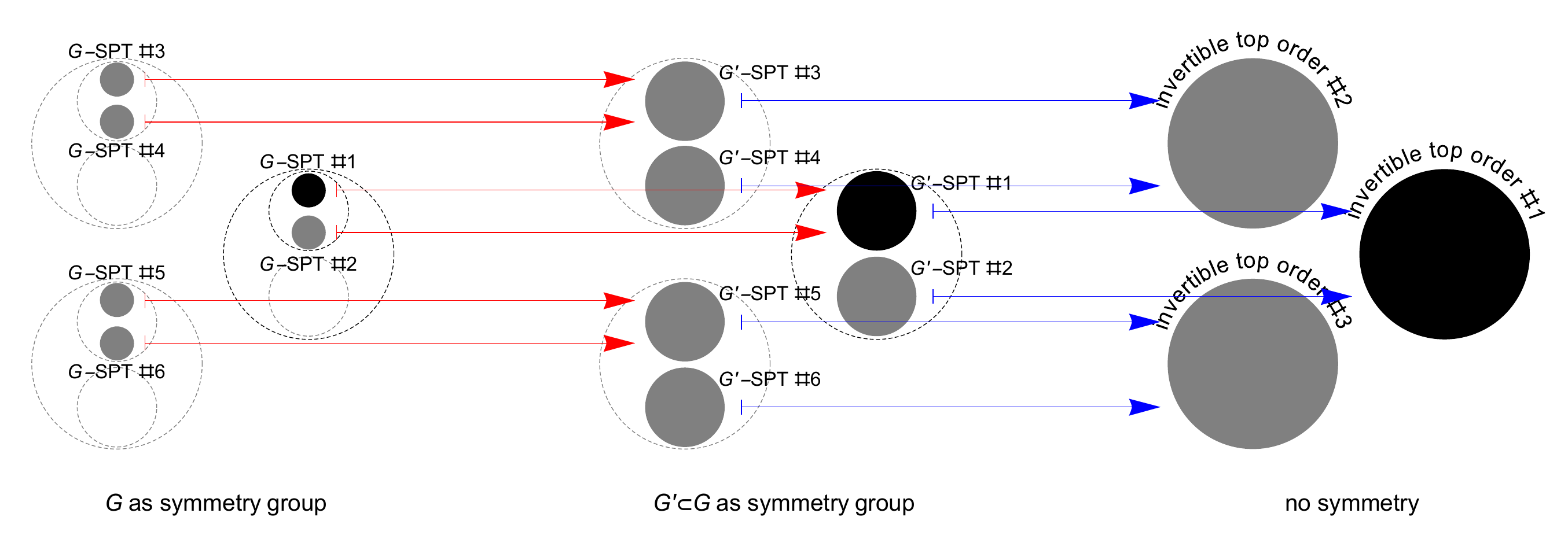}
\caption{(color online). Given $G'\subset G$, a representative of a $d$-dimensional $G$-protected SPT phase can also be viewed as a representative of a $d$-dimensional $G'$-protected SPT phase, which in turn can be viewed as a representative of a $d$-dimensional invertible topological order, by forgetting first the symmetry operations outside $G'$ and then $G'$ itself. This defines a map from the set of $d$-dimensional $G$-protected SPT phases to the discrete abelian group of $d$-dimensional $G'$-protected SPT phases, and then to the set of $d$-dimensional invertible topological orders.}
\label{fig:functoriality}
\end{figure}

Let us first understand this in the special case where $G'$ is a subgroup of $G$ and $\varphi$ is the inclusion. A $d$-dimensional $G$-protected SPT phase is represented by a $d$-dimensional, local, gapped, $G$-symmetric, non-symmetry-breaking system. By forgetting all symmetry operations outside the subgroup $G'$, we can view this same system as a representative of a $d$-dimensional $G'$-protected SPT phase. Since this applies to paths of systems as well, we get a well-defined map from the set of $d$-dimensional $G$-protected SPT phases to the set of $d$-dimensional $G'$-protected SPT phases. This is the induced map $\varphi^\ast$. It is easy to check that $\varphi^\ast$ preserves discrete abelian group structure. Moreover, such maps can be composed. For instance, we can further forget $G'$ entirely to obtain a map into the set of $d$-dimensional invertible topological orders. Forgetting symmetry operations in two steps is clearly equivalent to forgetting them all at once, which is the origin of the coherence relation $(\varphi \circ \phi)^\ast = \phi^\ast \circ \varphi^\ast$. These are illustrated in Fig.\,\ref{fig:functoriality}.

The general case where $\varphi: G' \fromto G$ is an arbitrary homomorphism only requires a small modification. A $d$-dimensional $G$-protected SPT phase is represented by a triple $\paren{\mathscr H, \rho, \hat H}$, where $\hat H$ is a Hamiltonian and $\rho: G \fromto U(\mathscr H)$ is a representation of $G$ on some Hilbert space $\mathscr H$. By precomposing $\varphi$, we obtain a representation $\rho \circ \varphi: G' \xfromto{\varphi} G \xfromto{\rho} U(\mathscr H)$ of $G'$. Then the triple $\paren{\mathscr H, \rho\circ \varphi, \hat H}$ represents a $d$-dimensional $G'$-protected SPT phase. This defines the map $\varphi^*$.

Note that the same argument also shows that every homomorphism $\varphi: G' \fromto G$ between any symmetry groups $G'$ and $G$ induces a homomorphism $\varphi^\ast: \M^d(G) \fromto \M^d(G')$ between the monoids of $d$-dimensional $G$- and $G'$-protected topological phases.

\section{The Generalized Cohomology Hypothesis\label{sec:generalized_cohomology_hypothesis}}

In this section, we will state the Generalized Cohomology Hypothesis, which is the foundation of our formalism.
Intuitively, the Hypothesis says that the classifications of SPT phases in different dimensions protected by different symmetry groups are intertwined in some intricate fashion, so that all information can be encoded into what is called an $\Omega$-spectrum. And just like proteins are produced from genes through the processes of transcription and translation, the classifications of $d$-dimensional $G$-protected SPT phases for varying $d$ and $G$ can be produced from the $\Omega$-spectrum through the classifying space construction and homotopy theory.

An $\Omega$-spectrum is by definition a sequence of pointed topological spaces $F_d$ indexed by integers $d\in \ZZZ$ together with pointed homotopy equivalences $F_d \homotopic \Omega F_{d+1}$, where $\Omega F_{d+1}$ is the loop space of $F_{d+1}$ (see App.\,\ref{subapp:notions_algebraic_topology}). As discussed in Sec.\,\ref{subsec:physical_interpretations_Omega_spectrum}, $F_d$ is believed to be the space of $d$-dimensional SRE states, and the pointed homotopy equivalences $F_d \homotopic \Omega F_{d+1}$ can be given physical interpretations as well.
Note that shifting $d$ turns an $\Omega$-spectrum into another $\Omega$-spectrum. This is responsible for the validity of the results in Sec.\,\ref{sec:consequences_hypothesis_physical_implications} in arbitrarily high dimensions.

\begin{dfn}
An (unreduced) generalized cohomology theory $h$ has an $\Omega$-spectrum $\paren{F_d}_{d\in \ZZZ}$ as its data. Given an integer $d$, it assigns to each topological space $X$ the discrete abelian group $h^d(X) \coloneq \brackets{X, F_d}$, i.e.\,the homotopy classes of maps from $X$ to $F_d$.%
\footnote{This differs from the standard definition in two ways, even when the Brown representability theorem is assumed: first, the representing $\Omega$-spectrum is part of the data; and second, we are not considering pairs of spaces. These differences, however, are completely innocuous.}
\label{dfn:unreduced_generalized_cohomology_theory}
\end{dfn}

\begin{dfn}
A reduced generalized cohomology theory $\tilde h$ has an $\Omega$-spectrum $\paren{F_d}_{d\in \ZZZ}$ as its data. Given an integer $d$, it assigns to each pointed topological space $X$ the discrete abelian group $\tilde h^d(X) \coloneq \angles{X, F_d}$, i.e.\,the homotopy classes of pointed maps from $X$ to $F_d$.\label{dfn:reduced_generalized_cohomology_theory}
\end{dfn}

Different choices of $\Omega$-spectrum can give wildly different generalized cohomology groups $h^d(X)$'s and $\tilde h^d(X)$'s. This is the degree of freedom that will allow us to encompass various inequivalent classification proposals. Furthermore, unreduced and reduced theories come hand in hand and can be recovered from each other.

The discrete abelian group structure on $h^d(X)$ is defined via the bijection $h^d(X) \coloneq \brackets{X, F_d} \approx \brackets{X, \Omega F_{d+1}}$.
Given two classes $\brackets{c_1}, \brackets{c_2}\in h^d(X)$ represented by maps $c_1, c_2: X \fromto \Omega F_{d+1}$, we define $\brackets{c_1}+\brackets{c_2}$ by concatenating the loops $c_1(x)$ and $c_2(x)$ for each $x$.
Further replacing $\Omega F_{d+1}$ by $\Omega^2 F_{d+2}$, one can show that $\brackets{c_1}+\brackets{c_2} = \brackets{c_2} + \brackets{c_1}$. The reduced case is similar.

$h^d$ is also functorial, in that every map $f: X \fromto Y$ induces a homomorphism $f^*: h^d(Y) \fromto h^d(X)$ so that $\paren{f\circ g}^\ast = g^\ast \circ f^\ast$ for all composable $f$ and $g$. Given a class $\brackets{c}\in h^d(Y)$ represented by a map $c: Y \fromto F_d$, we define $f^\ast\paren{\brackets c}$ by precomposing $f$ with $c$. The reduced case is similar.

Before stating the Generalized Cohomology Hypothesis, we recall there is a so-called classifying space functor $B$ (see App.\,\ref{subapp:generalized_cohomology_theories}). It assigns a pointed topological space $BG$ to each group $G$, and a pointed map $B\varphi: BG'\fromto BG$ to each homomorphism $\varphi: G'\fromto G$. As a result, the composition $h^d(B-)$ of $B$ and $h^d$ assigns a discrete abelian group $h^d(BG)$ to each group $G$, and a homomorphism $\varphi^*: h^d(BG)\fromto h^d(BG')$ to each homomorphism $\varphi:G'\fromto G$. The reduced case is similar.
We are now ready to state the

\begin{framed}
\begin{framednameddef}[Generalized Cohomology Hypothesis]
There exists an (unreduced) generalized cohomology theory $h$ such that, given any dimension $d\geq 0$ and symmetry group $G$, $h^d(BG)$ classifies $d$-dimensional $G$-protected SPT phases (see Sec.\,\ref{subsubsec:new_definition_SPT_phases}), with its discrete abelian group structure corresponding to stacking (see Sec.\,\ref{subsubsec:additivity}) and its functorial structure corresponding to replacing symmetry groups (see Sec.\,\ref{subsubsec:functoriality}).
\end{framednameddef}
\end{framed}

\section{Existing Classification Proposals\label{sec:comparison_different_proposals}}

In order to validate the minimalist approach of this paper, it is important at this point to recognize that various existing proposals for the classification of SPT phases are examples of generalized cohomology theories \cite{Kitaev_Stony_Brook_2011_SRE_1, Kitaev_Stony_Brook_2013_SRE, Kitaev_IPAM}. These include the Borel group cohomology proposal \cite{Wen_Boson}, the oriented cobordism proposal \cite{Kapustin_Boson}, Kitaev's proposal \cite{Kitaev_Stony_Brook_2011_SRE_1, Kitaev_Stony_Brook_2013_SRE}, and the Freed-Hopkins proposal \cite{Freed_SRE_iTQFT, Freed_ReflectionPositivity} in the bosonic case; and the group supercohomology proposal \cite{Wen_Fermion}, the spin cobordism proposal \cite{Kapustin_Fermion}, Kitaev's proposal \cite{Kitaev_Stony_Brook_2013_SRE, Kitaev_IPAM}, and the Freed-Hopkins proposal \cite{Freed_SRE_iTQFT, Freed_ReflectionPositivity} in the fermionic case. Each of these proposals corresponds to a distinct generalized cohomology theory and a distinct $\Omega$-spectrum. In fact, these $\Omega$-spectra can be explicitly written down, which we briefly summarize in Table \ref{table:existing_proposals} and will elaborate upon in App.\,\ref{app:existing_proposals_generalized_cohomology_theories}.

We highlight some of the differences in the predictions of different proposals. In the bosonic case with time-reversal symmetry, the Borel group cohomology proposal \cite{Wen_Boson} predicts a $\ZZZ_2$ classification in 3 spatial dimensions. In contrast, the cobordism proposal \cite{Kapustin_Boson} predicts a $\ZZZ_2^2$ classification, with the extra $\ZZZ_2$ conjectured to correspond to the ``3D $E_8$ phase" studied in Refs.\,\cite{3dBTScVishwanathSenthil, 3dBTScWangSenthil, 3dBTScBurnell}. Furthermore, there are certain phases in 6 dimensions that are nontrivial according to the Borel group cohomology proposal but are trivial according to the cobordism proposal. In the fermionic case, with a time-reversal symmetry that squares to the identity (as opposed to fermion parity), the group supercohomology proposal \cite{Wen_Fermion} predicts a $\ZZZ_4$ classification of SPT phases in 1 dimension. In contrast, the spin cobordism \cite{Kapustin_Fermion} proposal predicts a $\ZZZ_8$ classification. We thus see that these proposals are inequivalent.

In general, these $\Omega$-spectra should be thought of as good approximations to the true, unknown $\Omega$-spectrum $\paren{F^{\rm (true)}_d}_{d\in \ZZZ}$ that gives the complete classification of SPT phases. More precisely, there will be one such $\Omega$-spectrum $\paren{F^{\rm (true,b)}_d}_{d\in \ZZZ}$ for bosonic SPT phases and one such $\Omega$-spectrum $\paren{F^{\rm (true,f)}_d}_{d\in \ZZZ}$ for fermionic SPT phases. Existing bosonic proposals are approximations to $\paren{F^{\rm (true,b)}_d}_{d\in \ZZZ}$ whereas fermionic proposals are approximations to $\paren{F^{\rm (true,f)}_d}_{d\in \ZZZ}$.




\begin{table}
\caption{Generalized cohomology theories that have been proposed to classify SPT phases, and spectra that represent them. Here, $K(A,n)$ denotes the $n$-th Eilenberg-Mac Lane space of $A$ (see App. \ref{subapp:generalized_cohomology_theories}), and $U$ denotes the infinite unitary group $U(\infty) = \bigcup_{i=1}^\infty U(i)$. $\CCC P^\infty = \bigcup_{i=1}^\infty \CCC P^i$ denotes the infinite projective space, and $\pi_i$ and $k_i$ denote the $i$-th homotopy group and the $i$-th $k$-invariant \cite{Hatcher}, respectively. The $\CCC P^\infty$ in $F_0$, $\ZZZ_2$ in fermionic $F_0$, $\ZZZ_2$ in fermionic $F_1$, and $\ZZZ$ in bosonic $F_2$ have to do with Berry's phase, fermion parity, the Majorana chain, and the $E_8$-model, respectively (cf.\,Table \ref{table:SPT_examples}) \cite{Kitaev_Stony_Brook_2011_SRE_1, Kitaev_Stony_Brook_2013_SRE, Kitaev_IPAM}. More details of these proposals can be found in App.\,\ref{app:existing_proposals_generalized_cohomology_theories}.}
\scriptsize
\resizebox{\columnwidth}{!}{
\begin{tabular}{p{4.8cm}p{3cm}p{6.2cm}}
\hline
\hline
Classification proposal & Spectrum & Further information \\
\hline
Borel group cohomology as in Ref.\,\cite{Wen_Boson} & Shifted $H\ZZZ$ & $F_d = \begin{cases} K\paren{\ZZZ, d+2}, & d\geq -2, \\\pt, & d<-2. \end{cases}$ \par
In particular, $F_0 \homotopic \CCC P^\infty$. \\
Group supercohomology as in Ref.\,\cite{Wen_Fermion} & ``Twisted product" of\par $H\ZZZ_2$ and shifted $H\ZZZ$ & $F_d$ can be constructed as a Postnikov tower:\par
$\pi_i(F_d) \isomorphic \begin{cases} \ZZZ_2, & i=d, \\\ZZZ, & i = d+2, \\0, & \text{otherwise}, \end{cases}$
~ $k_{d+1} = \beta \circ Sq^2$,\par
where $Sq^2$ is the Steenrod square and $\beta$ is the Bockstein homomrphism associated with $0 \fromto \ZZZ \xfromto{2} \ZZZ \fromto \ZZZ_2 \fromto 0$ \cite{Hatcher}. In particular, $F_0 \homotopic \CCC P^\infty\times \ZZZ_2$ and $F_1 \homotopic K(\ZZZ, 3) \times K(\ZZZ_2, 1)$.\\
Oriented cobordism as in Ref.\,\cite{Kapustin_Boson} & A variation of the Thom spectrum $MSO$ & See App.\,\ref{app:existing_proposals_generalized_cohomology_theories}. \\
Spin cobordism as in Ref.\,\cite{Kapustin_Fermion} & A variation of the Thom spectrum $MSpin$ & See App.\,\ref{app:existing_proposals_generalized_cohomology_theories}. \\
Kitaev's bosonic proposal \cite{Kitaev_Stony_Brook_2011_SRE_1, Kitaev_Stony_Brook_2013_SRE} & Constructed from physical knowledge & $F_d$ is uniquely determined in low dimensions:\par
$F_d = \begin{cases} K(\ZZZ,2) \homotopic \CCC P^\infty, & d=0, \\ K(\ZZZ,3), & d = 1, \\K(\ZZZ, 4)\times \ZZZ, & d=2, \\K(\ZZZ, 5) \times K(\ZZZ,1) \homotopic K(\ZZZ, 5) \times \SS^1, & d=3. \end{cases}$\par
See App.\,\ref{app:existing_proposals_generalized_cohomology_theories}. \\
Kitaev's fermionic proposal \cite{Kitaev_Stony_Brook_2013_SRE, Kitaev_IPAM} & Constructed from physical knowledge & $F_0 = K(\ZZZ, 2)\times \ZZZ_2 \homotopic \CCC P^\infty \times \ZZZ_2$ is uniquely determined, and $F_{d>0}$ are partially determined. See App.\,\ref{app:existing_proposals_generalized_cohomology_theories}.\\
\hline
\hline
\end{tabular}
}
\label{table:existing_proposals}
\end{table}

\section{Justification of the Hypothesis\label{sec:justification_hypothesis}}

In this section we provide justifications for the Hypothesis from several different angles. We will first give some empirical evidence in Secs.\,\ref{subsec:additivity_functoriality}, \ref{subsec:existing_proposals_special_cases}, \ref{subsec:ubiquity_generalized_cohomology_theories}, and \ref{subsec:rationale_classifying_spaces}. We will then present physical interpretations of $\Omega$-spectrum in Sec.\,\ref{subsec:physical_interpretations_Omega_spectrum}, and comment on the realizability of SPT phases described by generalized cohomology theories in Sec.\,\ref{subsec:realizability}.

\subsection{Additivitiy and functoriality\label{subsec:additivity_functoriality}}

We have seen at the end of Sec.\,\ref{sec:generalized_cohomology_hypothesis} that every generalized cohomology theory is additive and functorial. That is, for any generalized cohomology theory $h$, the set $h^d(BG)$ always has an abelian group structure, and every homomorphism $G_1 \fromto G_2$ induces a map $h^d(BG_2) \fromto h^d(BG_1)$. This is consistent with the abelian group structure of SPT phases defined by stacking (see Sec.\,\ref{subsubsec:additivity} and also Sec.\,\ref{subsubsec:new_definition_SPT_phases}), and the functoriality of SPT phases defined by symmetry forgetting (see Sec.\,\ref{subsubsec:functoriality}).

\subsection{Existing proposals as special cases\label{subsec:existing_proposals_special_cases}}

As noted in Sec.\,\ref{sec:comparison_different_proposals} and elaborated upon in App.\,\ref{app:existing_proposals_generalized_cohomology_theories}, the Hypothesis is satisfied by various existing proposals for the classification of SPT phases \cite{Kitaev_Stony_Brook_2011_SRE_1, Kitaev_Stony_Brook_2013_SRE, Kitaev_IPAM}. These include the Borel group cohomology proposal \cite{Wen_Boson}, the oriented cobordism proposal \cite{Kapustin_Boson}, Kitaev's proposal \cite{Kitaev_Stony_Brook_2011_SRE_1, Kitaev_Stony_Brook_2013_SRE}, and the Freed-Hopkins proposal \cite{Freed_SRE_iTQFT, Freed_ReflectionPositivity} in the bosonic case; and the group supercohomology proposal \cite{Wen_Fermion}, the spin cobordism proposal \cite{Kapustin_Fermion}, Kitaev's proposal \cite{Kitaev_Stony_Brook_2013_SRE, Kitaev_IPAM}, and the Freed-Hopkins proposal \cite{Freed_SRE_iTQFT, Freed_ReflectionPositivity} in the fermionic case.
Furthermore, we have explicitly checked, at least for the Borel group cohomology proposal, that the additive and functorial structures of the generalized cohomology theory $h$ indeed correspond to the stacking and symmetry forgetting of SPT phases (see App.\,\ref{app:additivity_functoriality_group_cohomology_construction}). The fact that existing proposals fit into our framework can be viewed as circumstantial evidence for the validity of the Hypothesis.





\subsection{Adaptation of generalized cohomology theories to incorporate new phases\label{subsec:ubiquity_generalized_cohomology_theories}}

As commented in Sec.\,\ref{sec:comparison_different_proposals}, existing proposals should be thought of as approximations to the true, unknown generalized cohomology theory that gives the complete classification of SPT phases. Suppose we have such an approximation $h$ and we discover afterwards that there is a new SPT phase not captured by $h$. Can we construct a new generalized cohomology theory to incorporate this new phase? Put differently, can a generalized cohomology theory evolve as new SPT phases are discovered?

The answer to these questions is affirmative. Suppose the original theory $h$ is represented by the $\Omega$-spectrum $F$, and that we have discover a new invertible topological order in $d$ dimensions with a $\ZZZ_n$ classification ($n$ can be infinite). Then, we can construct a new generalized cohomology theory $h'$ to incorporate this phase by extending the $\Omega$-spectrum as follows. We define a new sequence of spaces
\begin{eqnarray}
F'_i = \begin{cases}
F_i \times K(\ZZZ_n, i-d), & i \geq d, \\
F_i, & i < d,
\end{cases}\label{product_spectrum}
\end{eqnarray}
for $i \in \ZZZ$, where $K(\ZZZ_n, i-d)$ denotes the $(i-d)$th Eilenberg-MacLane space of $\ZZZ_n$ (see App.\,\ref{subapp:generalized_cohomology_theories}). It can be shown that the spaces $F'_i$ still form an $\Omega$-spectrum, which we use to define the new theory $h'$ that incorporates the new invertible topological order. More generally, we can define $F'_i$ for $i\geq d$ to be the total space of a fiber bundle with base space $K(\ZZZ_n, i-d)$ and fiber $F_i$. Whether we have Eq.\,(\ref{product_spectrum}) or a fiber bundle can be determined from the stacking properties of the new phases we are incorporating. The adaptability of generalized cohomology theories is another hint at the validity of the Hypothesis.

\subsection{Classifying spaces as universal target spaces for gauge theories\label{subsec:rationale_classifying_spaces}}

To further rationalize the Hypothesis, we note that the appearance of classifying spaces $BG$ in the Hypothesis simply reflects the belief that SPT phases can be described using gauge fields at low energies \cite{Wen_Boson}. To understand this, we need to appeal to the following mathematical fact: if $X$ is a topological space, then gauge field configurations over $X$ with gauge group $G$ are in one-to-one correspondence with homotopy classes of maps from $X$ to $BG$. Therefore, $BG$ can be thought of as the universal target space for gauge theories with gauge group $G$. The specification of a cohomology class $[c]$ of $h^d(BG)$ then tells us how each gauge field configuration is to be converted to an amplitude.

As an example, we set $G$ to be $U(1)$ and consider $U(1)$ Chern-Simons theories \cite{2dChiralBosonicSPT}. In this case, $X$ would be a spacetime manifold. The classifying space $BG$ would be $BU(1) = \CCC P^\infty$. Different gauge field configurations $A^\mu(x)$ over the spacetime manifold would be in one-to-one correspondence with maps from $X$ to $\CCC P^\infty$. The specification of a cohomology class $[c]$ of $h^d(BU(1))$ would be the specification of a differential form, e.g. $A \wedge dA$. This form tells us how to convert a gauge field configuration into a number; namely, we compute the integral $\int_X A \wedge dA$.

\subsection{Physical interpretations of $\Omega$-spectrum\label{subsec:physical_interpretations_Omega_spectrum}}

\begin{figure}
\centering
\includegraphics[width=6in]{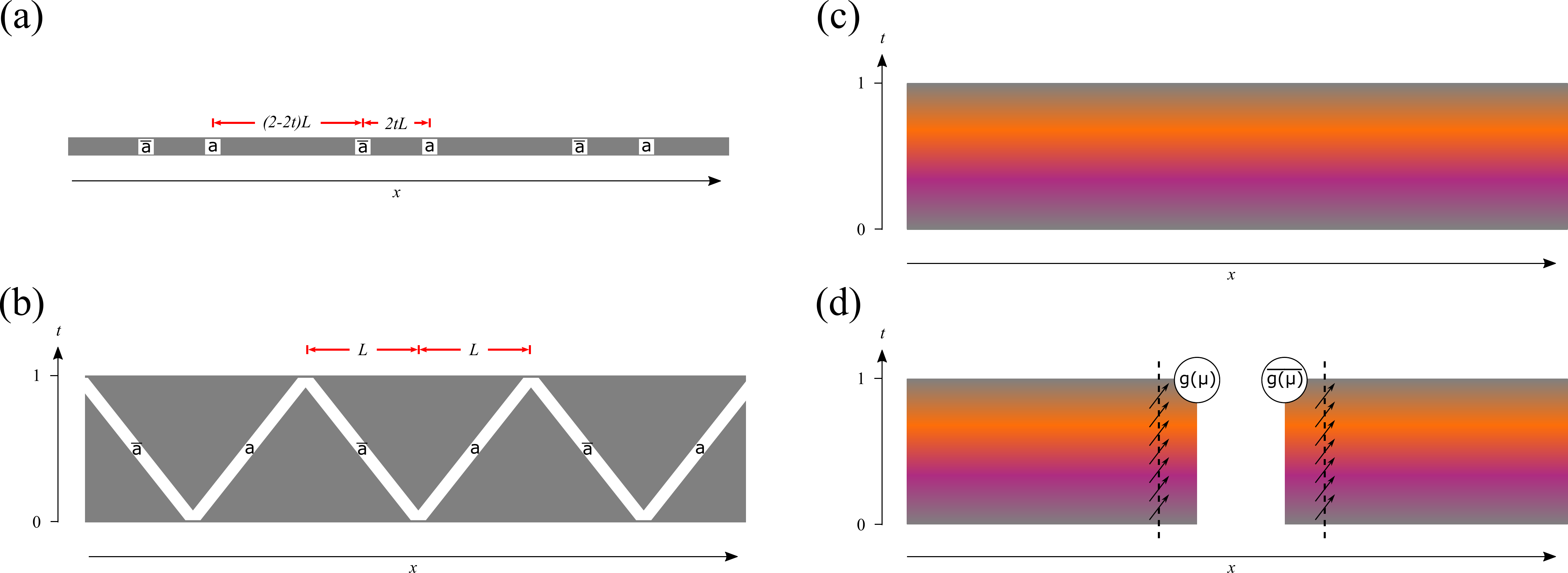}
\caption{(color online). The pumping interpretation of $\Omega$-spectrum. (a) A $(d+1)$-dimensional SRE state $f(a)_t$ constructed from a $d$-dimensional SRE state $a$. (b) The evolution of $f(a)_t$ as $t$ varies from 0 to 1. (c) An arbitrary one-parameter family of $(d+1)$-dimensional SRE states, $\mu(t)$ for $0 \leq t \leq 1$. (d) The pumping of a $d$-dimensional SRE state to the boundary of a $(d+1)$-dimensional system that is cut open, in the adiabatic evolution defined by $\mu$.}
\label{fig:pumping}
\end{figure}

To justify the Hypothesis more concretely, we now explain the physical meaning of the $\Omega$-spectrum $(F_d)_{d\in \ZZZ}$ \cite{Kitaev_Stony_Brook_2011_SRE_1, Kitaev_Stony_Brook_2013_SRE, Kitaev_IPAM}. We recall that an $\Omega$-spectrum is a sequence of spaces $F_d$ with $d\in \ZZZ$ such that the homotopy equivalence condition
\begin{equation}
F_d \homotopic \Omega F_{d+1}
\end{equation}
holds for all $d$. Concretely, this means there exists a pair of maps
\begin{eqnarray}
&&f: F_d \fromto \Omega F_{d+1}, \\
&&g: \Omega F_{d+1} \fromto F_d,
\end{eqnarray}
such that both $f \circ g$ and $g \circ f$ are homotopic to the identity. Therefore, there are two parts to the interpretation of $\Omega$-spectrum. First, we need to interpret the spaces $F_d$ themselves. Second, we need to interpret the maps $f$ and $g$.

According to Refs.\,\cite{Kitaev_Stony_Brook_2011_SRE_1, Kitaev_Stony_Brook_2013_SRE, Kitaev_IPAM}, for a given $d$, $F_d$ can be interpreted as the space of $d$-dimensional SRE states. The basepoint of $F_d$ represents the trivial $d$-dimensional SRE state, i.e. a product state. As for the maps $f$ and $g$, there are two ways to interpret them. One of them is based on pumping a $d$-dimensional state to the boundary of a $(d+1)$-dimensional state \cite{Kitaev_Stony_Brook_2011_SRE_1, Kitaev_Stony_Brook_2013_SRE}. The other is based on viewing a texture of $(d+1)$-dimensional states as a $d$-dimensional domain wall \cite{Kitaev_IPAM}. The two interpretations can be shown to be equivalent (App.\,\ref{subapp:homotopy_equivalence_equivalence}), and compatible with the additivity structure of SPT phases (App.\,\ref{subapp:homotopy_equivalence_compatibility}). In Sec.\,\ref{subsubsec:pumping}, we present the first interpretation. In Sec.\,\ref{subsubsec:domain_wall}, we present the second.


\subsubsection{Pumping interpretation\label{subsubsec:pumping}}

With $F_d$ being the space of $d$-dimensional SRE states, $f$ will now be a map that sends a $d$-dimensional SRE state $a$ to a one-parameter family of $(d+1)$-dimensional SRE states (recall $\Omega F_{d+1}$ is the loop space of $F_{d+1}$). We denote this family of $(d+1)$-dimensional SRE states by $f(a)_t$, for $0 \leq t \leq 1$, which satisfies $f(a)_0 = f(a)_1$. In the pumping interpretation \cite{Kitaev_Stony_Brook_2011_SRE_1, Kitaev_Stony_Brook_2013_SRE}, we set $f(a)_t$ to be the state shown in Fig.\,\ref{fig:pumping}(a). It is obtained by putting copies of $a$ at $x = (2n+t)L$ and copies of the inverse $\bar a$ at $x = (2n-t)L$ for all $n\in \ZZZ$, where $x$ is the additional coordinate the $(d+1)$-dimensional system has compared to $d$-dimensional systems, and $L$ is a length scale much greater than the correlation length $\xi$. As $t$ increases, the separation between $a$'s and $\bar a$'s changes. The evolution of $f(a)_t$ with $t$ is illustrated in Fig.\,\ref{fig:pumping}(b).

Conversely, given a one-parameter family of $(d+1)$-dimensional SRE states, $\mu(t)$ with $0\leq t \leq 1$, the map $g$ will send it to a $d$-dimensional SRE state, $g(\mu)$. To define this state, we take $t$ to be the time coordinate and regard $\mu(t)$ as defining an adiabatic evolution of a $(d+1)$-dimensional system. We will then set $g(\mu)$ to be the $d$-dimensional state that is pumped across the $(d+1)$-dimensional system in the adiabatic evolution. Put differently, if the $(d+1)$-dimensional system is cut open, the $g(\mu)$ will be the $d$-dimensional state that is pumped to the boundary of the $(d+1)$-dimensional system. This is illustrated in Fig.\,\ref{fig:pumping}(c)(d).

In App.\,\ref{subapp:homotopy_equivalence_pumping}, we will show that $f$ and $g$ are indeed homotopy inverses of each other, that is, both $f\circ g$ and $g \circ f$ are homotopic to the identity.

\subsubsection{Domain wall interpretation\label{subsubsec:domain_wall}}

\begin{figure}
\centering
\includegraphics[width=4.5in]{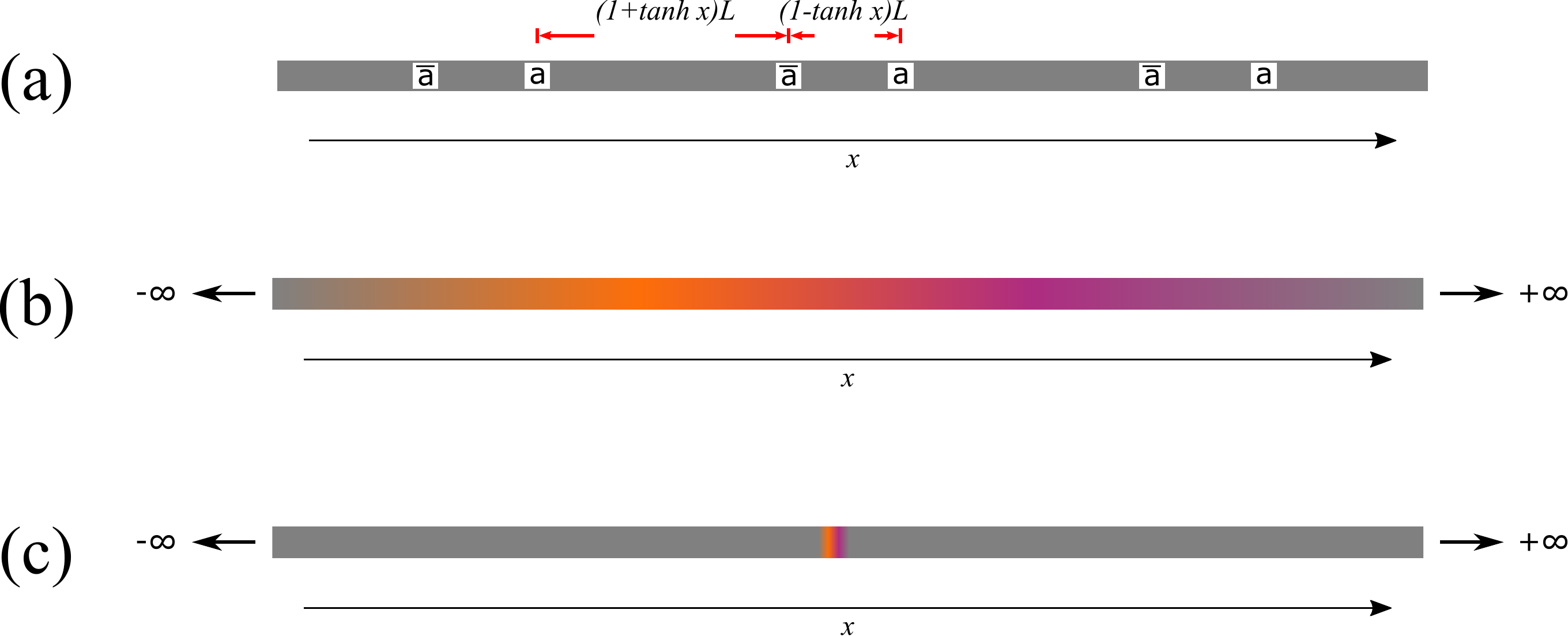}
\caption{(color online). The domain wall interpretation of $\Omega$-spectrum. (a) A $(d+1)$-dimensional SRE state $f(a)_x$ constructed from a $d$-dimensional state $a$. (b) A texture of $(d+1)$-dimensional SRE states, $\mu(x)$ for $-\infty < x < \infty$. (c) The $d$-dimensional domain wall obtained by squeezing the texture in (b).}
\label{fig:domain_wall}
\end{figure}

Just like in the pumpin interpretation, $f$ is again a map that sends a $d$-dimensional SRE state $a$ to a one-parameter family of $(d+1)$-dimensional SRE states. This time, however, we will denote the family of $(d+1)$-dimensional SRE states by $f(a)_x$ for $-\infty < x < \infty$, which satisfies $f(a)_{-\infty} = f(a)_{+\infty}$. We will set $f(a)_x$ to be the state shown in Fig.\,\ref{fig:domain_wall}(a). This is equivalent to the state in Fig.\,\ref{fig:pumping}(a) via the formal change of variable $t = \frac{1 - \tanh x}{2}$.

Conversely, given a one-parameter family of $(d+1)$-dimensional SRE states, $\mu(x)$ for $-\infty < x < \infty$, the map $g$ will send it to a $d$-dimensional SRE state, $g(\mu)$. This time, we will treat $x$ as a spatial coordinate and regard $\mu$ as defining a texture of $(d+1)$-dimensional SRE states, which varies spatially with $x$ and is locally indistinguishable from the $(d+1)$-dimensional SRE state $\mu(x)$ in the vicinity of $x$; see Fig.\,\ref{fig:domain_wall}(b). To define $g(\mu)$, we will then squeeze the texture in the $x$-direction; see Fig.\,\ref{fig:domain_wall}(c). This results in a $d$-dimensional domain wall within the $(d+1)$-dimensional system, and we set $g(\mu)$ to be the $d$-dimensional state that lives on the domain wall.

In App.\,\ref{subapp:homotopy_equivalence_domain_wall}, we will show that $f$ and $g$ are indeed homotopy inverses of each other, that is, both $f\circ g$ and $g \circ f$ are homotopic to the identity.

\subsection{Realizability of phases described by generalized cohomology theories\label{subsec:realizability}}

To see that SPT phases described by generalized cohomology theories are realizable by local Hamiltonians, let us sketch a general construction of such phases for an arbitrary theory $h$. For definiteness, we shall focus on internal symmetries (the issue of spatiotemporal symmetries will be discussed in Sec.\,\ref{subsec:spatiotemporal}). The key to the construction is a decomposition of $h^d(BG)$ via the so-called Atiyah-Hirzebruch spectral sequence \cite{Adams2}, which indicates that $h^d(BG)$ can be decomposed into individual terms that correspond to decorated domain walls \cite{decorated_domain_walls} of various dimensions.

By definition, the Atiyah-Hirzebruch spectral sequence \cite{Adams2} is a family of abelian groups $E^{p,q}_r$ indexed by three integers, $p$, $q$, and $r$, that converge to a graded quotient of $h$ in the limit of large $r$. Concretely, there exists a sequence of subgroups of $h^d(BG)$,
\begin{equation}
0 \subset \cdots \subset A_2 \subset A_1 \subset A_0 \coloneq h^d(BG), \label{filtration}
\end{equation}
with quotients given by
\begin{equation}
A_n / A_{n+1} \isomorphic E_\infty^{n, d-n}, \label{quotient}
\end{equation}
where $E_\infty^{p,q} \coloneq \lim_{r\fromto \infty} E_r^{p,q}$ is the limit of $E_r^{p,q}$ when $r$ is large. Explicitly, the value $E_r^{p,q}$ for $r=2$ is
\begin{equation}
E_2^{p,q} = H^p_{\rm group}\paren{G; h^q(\pt)}. \label{E2}
\end{equation}
To determine $E_r^{p,q}$ for higher $r$, one take the homology of $E_{r-1}^{p,q}$ with respect to a certain differential and proceed inductively in $r$.

\begin{figure}
\centering
\includegraphics[width=5in]{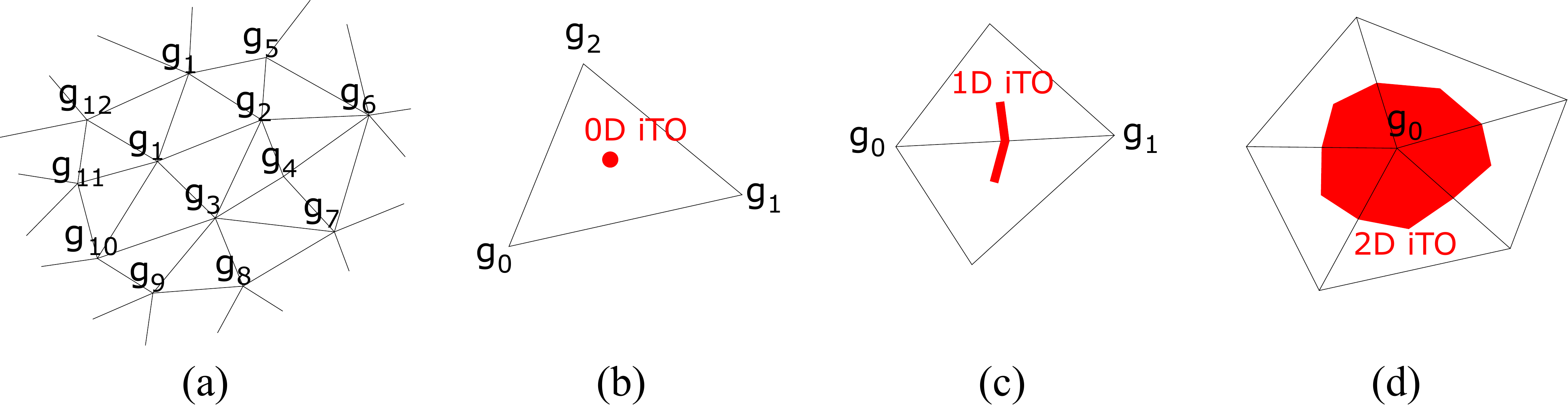}
\caption{(color online). Construction of SPT phases described by generalized cohomology theories. (a) A triangulated spatial manifold with Hilbert space (\ref{CG}) on each vertex, depicted for $d=2$. (b)-(d) A $p$-simplex (vertices labeled $g_0$, $g_1$, etc.) decorated by a $(d-p)$-dimensional invertible topological order (red dot, line, patch, etc.), for (b) $p=2$, (c) $p=1$, and (d) $p=0$, respectively. }
\label{fig:decorated_domain_walls}
\end{figure}

Now, for the sake of argument, let us assume $E_\infty^{p,q} = E_2^{p,q}$, as is often true. (In general, elements of $E_\infty^{p,q}$ can always be represented by elements of $E_2^{p,q}$ in a possibly non-unique fashion.) Eqs.\,(\ref{filtration})(\ref{quotient})(\ref{E2}) would then imply that, as a set,
\begin{equation}
h^d(BG) \overset{\text{as~a~set}}{\homeomorphic} \prod_{p \geq 0} H^p_{\rm group}\paren{G; h^{d-p}(\pt)},
\end{equation}
where $\pt$ is the one-point topological space. Thus to specify an element of $[c] \in h^d(BG)$, it suffices to specify an element
\begin{equation}
[c_p] \in H^p_{\rm group}\paren{G; h^{d-p}(\pt)}
\end{equation}
for each $p$, and each $[c_p]$ can in turn be represented by a function
\begin{equation}
c_p: \underbrace{G \times \cdots \times G}_{p+1\text{~times}} \fromto  h^{d-p}(\pt).
\end{equation}
Crucially, we recognize that $h^{d-p}(\pt)$ is the classification of $(d-p)$-dimensional invertible topological orders; this can be seen by setting $G$ to be the trivial group $0$ in the Hypothesis and noting that $B0 = \pt$ (see Sec.\,\ref{subsec:unification_old_new_definitions_SPT_phases}). Thus for any given $(g_0, g_1, \ldots, g_p)$, $c_p(g_0, g_1, \ldots, g_p)$ is a $(d-p)$-dimensional invertible topological order.
To construct the SPT phase described by the tuple $\paren{c_0, c_1, c_2, \ldots}$, we consider a triangulated spatial manifold $X$ with the Hilbert space
\begin{eqnarray}
\mathscr H = {\rm span} \braces{ \ket{g} | g\in G }, ~~ \braket{g_1}{g_2} = \delta_{g_1, g_2} \label{CG}
\end{eqnarray}
assigned to each vertex; see Fig.\,\ref{fig:decorated_domain_walls}(a). The ground state of the SPT phase will have the form
\begin{eqnarray}
\ket{\Psi} = \sum_{\braces{g_i}} \ket{\braces{g_i}} \otimes \paren{ \bigotimes_{p\geq 0} \ket{\psi_p\paren{\braces{g_i}}} } , \label{ground_state_DDW}
\end{eqnarray}
where the $p$- and $\braces{g_i}$-dependent state $\ket{\psi_p\paren{\braces{g_i}}}$ is constructed as follows. Namely, for each $p$-simplex of $X$ with vertices valued at $(g_0, g_1, \ldots, g_p)$, we place a copy of $c_p(g_0, g_1, \ldots, g_p) \in h^{d-p}(\pt)$ on the $(d-p)$-cell dual to the $p$-simplex. In particular, $\ket{\psi_d\paren{\braces{g_i}}}$ is constructed by placing $0$-dimensional invertible topological orders at the centers of $d$-simplices of $X$; $\ket{\psi_0\paren{\braces{g_i}}}$ is constructed by placing $d$-dimensional invertible topological orders in neighborhoods of vertices of $X$; and $\ket{\psi_p\paren{\braces{g_i}}}$ for $0 < p < d$ is constructed by placing $(d-p)$-dimensional invertible topological orders on $(d-p)$-cells that intersect $p$-simplices of $X$. This is illustrated in Fig.\,\ref{fig:decorated_domain_walls}(b)-(d).

Thus, we see that $H^p_{\rm group}\paren{G; h^{d-p}(\pt)}$ corresponds to decorated domain walls of $d-p$ dimensions. By letting the domain walls fluctuate via the summation in Eq.\,(\ref{ground_state_DDW}), we arrive at a state that respects the symmetry $G$. This state is realizable by local Hamiltonian because the invertible topological orders $h^{d-p}(\pt)$ by assumption are.

\subsection{Applicability to spatiotemporal symmetries\label{subsec:spatiotemporal}}

Kitaev's original proposal to describe SPT phases using generalized cohomology theories was focused on internal symmetries \cite{Kitaev_Stony_Brook_2011_SRE_1, Kitaev_Stony_Brook_2013_SRE, Kitaev_IPAM}.  However, there has been a growing body of evidence \cite{Keyserlingk_Floquet, Else_Floquet, Potter_Floquet, Wen_1d, Cirac, Wen_sgSPT_1d, Hsieh_sgSPT, Hsieh_sgSPT_2, You_sgSPT, SPt, Cho_sgSPT, Yoshida_sgSPT, Jiang_sgSPT, Thorngren_sgSPT} that the Generalized Cohomology Hypothesis should apply to spatiotemporal symmetries, too. Specifically, if $h$ is the generalized cohomology theory that classifies SPT phases with internal symmetries, then the classification of $d$-dimensional SPT phases with spatiotemporal symmetry $G$ will be given by $h^d(BG)$ for the same theory $h$, where elements of $G$ that reverse the orientation of spacetime are treated antiunitarily.

The above claim is supported by various known examples. Indeed, in 1 dimension, Refs.\,\cite{Wen_1d, Cirac, Wen_sgSPT_1d} classified bosonic SPT phases with internal symmetry $G$ (time reversal allowed) and translation symmetry $\ZZZ$, and the results are consistent with the second group cohomology $H^2_{\rm Borel}(G \times \ZZZ;U(1))$. Ref.\,\cite{SPt} classified bosonic SPT phases with composite inversion-spin rotation symmetry, and the result is consistent with the second group cohomology $H^2_{\rm Borel}(\ZZZ_2^T; U(1))$. In 2 dimensions, Ref.\,\cite{You_sgSPT} classified bosonic SPT phases with $T \times I$ and $U(1) \times I$ symmetries, where $T$ is time-reversal and $I$ is inversion, and the results are consistent with $H^3_{\rm Borel}(\ZZZ_2^T \times \ZZZ_2; U(1))$ and $H^3_{\rm Borel}(U(1) \times \ZZZ_2; U(1))$, respectively. Ref.\,\cite{Yoshida_sgSPT} classified bosonic SPT phases with $\ZZZ_n \times R$, $\ZZZ_n \rtimes R$, $U(1) \times R$, and $U(1) \rtimes R$ (also considered by Refs.\,\cite{Hsieh_sgSPT, Cho_sgSPT}) symmetries, where $\ZZZ_n$ or $U(1)$ is an internal symmetry and $R$ is a mirror reflection, and the results are consistent with the third group cohomology $H^3_{\rm Borel}(\ZZZ_n \times \ZZZ_2^T;U(1))$, $H^3_{\rm Borel}(\ZZZ_n \rtimes \ZZZ_2^T;U(1))$, $H^3_{\rm Borel}(U(1) \times \ZZZ_2^T;U(1))$, and $H^3_{\rm Borel}(U(1) \rtimes \ZZZ_2^T;U(1))$, respectively. In 3 dimensions, Ref.\,\cite{Hermele_torsor} classified bosonic SPT phases with one or two mirror symmetries, where $E_8$ phases were excluded, and the results are consistent with $H^4_{\rm Borel}(\ZZZ_2^T;U(1))$ and $H^4_{\rm Borel}(\ZZZ_2^T \times \ZZZ_2^T;U(1))$, respectively. 

More recently, there has been significant progress in establishing the claim in full generality. In particular, Ref.\,\cite{Jiang_sgSPT} presented explicit tensor network constructions for bosonic SPT phases in $d=$1, 2, and 3 dimensions with symmetry $G$ corresponding to all elements of $H^{d+1}_{\rm Borel}(G;U(1))$, where $G$ can be either internal or spatial. Furthermore, Ref.\,\cite{Thorngren_sgSPT} introduced a notion of crystalline gauge field and argued that the classification of bosonic or fermionic SPT phases with symmetry $G$ is the same for internal symmetries and spatial symmetries. Finally, Refs.\,\cite{Keyserlingk_Floquet, Else_Floquet, Potter_Floquet} showed that Floquet eigenstates of Floquet systems have an additional $\ZZZ$ symmetry generated by the Floquet unitary that their classification are the same as if $\ZZZ$ is an internal symmetry.

The aforementioned results strongly suggest that we can dispense with the restriction that $G$ is an internal symmetry, which is what we will do in the rest of the paper.

\section{Consequences of the Hypothesis and Comparison with the Literature \label{sec:consequences_hypothesis_physical_implications}}

In this section, we will present some simple physical consequences of the Hypothesis, and verify that they are consistent with the literature. These physical consequences follow readily from mathematical theorems that can be derived from the Hypothesis, which have been collected in App.\,\ref{app:proofs}.
In what follows, we will denote by $h$ the generalized cohomology theory that classifies SPT phases, by $\tilde h$ the corresponding reduced theory, and by $\paren{F_d}_{d\in \ZZZ}$ the representing $\Omega$-spectrum.



\subsection{Unification of old and new definitions of SPT phases \label{subsec:unification_old_new_definitions_SPT_phases}}

In Sec.\,\ref{subsec:definition_SPT_phases}, we reviewed the old definition of SPT phases \cite{Wen_Definition, Cirac}, and formalized a new definition of SPT phases based on ideas in Refs.\,\cite{Kitaev_Stony_Brook_2011_SRE_2, Kitaev_Stony_Brook_2013_SRE, Kapustin_Boson, Freed_SRE_iTQFT, Freed_ReflectionPositivity, McGreevy_sSourcery}. The old definition is in terms of deformability to product states, whereas the new one is in terms of invertibility of phases, which is closely related and potentially equivalent \cite{Kitaev_Stony_Brook_2011_SRE_2, Kitaev_Stony_Brook_2013_SRE, Kapustin_Boson, Freed_SRE_iTQFT, Freed_ReflectionPositivity, McGreevy_sSourcery} to the condition of unique ground state on arbitrary spatial slice and, in two dimensions, the condition of no nontrivial anyonic excitations.

We have seen in Sec.\,\ref{subsubsec:comparison_definition_SPT_phases} that $d$-dimensional $G$-protected SPT phases in the old sense form a subset of those in the new sense.
Here we would like to make their relationship more explicit.

\begin{framed}\begin{rslt}
If SPT phases (in the new sense) are classified by a generalized cohomology theory $h$ as in the Hypothesis, then $d$-dimensional invertible topological orders (i.e.\,$d$-dimensional SPT phases protected by the trivial symmetry group) are classified by $h^d(\pt)$.\label{rslt:invertible_topological_orders}
\end{rslt}\end{framed}

\begin{pf}
This is a simple application of the Hypothesis: set $G$ to be the trivial group $0$ and recall that the classifying space of the trivial group, $B0$, is homotopy equivalent to the one-point set, $\pt$.
\qed\end{pf}

The merit of Physical Result \ref{rslt:invertible_topological_orders} lies in the fact that the value on a point, $h^d\paren{\pt}$, is basic to any generalized cohomology theory $h$. Given an $h$, $h^d\paren{\pt}$ is usually the simplest to compute. Conversely, from $h^d\paren{\pt}$, one can deduce important information about $h^d\paren{X}$ for any $X$ (which was the basis of the approach in Refs.\,\cite{Kitaev_Stony_Brook_2011_SRE_1, Kitaev_Stony_Brook_2013_SRE, Kitaev_IPAM}; see Apps.\,\ref{subapp:Kitaev_bosonic_proposal} and \ref{subapp:Kitaev_fermionic_proposal}).

\begin{framed}\begin{rslt}\label{rslt:SPT_phases_old_sense}
If SPT phases in the new sense are classified by an unreduced generalized cohomology theory $h$ as in the Hypothesis, then SPT phases in the old sense are classified by the corresponding reduced theory $\tilde h$, where the same remarks about additivity and functoriality apply.
\end{rslt}\end{framed}

\begin{pf}
As remarked in Sec.\,\ref{subsubsec:comparison_definition_SPT_phases}, SPT phases in the old sense are precisely those SPT phases in the new sense that, by forgetting the symmetry, represent the trivial topological order. Thus, by the functoriality part of the Hypothesis, they are precisely the kernel of the map $p$ in Lemma \ref{lem:relationship_reduced_unreduced_generalized_cohomology_theories}, which is naturally isomorphic to $\tilde h^d\paren{BG}$ by exactness.
\qed\end{pf}

We would like to point out that the converse of Physical Result \ref{rslt:SPT_phases_old_sense} is not automatic. That is, had we formulated the Hypothesis for SPT phases in the old sense in terms of $\tilde h$, then it would not have been nearly as easy, if not impossible, to deduce that SPT phases in the new sense are classified by $h$.

\begin{framed}\begin{rslt}\label{rslt:reduced_unreduced_isomorphism}
There is a natural isomorphism of abelian groups,
\begin{eqnarray}
\braces{\parbox{3.4cm}{$d$-dimensional $G$-protected SPT phases in the new sense}} 
\isomorphic
\braces{\parbox{3.4cm}{$d$-dimensional $G$-protected SPT phases in the old sense}} 
\oplus
\braces{\parbox{2.4cm}{$d$-dimensional invertible topological orders}}.
\end{eqnarray}
\end{rslt}\end{framed}

\begin{pf}
We have seen in Physical Result \ref{rslt:invertible_topological_orders} that $h^d\paren{\pt} \isomorphic h^d\paren{B0}$ classifies $d$-dimensional invertible topological orders, and in Physical Result \ref{rslt:SPT_phases_old_sense} that $\tilde h^d\paren{BG}$ classifies $d$-dimensional $G$-protected SPT phases in the old sense. The desired natural isomorphism then follows from Corollary \ref{cor:relationship_reduced_unreduced_generalized_cohomology_theories}.
\qed\end{pf}

We note that Physical Result \ref{rslt:reduced_unreduced_isomorphism} is consistent with Ref.\,\cite{Kapustin_Fermion}. The latter considered specifically the spin cobordism proposal $\Omega^d_{\rm Spin}$ for the classification of fermionic SPT phases, which is an example of generalized cohomology theories. Ref.\,\cite{Kapustin_Fermion} pointed out that $\Omega^d_{\rm Spin}(BG) \isomorphic \tilde \Omega^d_{\rm Spin}(BG) \oplus \Omega^d_{\rm Spin}(\pt)$, which translates precisely to Physical Result \ref{rslt:reduced_unreduced_isomorphism}.

The next result gives more information about the isomorphism in Physical Result \ref{rslt:reduced_unreduced_isomorphism}.

\begin{framed}\begin{rslt}
The isomorphism in Physical Result \ref{rslt:reduced_unreduced_isomorphism} is such that the canonical injection
\begin{eqnarray}
i: \braces{\parbox{4.3cm}{$d$-dimensional $G$-protected SPT phases in the old sense}} 
\oneone
\braces{\parbox{4.3cm}{$d$-dimensional $G$-protected SPT phases in the new sense}}
\end{eqnarray}
is given by inclusion, and that the canonical projection
\begin{eqnarray}
p:
\braces{\parbox{4.3cm}{$d$-dimensional $G$-protected SPT phases in the new sense}} 
\onto
\braces{\parbox{3.7cm}{$d$-dimensional invertible topological orders}}
\end{eqnarray}
is given by forgetting symmetry $G$.\label{rslt:reduced_unreduced_maps}
\end{rslt}\end{framed}

\begin{pf}
Recall that Corollary \ref{cor:relationship_reduced_unreduced_generalized_cohomology_theories} came from Lemma \ref{lem:relationship_reduced_unreduced_generalized_cohomology_theories}.
We have seen in Physical Result \ref{rslt:invertible_topological_orders} that $h^d\paren{B0}$ classifies $d$-dimensional invertible topological orders, and in Physical Result \ref{rslt:SPT_phases_old_sense} that $\kernel p$ classifies $d$-dimensional $G$-protected SPT phases in the old sense.
The first half of Physical Result \ref{rslt:reduced_unreduced_maps} is then trivial, whereas the second half follows from the functoriality part of the Hypothesis.
\qed\end{pf}

\subsection{Strong and weak topological indices in the interacting world\label{subsec:strong_weak_topological_indices_interacting_world}}

As observed already in the 1-dimensional bosonic case \cite{Wen_1d, Cirac}, the classification of SPT phases can be modified by an additionally imposed discrete spatial translational symmetry. Two translationally invariant systems that are inequivalent in the presence of translational symmetry may be deformable to each other via non-translationally invariant paths. A priori, it is also not obvious that there are no intrinsically non-translationally invariant SPT phases.

Here we would like to clarify the relationship between classifications in the presence and absence of discrete translational symmetry. We will begin with discrete translation $\ZZZ$ in only one direction and take $G$ to be a symmetry it commutes with (hence forming $\ZZZ\times G$).

\begin{framed}\begin{rslt}\label{rslt:strong_weak_isomorphism}
Let $\ZZZ$ act as discrete spatial translations. Then there is a natural isomorphism of abelian groups,
\begin{eqnarray}
\braces{\parbox{2.8cm}{$d$-dimensional $(\ZZZ \times G)$-protected SPT phases}}
\isomorphic
\braces{\parbox{2.9cm}{$(d-1)$-dimensional $G$-protected SPT phases}}
\oplus
\braces{\parbox{2.2cm}{$d$-dimensional $G$-protected SPT phases}}.\label{strong_weak_topological_indices_interacting_world_isomorphism}
\end{eqnarray}
\end{rslt}\end{framed}

\begin{pf}
This is an immediate consequence of the second isomorphism in Corollary \ref{cor:generalized_Kunneth_formula}.
\qed\end{pf}

We note that Physical Result \ref{rslt:strong_weak_isomorphism} is consistent with Refs.\,\cite{Wen_1d, Cirac, Wen_sgSPT_1d}. These references considered 1-dimensional bosonic SPT phases with an internal symmetry $G$ and possibly a translational symmetry $\ZZZ$. It was concluded therein that the classification without translational symmetry is $H^2_{\rm Borel}\paren{G; U(1)}$, whereas the classification with translation symmetry is $H^1_{\rm Borel}\paren{G;U(1)} \oplus H^2_{\rm Borel}\paren{G; U(1)}$. We see that this agrees with Physical Result \ref{rslt:strong_weak_isomorphism} because the additional term $H^1_{\rm Borel}\paren{G;U(1)}$ is precisely the classification of 0-dimensional bosonic SPT phases, which are labeled by isomorphism classes of 1-dimensional (non-projective) representations of $G$.


The next two results give more information about the isomorphism.

\begin{framed}\begin{rslt}
The isomorphism in Physical Result \ref{rslt:strong_weak_isomorphism} is such that the canonical projection
\begin{eqnarray}
\beta: \braces{\parbox{3.6cm}{$d$-dimensional $(\ZZZ \times G)$-protected SPT phases}}
\onto
\braces{\parbox{3.4cm}{$d$-dimensional $G$-protected SPT phases}}.
\end{eqnarray}
is given by forgetting translational symmetry.\label{rslt:strong_weak_projection}
\end{rslt}\end{framed}

\begin{pf}
Recall that Corollary \ref{cor:generalized_Kunneth_formula} came from Proposition \ref{prp:generalized_Kunneth_formula}. The claim then follows from the funtoriality part of the Hypothesis.
\qed\end{pf}

\begin{framed}\begin{rslt}
It seems plausible that the isomorphism in Physical Result \ref{rslt:strong_weak_isomorphism} is such that the canonical injection
\begin{eqnarray}
\alpha:
\braces{\parbox{3.5cm}{$(d-1)$-dimensional $G$-protected SPT phases}}
\oneone
\braces{\parbox{3.6cm}{$d$-dimensional $(\ZZZ \times G)$-protected SPT phases}}
\end{eqnarray}
is given by the layering construction where one produces a $d$-dimensional $\paren{\ZZZ \times G}$-symmetric system by layering identical copies of a $(d-1)$-dimensional $G$-symmetric system.\label{rslt:strong_weak_injection}
\end{rslt}\end{framed}

\begin{ag}
A special case of Physical Result \ref{rslt:strong_weak_isomorphism} has been observed in the group cohomology classification of 1-dimensional bosonic SPT phases, where $\alpha$ is indeed given by such a layering construction; see Sec.\,VB4 of Ref.\,\cite{Wen_1d} and Sec.\,IVC3 of Ref.\,\cite{Cirac}. As for arbitrary generalized cohomology theories in arbitrary dimensions, a field-theoretic construction is proposed in App.\,\ref{app:field_theoretic_argument_weak_index_interpretation} to justify this interpretation of $\alpha$.
\qed\end{ag}

Therefore, in parallel with the notions of strong and weak topological insulators \cite{weak_TI}, we can divide $d$-dimensional $\paren{\ZZZ \times G}$-protected SPT phases into strong ones and weak ones, according to whether they can be produced through the layering construction, or equivalently whether they become trivial upon forgetting the translational symmetry.
We shall call the first and second direct summands in the right-hand side of Eq.\,(\ref{strong_weak_topological_indices_interacting_world_isomorphism}) the \emph{weak topological index} and the \emph{strong topological index}, respectively.
Their counterparts in Ref.\,\cite{weak_TI} would be $\ZZZ_2 \oplus \ZZZ_2 \oplus \ZZZ_2$ and $\ZZZ_2$, respectively.
Despite the similarities, there is a crucial distinction between our Physical Results \ref{rslt:strong_weak_isomorphism}-\ref{rslt:strong_weak_injection} and Ref.\,\cite{weak_TI}: the former deal with possibly interacting bosonic or fermionic systems, whereas the latter dealt with free fermion systems.

The next two addenda tell us how Physical Result \ref{rslt:strong_weak_isomorphism} interacts with Physical Result \ref{rslt:reduced_unreduced_isomorphism}.

\begin{framed}\begin{framednameddef}[Addendum to Physical Result \ref{rslt:strong_weak_projection}]
$\beta$ does not mix different invertible topological orders.
In particular, it takes SPT phases in the old sense to SPT phases in the old sense.
\end{framednameddef}\end{framed}

\begin{pf}
The invertible topological order an SPT phase represents is obtained by forgetting all symmetry operations. We have seen that $\beta$ is given by forgetting $\ZZZ$. Since forgetting first $\ZZZ$ and then $G$ is equivalent to forgetting $\ZZZ \times G$ in one step, $\beta$ must preserve invertible topological orders.
The second half of the addendum also follows independently from the commutativity of the second square in Eq.\,(\ref{generalized_Kunneth_formula_diagram}).
\qed\end{pf}

\begin{framed}\begin{framednameddef}[Addendum to Physical Result \ref{rslt:strong_weak_injection}]
$\alpha$ can never produce $d$-dimensional $(\ZZZ \times G)$-protected SPT phases with nontrivial invertible topological orders.
\end{framednameddef}\end{framed}

\begin{pf}
This follows from the commutativity of the first square in Eq.\,(\ref{generalized_Kunneth_formula_diagram}).
\qed\end{pf}

This addendum is independent of the arguments for Physical Result \ref{rslt:strong_weak_injection}. If one believes in those arguments, however, then what the addendum is saying is that the layering construction can never produce nontrivial invertible topological orders.

Now, let us spell out the implications of Physical Results \ref{rslt:strong_weak_isomorphism}-\ref{rslt:strong_weak_injection} in detail.

\begin{framed}\begin{rslt}\label{rslt:strong_weak_detail}
Let $\ZZZ$ act as discrete spatial translations and assume the interpretation of $\alpha$ in Physical Result \ref{rslt:strong_weak_injection} is valid. Then we have the following:
\begin{enumerate}[(i)]
\item Every $d$-dimensional $G$-protected SPT phase can be canonically represented by a $d$-dimensional $\paren{\ZZZ\times G}$-protected SPT phase.

\item The layering construction turns equivalent $(d-1)$-dimensional systems into equivalent $d$-dimensional systems, and is hence well-defined at the level of phases.

\item The layering construction commutes with addition of phases and replacement of $G$.

\item The layering construction turns trivial, nontrivial, or distinct $(d-1)$-dimensional $G$-protected SPT phases into trivial, nontrivial, or distinct $d$-dimensional $(\ZZZ\times G)$-protected SPT phases, respectively.

\item Every $d$-dimensional $(\ZZZ\times G)$-protected SPT phase obtained through the layering construction becomes trivial upon forgetting $\ZZZ$.

\item Every $d$-dimensional $(\ZZZ\times G)$-protected SPT phase that becomes trivial upon forgetting $\ZZZ$ can be obtained through the layering construction.

\item If two $d$-dimensional $(\ZZZ\times G)$-protected SPT phases become the same phase upon forgetting $\ZZZ$, then their difference can be obtained through the layering construction.

\item A $(\ZZZ\times G)$-protected SPT phase is uniquely determined by its strong and weak topological indices, and every combination of strong and weak topological indices is allowed.
\end{enumerate}
\end{rslt}\end{framed}

\begin{pf}
All statements follow from the exactness of the second row of Eq.\,(\ref{generalized_Kunneth_formula_diagram}), except for the one about replacement of $G$, which depends on naturality, and the one about canonical representative, which depends on splitting.
\qed\end{pf}

\subsection{Hierarchy of strong and weak topological indices\label{subsec:hierarchy_strong_weak_topological_indices}}

\begin{figure}
    \centering
    \includegraphics[width=5in]{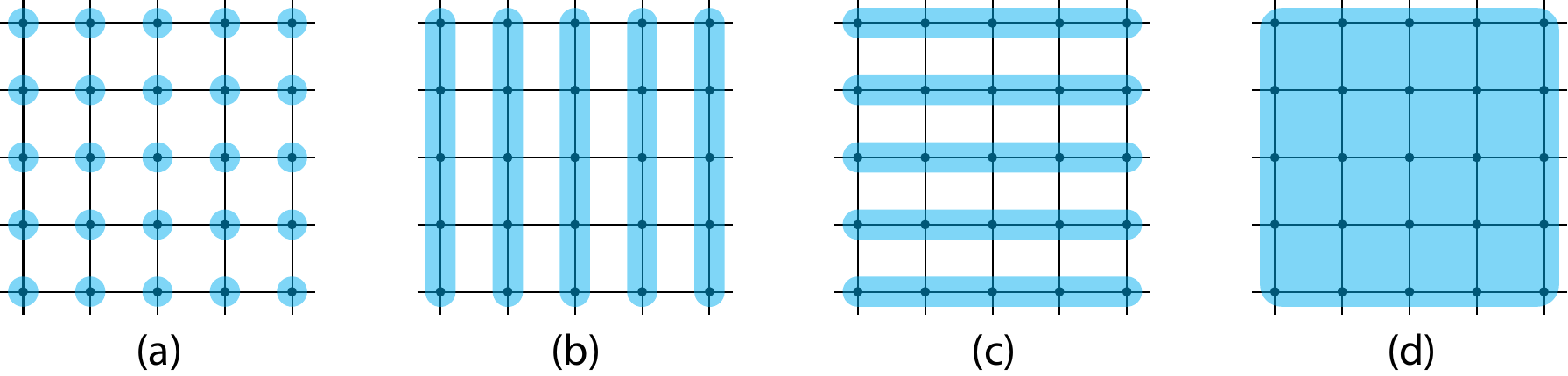}
    \caption{$d$-dimensional SPT phases with internal symmetry $G$ and translation symmetry $\ZZZ \times \ZZZ$ in two linearly independent directions, depicted for $d=2$. (a) A very weak $d$-dimensional SPT phase arising from layering identical copies of a $(d-2)$-dimensional SPT phase two-dimensionally. (b)(c) A weak $d$-dimensional SPT phase arising from layering identical copies of a $(d-1)$-dimensional SPT phase in the (b) first and (c) second directions of translation, respectively. (d) A strong $d$-dimensional SPT phase that is nontrivial independently of the translation symmetry.}
    \label{fig:hierarchy}
\end{figure}

We now perform a sanity check on Physical Results \ref{rslt:strong_weak_isomorphism}-\ref{rslt:strong_weak_detail} by imposing discrete spatial translational symmetry in multiple linearly independent directions. With translational symmetry in two directions, for example, we have
\begin{eqnarray}
\braces{\parbox{2.4cm}{$d$-dim $(\ZZZ \times \ZZZ \times G)$-SPT phases}}
&\isomorphic&
\braces{\mbox{$(d-1)$-dim $\paren{\ZZZ \times G}$-SPT phases}}
\oplus
\braces{\mbox{$d$-dim $\paren{\ZZZ \times G}$-SPT phases}}~~~ \nonumber\\
&\isomorphic&
\braces{\parbox{1.7cm}{$(d-2)$-dim $G$-SPT phases}} 
\oplus
\braces{\parbox{1.7cm}{$(d-1)$-dim $G$-SPT phases}}
\oplus
\braces{\parbox{1.7cm}{$(d-1)$-dim $G$-SPT phases}} 
\oplus
\braces{\parbox{1.1cm}{$d$-dim $G$-SPT phases}}.
\end{eqnarray}
We note that this is consistent with one's physical intuition. In the last line, the last direct summand is a strong index arising from forgetting translational symmetry in both directions; it corresponds to $d$-dimensional SPT phases that are nontrivial independently of translations [Fig.\,\ref{fig:hierarchy}(d)]. The second and third direct summands are weak indices arising from layering identical copies of $(d-1)$-dimensional SPT phases in two ways; they correspond to $d$-dimensional SPT phases that are nontrivial due to the existence of one of the translations [Fig.\,\ref{fig:hierarchy}(b)(c)]. The first direct summand is a ``very weak" index arising from layering identical copies of $(d-2)$-dimensional SPT phases two-dimensionally; it corresponds to $d$-dimensional SPT phases that are nontrivial due to the existence of both translations [Fig.\,\ref{fig:hierarchy}(a)].


The decomposition above can be generalized to translation in $n$ directions in a straightforward fashion.

\begin{framed}\begin{rslt}\label{rslt:hierarchy}
Let $\ZZZ^n$ act as discrete spatial translations in $n$ linearly independent directions. Then there is a natural isomorphism of abelian groups,
\begin{eqnarray}
\braces{\mbox{$d$-dim $(\ZZZ^n \times G)$-SPT phases}} 
&\isomorphic&
\braces{\mbox{$(d-n)$-dim $G$-SPT phases}} \nonumber\\
&\oplus&
\underbrace{
\braces{\parbox{2.2cm}{$(d-n+1)$-dim $G$-SPT phases}} \oplus \cdots \oplus \braces{\parbox{2.2cm}{$(d-n+1)$-dim $G$-SPT phases}}
}_{\text{${n \choose n-1} = n$~times}}
\nonumber\\
&\cdots&
\nonumber\\
&\oplus&
\underbrace{
\braces{\parbox{2.1cm}{$(d-k)$-dim $G$-SPT phases}} \oplus \cdots \oplus \braces{\parbox{2.1cm}{$(d-k)$-dim $G$-SPT phases}}
}_{\text{$n \choose k$~times}}
\nonumber\\
&\cdots&
\nonumber\\
&\oplus&
\underbrace{
\braces{\parbox{2.1cm}{$(d-1)$-dim $G$-SPT phases}} \oplus \cdots \oplus \braces{\parbox{2.1cm}{$(d-1)$-dim $G$-SPT phases}}
}_{\text{${n \choose 1} = n$~times}}
\nonumber\\
&\oplus&
\braces{\mbox{$d$-dim $G$-SPT phases}},
\end{eqnarray}
where ${n \choose k} \coloneq \frac{n!}{k! (n-k)!}$.
\end{rslt}\end{framed}

\begin{pf}
Iterate Physical Result \ref{rslt:strong_weak_isomorphism}.
\qed\end{pf}

We thus see a hierarchy of topological indices in different codimensions. There is a single strong topological index, in $0$ codimension (i.e.\,$d$ dimensions), which arises from forgetting translational symmetry in all $n$ directions. There are $n \choose k$ weak topological indices in $k$ codimensions (i.e.\,$d-k$ dimensions), which correspond to layering identical copies of $k$-codimensional phases in $n \choose k$ different ways.

\subsection{Pumping, Floquet eigenstates, and classification of Floquet SPT phases\label{subsec:pumping_Floquet_eigenstates_classification_Floquet_SPT_phases}}

Now, let us reinterpret the $\ZZZ$ in Physical Results \ref{rslt:strong_weak_isomorphism}-\ref{rslt:strong_weak_detail} as a discrete temporal translational symmetry. Accordingly, we will reinterpret a $\paren{\ZZZ\times G}$-protected SPT phase to be a $G$-protected (interacting) Floquet SPT phase. On the other hand, a $G$-protected SPT phase will mean a stationary SPT phase with symmetry $G$.

To clarify our terminology, we define a $G$-protected Floquet SPT phase to be a deformation class of Floquet eigenstates \cite{Else_Floquet}; this is to be contrasted with a deformation class of periodic Hamiltonians. Here, a Floquet eigenstate is an eigenstate of the Floquet unitary $\exp\brackets{-i \int_0^T \hat H(t) dt}$. In principle, different periodic Hamiltonians can have common Floquet eigenstates, and it is irrelevant to us which periodic Hamiltonian a Floquet eigenstate comes from. Also, each periodic Hamiltonian has an ensemble of Floquet eigenstates, but we are only interested in individual Floquet eigenstates not the whole ensemble of Floquet eigenstates. An individual Floquet eigenstate is invariant under both the Floquet unitary $\exp\brackets{-i \int_0^T \hat H(t) dt}$ and the $G$-action, which is why it effectively has a $\ZZZ \times G$ symmetry \cite{Else_Floquet}.


\begin{framed}\begin{rslt}\label{rslt:Floquet_isomorphism}
Let $G$ act in a way that commutes with the group $\ZZZ$ of discrete temporal translations. There is a natural isomorphism of abelian groups,
\begin{eqnarray}
\braces{\parbox{4.1cm}{$d$-dimensional $G$-protected Floquet SPT phases}}
\isomorphic
\braces{\parbox{3.5cm}{$(d-1)$-dimensional $G$-protected SPT phases}}
\oplus
\braces{\parbox{3.4cm}{$d$-dimensional $G$-protected SPT phases}}.\label{pumping_Floquet_eigenstates_classification_Floquet_SPT_phases_isomorphism}
\end{eqnarray}
\end{rslt}\end{framed}

\begin{pf}
Same as Physical Result \ref{rslt:strong_weak_isomorphism}.
\qed\end{pf}

\begin{framed}\begin{rslt}\label{rslt:Floquet_projection}
The isomorphism in Physical Result \ref{rslt:Floquet_isomorphism} is such that the canonical projection
\begin{eqnarray}
\beta: \braces{\parbox{4.1cm}{$d$-dimensional $G$-protected Floquet SPT phases}}
\onto
\braces{\parbox{3.4cm}{$d$-dimensional $G$-protected SPT phases}}.
\end{eqnarray}
is given by forgetting temporal translational symmetry.
\end{rslt}\end{framed}

\begin{pf}
Same as Physical Result \ref{rslt:strong_weak_projection}.
\qed\end{pf}

\begin{framed}\begin{rslt}\label{rslt:Floquet_projection_2}
It seems plausible that the isomorphism in Physical Result \ref{rslt:Floquet_isomorphism} is such that the canonical projection
\begin{eqnarray}
\gamma: 
\braces{\parbox{4.1cm}{$d$-dimensional $G$-protected Floquet SPT phases}}
\onto
\braces{\parbox{3.5cm}{$(d-1)$-dimensional $G$-protected SPT phases}}
\end{eqnarray}
is given by measuring what $(d-1)$-dimensional $G$-protected SPT phase is pumped across an imaginary cut in a $d$-dimensional system in one Floquet cycle.
\end{rslt}\end{framed}

\begin{ag}
A special case of Physical Result \ref{rslt:Floquet_isomorphism} has been observed in the classification of 1-dimensional Floquet SPT phases within the group cohomology framework \cite{Keyserlingk_Floquet, Else_Floquet, Potter_Floquet}, where it was argued that $\gamma$ should have such a pumping interpretation, at least when $G$ is finite abelian.
\qed\end{ag}

We note that Physical Results \ref{rslt:Floquet_isomorphism}-\ref{rslt:Floquet_projection_2} are consistent with Refs.\,\cite{Keyserlingk_Floquet, Else_Floquet, Potter_Floquet}. It was found in those references that the classification of 1-dimensional bosonic $G$-protected Floquet SPT phases are classified by $H^2_{\rm Borel}\paren{G;U(1)} \oplus H^1_{\rm Borel}\paren{G;U(1)}$. We see that this agrees with Physical Result \ref{rslt:Floquet_isomorphism} because $H^2_{\rm Borel}\paren{G;U(1)}$ is the classification of 1D stationary bosonic SPT phases \cite{Wen_Boson}, whereas $H^1_{\rm Borel}\paren{G;U(1)}$ is the classification of 0-dimensional sataionary bosonic SPT phases, which are labeled by isomorphism classes of 1-dimensional (non-projective) representations of $G$. Furthermore, Refs.\,\cite{Keyserlingk_Floquet, Else_Floquet, Potter_Floquet} argued that the two terms $H^2_{\rm Borel}\paren{G;U(1)}$ and $H^1_{\rm Borel}\paren{G;U(1)}$ can be interpreted as the stationary SPT phase obtained by forgetting time-translation symmetry, and the 0-dimensional stationary SPT phase pumped to the boundary of the 1-dimensional system in one Floquet cycle. We see that these interpretations agree with Physical Results \ref{rslt:Floquet_projection} and \ref{rslt:Floquet_projection_2}. Finally, Ref.\,\cite{Else_Floquet} noted that the arguments that led to their conclusions generalize to fermionic systems and higher dimensions. This is consistent with Physical Result \ref{rslt:Floquet_isomorphism}, which claims that the decomposition holds for arbitrary $d$ and fermionic as well as bosonic systems.

The next two addenda tell us how Physical Result \ref{rslt:Floquet_isomorphism} interacts with Physical Result \ref{rslt:reduced_unreduced_isomorphism}.

\begin{framed}\begin{framednameddef}[Addendum to Physical Result \ref{rslt:Floquet_projection}]
$\beta$ does not mix different invertible topological orders.
In particular, it takes Floquet SPT phases in the old sense to SPT phases in the old sense.
\end{framednameddef}\end{framed}

\begin{pf}
Same as Addendum to Physical Result \ref{rslt:strong_weak_projection}.
\qed\end{pf}

\begin{framed}\begin{framednameddef}[Addendum to Physical Result \ref{rslt:Floquet_projection_2}]
Every $(d-1)$-dimensional $G$-protected SPT phase, with trivial or nontrivial invertible topological order, can be produced via $\gamma$ from a $d$-dimensional $G$-protected Floquet SPT phase with trivial invertible topological order.
\end{framednameddef}\end{framed}

\begin{pf}
Same as Addendum to Physical Result \ref{rslt:strong_weak_injection}.
This addendum is independent of the arguments for Physical Result \ref{rslt:Floquet_projection_2}.
\qed\end{pf}


Now, let us spell out the implications of Physical Results \ref{rslt:Floquet_isomorphism}-\ref{rslt:Floquet_projection_2} in detail.

\begin{framed}\begin{rslt}
Let $G$ act in a way that commutes with the group $\ZZZ$ of discrete temporal translations and assume the interpretation of $\gamma$ in Physical Result \ref{rslt:Floquet_projection_2} is valid. Then we have the following:
\begin{enumerate}[(i)]
\item Equivalent Floquet systems pump equivalent stationary systems across the cut. That is, pumping is well-defined at the level of phases.

\item Every $d$-dimensional $G$-protected SPT phase can be obtained by forgetting the discrete temporal translational symmetry of some canonical $d$-dimensional $G$-protected Floquet SPT phase, which pumps the trivial $(d-1)$-dimensional $G$-protected SPT phase across the cut.

\item Every $(d-1)$-dimensional $G$-protected SPT phase can be obtained through pumping from some canonical $d$-dimensional $G$-protected Floquet SPT phase, which becomes trivial upon forgetting the discrete temporal translational symmetry.

\item Pumping commutes with addition of phases. That is, the $(d-1)$-dimensional $G$-protected SPT phase pumped across the cut by the sum of two $d$-dimensional $G$-protected Floquet SPT phases is equal to the sum of the $(d-1)$-dimensional $G$-protected SPT phases that are pumped across the cut by the two $d$-dimensional $G$-protected Floquet SPT phases respectively.

\item Pumping commutes with replacement of $G$. That is, given a homomorphism $\varphi: G' \fromto G$ and a $d$-dimensional $G$-protected Floquet SPT phase $\brackets{c}$, if we write $\varphi^* \brackets{c}$ for the $d$-dimensional $G'$-protected Floquet SPT phase induced from $\brackets{c}$ via $\varphi$, then the $(d-1)$-dimensional $G'$-protected SPT phase pumped across the cut by $\varphi^* \brackets{c}$ is equal to the one induced via $\varphi$ from the $(d-1)$-dimensional $G$-protected SPT phase that is pumped across the cut by $\brackets{c}$.

\item A $d$-dimensional $G$-protected Floquet SPT phase is uniquely determined by
\begin{enumerate}
\item the $d$-dimensional $G$-protected SPT phase obtained by forgetting the discrete temporal translational symmetry and

\item the $(d-1)$-dimensional $G$-protected SPT phase pumped across the cut,
\end{enumerate}
and every combination of $d$- and $(d-1)$-dimensional $G$-protected SPT phases is allowed.
\end{enumerate}
\end{rslt}\end{framed}

\begin{pf}
All statements follow from the exactness and splitting of the second row of Eq.\,(\ref{generalized_Kunneth_formula_diagram}), except for the one about replacement of $G$, which depends on naturality.
\qed\end{pf}

One can imagine combining ideas in Secs.\,\ref{subsec:strong_weak_topological_indices_interacting_world} and \ref{subsec:pumping_Floquet_eigenstates_classification_Floquet_SPT_phases} to treat cases where both spatial and temporal translational symmetries are present, cases where only a combination of spatial and temporal translations is a symmetry, etc., i.e., in a loose sense, spacetime crystals \cite{classicalTC, quantumTC, Li}.

\subsection{Some implications on space group-protected SPT phases \label{subsec:applications_space_group_protected_SPT_phases}}

\begin{framed}\begin{rslt}\label{rslt:space_group_TG_rtimes_PG}
Let $SG$ be a space group with all orientation-reversing elements removed. Let $PG$ be its point group. If $SG$ is symmorphic, then every $d$-dimensional $PG$-protected SPT phase can be canonically represented by a $d$-dimensional $SG$-protected SPT phase.
\end{rslt}\end{framed}

\begin{pf}
When symmorphic, $SG \isomorphic TG \rtimes PG$, where $TG$ is the translational group. Apply Proposition \ref{prp:generalization_arbitrary_semidirect_product}.
\qed\end{pf}

Put differently, when a space group is symmorphic, lifting the translational symmetry can never lead to ``intrinsically new" phases. Note that one is not obligated to retain all orientation-preserving elements in the symmetry group of a physical lattice. It is perfectly fine to let $G$ contain only rotations about a particular axis, for instance.

\begin{framed}\begin{rslt}\label{rslt:space_group_G_0_rtimes_SG}
Let $G_0$ be a group that acts in an on-site fashion and $SG$ be a space group with all orientation-reversing elements removed. Then every $d$-dimensional $SG$-protected SPT phase can be canonically represented by a $d$-dimensional $\paren{G_0 \rtimes SG}$-protected SPT phase.
\end{rslt}\end{framed}

\begin{pf}
Apply Proposition \ref{prp:generalization_arbitrary_semidirect_product}.
\qed\end{pf}

Again, Physical Result \ref{rslt:space_group_G_0_rtimes_SG} says, given an on-site symmetry and a space group symmetry, that lifting the former can never lead to ``intrinsically new" phases. Note that there is no condition on symmorphism.
An example of $G_0$ and $SG$ that do not commute is this: suppose $G_0 = \ZZZ_n$ is generated by spin rotation about the $y$-axis by an angle of $\frac{2\pi}{n}$, and $SG=\ZZZ_2$ is generated by spatial rotation about the $z$-axis by an angle of $\pi$; then the two does not commute as long as $n>2$.

When $G_0$ happens to commute with $SG$, we have the following additional result.

\begin{framed}\begin{rslt}\label{rslt:space_group_G_0_times_SG}
With the same set-up as in Physical Result \ref{rslt:space_group_G_0_rtimes_SG}, if $SG$ commutes with $G_0$, then every $d$-dimensional $G_0$-protected SPT phase can be canonically represented by a $d$-dimensional $\paren{G_0 \times SG}$-protected SPT phase.
\end{rslt}\end{framed}

\begin{pf}
Apply Proposition \ref{prp:generalization_arbitrary_product} or \ref{prp:generalization_arbitrary_semidirect_product}.
\qed\end{pf}

On the other hand, if $SG$ happens to be symmorphic, we have the following result.

\begin{framed}\begin{rslt}\label{rslt:space_group_G_0_times_TG_rtimes_PG}
With the same set-up as in Physical Result \ref{rslt:space_group_G_0_rtimes_SG}, if $SG$ is symmorphic, then every $d$-dimensional $PG$-protected SPT phase can be canonically represented by a $d$-dimensional $\paren{G_0 \rtimes PG}$-protected SPT phase, in fact a $d$-dimensional $\paren{G_0 \rtimes SG}$-protected one, where $PG$ is the point group.
\end{rslt}\end{framed}

\begin{pf}
When $SG$ is symmorphic, the total symmetry group is $G_0 \rtimes SG \isomorphic \paren{G_0 \times TG} \rtimes PG$, where $TG$ is the translational group. Apply Proposition \ref{prp:generalization_arbitrary_semidirect_product}.
\qed\end{pf}

This says, given an on-site symmetry and a symmorphic space group symmetry, that lifting the on-site symmetry and the translational symmetry can never lead to ``intrinsically new" phases.

Finally, let us see how Physical Results \ref{rslt:space_group_TG_rtimes_PG}-\ref{rslt:space_group_G_0_times_TG_rtimes_PG} interact with Physical Result \ref{rslt:reduced_unreduced_isomorphism}.

\begin{framed}\begin{framednameddef}[Addendum to Physical Results \ref{rslt:space_group_TG_rtimes_PG}-\ref{rslt:space_group_G_0_times_TG_rtimes_PG}]
If the phase being represented has trivial invertible topological order, then so does the canonical phase that represents it.
\end{framednameddef}\end{framed}

\begin{pf}
This follows from the commutativity of the second square in Eq.\,(\ref{generalization_semidirect_product_diagram}).
\qed\end{pf}

\subsection{Obstruction-free enlargement of symmetry group \label{subsec:obstruction_free_enlargement_symmetry_group}}

Here we would like to discuss the enlargement of symmetry groups in general. Let $G' \subset G$ be a subgroup. As one replaces $G'$ by $G$, one expects to refine the classification of SPT phases. It is also possible, however, for certain $G'$-protected SPT phases to be eliminated, for a priori there may be obstructions to lifting an action of $G'$ over to $G$. Here we give a sufficient condition for the absence of such obstructions.

\begin{framed}\begin{rslt}
Given $G' \subset G$, if there exists a subgroup $G''\subset G$ such that $G$ is a semidirect product $G'' \rtimes G'$, then every $d$-dimensional $G'$-protected SPT phase is representable by a $d$-dimensional $G$-protected SPT phase.
\end{rslt}\end{framed}

\begin{pf}
The condition is equivalent to the existence of a homomorphism $\pi: G \fromto G'$ such that $\pi \circ \iota= \identity$, where $\iota: G' \oneone G$ is the inclusion. This implies that $\iota^*\circ \pi^*: h^d\paren{BG'} \fromto h^d\paren{BG} \fromto h^d\paren{BG'}$ is the identity. In particular, $\iota^*: h^d\paren{BG} \fromto h^d\paren{BG'}$ is surjective.
\qed\end{pf}

Note that direct products are considered to be special cases of semidirect products. Moreover, there are many equivalent criteria for when $G$ is such a semidirect product:
\begin{enumerate}[(i)]
\item There exists a normal subgroup $G''\subset G$ such that every element $g\in G$ can be written as $g= g''g'$ for some unique $g''\in G''$ and $g'\in G'$.

\item There exists a normal subgroup $G''\subset G$ such that every element $g\in G$ can be written as $g= g'g''$ for some unique $g'\in G'$ and $g''\in G''$.

\item There exists a surjective homomorphism $G \onto G'$ that is the identity on $G'$.
\end{enumerate}

As a special case, the enlargement from the trivial symmetry group to any symmetry group $G$ is always obstruction-free. That is, every invertible topological order can be represented some $G$-protected SPT phase. This fact has been surreptitiously incorporated into Fig.\,\ref{fig:functoriality}.

\section{An Application: 3D Bosonic Crystalline SPT Phases Beyond Group Cohomology\label{sec:an_application}}

In the previous section, we have seen some simple results of the minimalist framework which are consistent with the literature. To demonstrate the applicability of the framework to concrete physical problems beyond the literature, we will make a prediction, in this section, for the complete classification of 3D bosonic SPT phases with space group symmetries. Currently, it is known that for a given space group symmetry $G$, there is a subset of 3D bosonic SPT phases one can construct that are in one-to-one correspondence with elements of the Borel group cohomology group $H^4_{\rm Borel}(G;U(1)) \isomorphic H^5(G;\ZZZ)$ \cite{Huang_dimensional_reduction, Thorngren_sgSPT}. These are the crystalline analogues of the internal SPT phases described by the group cohomology proposal \cite{Wen_Boson}. Unfortunately, just like in the case of internal symmetries, $H^4_{\rm Borel}(G;U(1))$ does not give a complete classification of 3D bosonic crystalline SPT phases. Exactly how the remaining, ``beyond-group cohomology" phases are classified by is an open question.

To answer this question, we will implement the idea in Sec.\,\ref{subsec:ubiquity_generalized_cohomology_theories} of adapting a given generalized cohomology theory to incorporate new phases. According to Ref.\,\cite{Huang_dimensional_reduction}, the new 3D bosonic crystalline SPT phases we wish to incorporate are those that are built from copies of the $E_8$ phase \cite{Kitaev_honeycomb, 2dChiralBosonicSPT, Kitaev_KITP}, which are embedded in the 3-dimensional space in a fashion that respects the space group symmetry. The $E_8$ phase is a 2-dimensional invertible topological order (SPT phase without symmetry) with a classification of $\ZZZ$. There, if $(F_d)_{d\in \ZZZ}$ is the $\Omega$-spectrum corresponding to the Borel group cohomology proposal, then the extended $\Omega$-spectrum $(F'_d)_{d\in \ZZZ}$ which incorporates the $E_8$ phase will read [see Eq.\,(\ref{product_spectrum})]
\begin{equation}
F'_d = \begin{cases}
F_d \times K(\ZZZ, d-2), & d \geq 2, \\
F_d, & d < 2.
\end{cases}
\end{equation}
In particular,
\begin{equation}
F'_3 = F_3 \times K(\ZZZ, 1).
\end{equation}
As mentioned in Sec.\,\ref{subsec:ubiquity_generalized_cohomology_theories}, in general $F'_d$ can be the total space of a fiber bundle $F_d \fromto F'_d \fromto K(\ZZZ, d-2)$ rather than the simple product $F_d \times K(\ZZZ, d-2)$. However, in the particular case of $F'_3$, $F_3 \times K(\ZZZ, 1)$ turns out to be mathematically the only possibility. To translate $F'_3$ into a classification of SPT phases, we take the homotopy classes of maps from $BG$ to $F'_3$:
\begin{eqnarray}
h'^3(BG) &=& [BG, F'_3] \nonumber\\
&\isomorphic& [BG, F_3] \times [BG, K(\ZZZ,1)].
\end{eqnarray}
By assumption, $[BG, F_3]$ corresponds to the group cohomology proposal $H^4_{\rm Borel}(G;U(1))$, so $[BG, K(\ZZZ,1)]$, which according to Table \ref{table:classic_generalized_cohomology_theories} is equal to the first group cohomology group $H^1(G;\ZZZ)$, must be the classification of the remaining phases. Therefore, we predict that the complete classification of 3D bosonic crystalline SPT phases is\footnote{The coefficient $\ZZZ$ is twisted. An element of $G$ acts on $\ZZZ$ nontrivially iff it reverses the orientation of spacetime \cite{Jiang_sgSPT, Thorngren_sgSPT}.}
\begin{equation}
H^4_{\rm Borel}(G;U(1)) \oplus H^1_{\rm group}(G;\ZZZ). \label{prediction}
\end{equation}

\begin{figure}[t]
\centering
\includegraphics[width=5in]{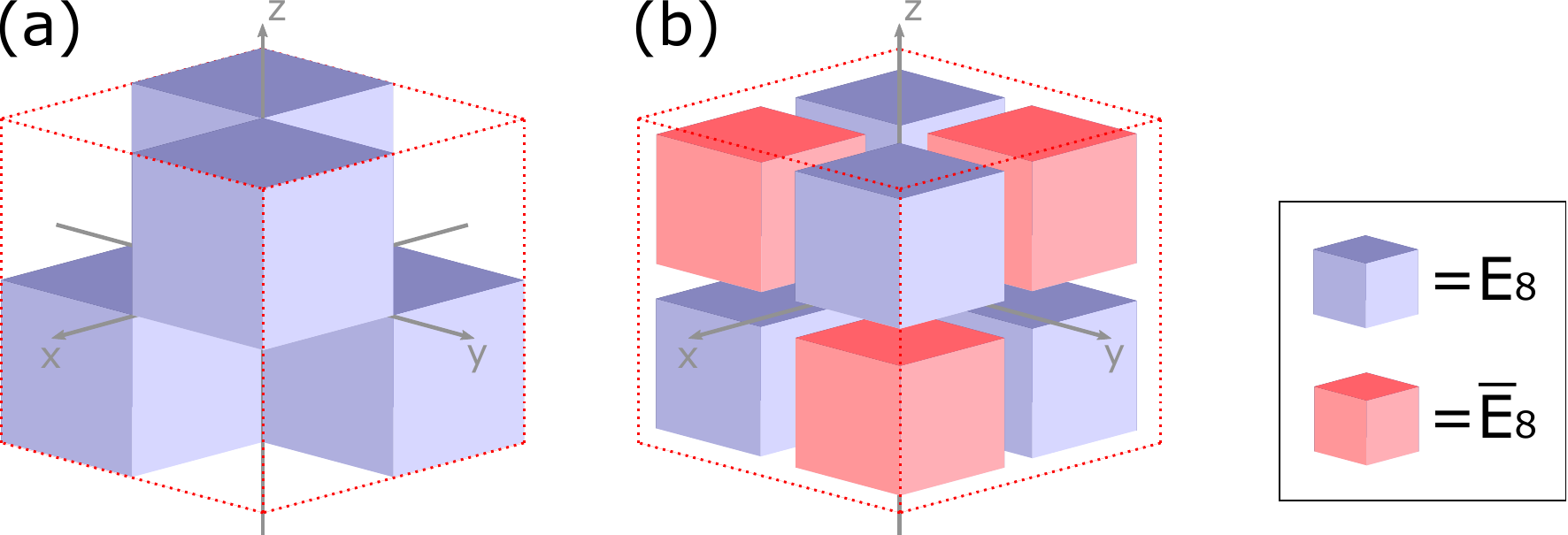}
\caption{(color online). The construction of a 3D bosonic SPT phase respecting the space group symmetry $Pmmm$ that is beyond the Borel group cohomology proposal. (a) The unit cell (indicated by dotted red lines) the 3D SPT phase, which has an $E_8$ phase with positive chirality living on the surface of each of the four cubes. (b) A configuration that has an $E_8$ phase with positive chirality living on the surface of each of the four purple cubes and an $E_8$ phase with negative chirality living on the surface of each of the four red cubes; as the size of the cubes changes, this configuration defines a path from the trivial state to two copies of the SPT phase in (a).}
\label{fig:3D_E8}
\end{figure}

Let us verify a particular case of the prediction (\ref{prediction}). Consider the space group $Pmmm$, which is generated by three orthogonal translations and three orthogonal mirror reflections. In this case, $H^1_{\rm group}(G;\ZZZ) \isomorphic \ZZZ_2$. The generator of this $\ZZZ_2$ is the SPT phase that has the configuration in Fig.\,\ref{fig:3D_E8}(a) as the unit cell; that is, it is obtained by repeating this configuration in all three directions. Since the chirality of the $E_8$ phase is invariant (resp.\,changes sign) under a reflection about a plane parallel (resp.\,perpendicular) to itself, we see that this SPT phase indeed respects the $Pmmm$ symmetry. Furthermore, two copies of this SPT phase is connected to the trivial state. This is achieved by varying the size of the cubes in the configuration shown in Fig.\,\ref{fig:3D_E8}(b). When the cubes in Fig.\,\ref{fig:3D_E8}(b) are large enough to fill up the entire unit cell, the configuration is equal to two copies of the configuration in Fig.\,\ref{fig:3D_E8}(a). When the size of the cubes goes to zero, the configuration gives the trivial state. This shows that the SPT phase has a $\ZZZ_2$ classification.

In an upcoming work, we will go on to construct 3D bosonic crystalline SPT phases that are described by $H^1_{\rm group}(G;\ZZZ)$ explicitly for all 230 space groups, thereby confirming our prediction Eq.\,(\ref{prediction}) in full generality. A simple corollary of the result (\ref{prediction}) is that there are phases beyond the group cohomology proposal \cite{Huang_dimensional_reduction} for all space groups that contain orientation-reversing elements, which generalizes the previous observation that there are additional crystalline phases that respect the space group $Pm$ \cite{Hermele_torsor}.

\section{Summary and Outlook\label{sec:summary_outlook}}

We have taken a novel, minimalist approach to the classification problem of SPT phases, where instead of directly classifying SPT phases, we looked for common ground among various existing classification proposals, which gave differing predictions in certain cases. The key in this approach was the formulation of a Generalized Cohomology Hypothesis (Sec.\,\ref{sec:generalized_cohomology_hypothesis}) that was satisfied by various existing classification proposals (Secs.\,\ref{sec:comparison_different_proposals} and \ref{subsec:existing_proposals_special_cases}) and captured essential aspects of SPT classification (Secs.\,\ref{subsec:additivity_functoriality}, \ref{subsec:ubiquity_generalized_cohomology_theories}, \ref{subsec:rationale_classifying_spaces}, \ref{subsec:physical_interpretations_Omega_spectrum}, and \ref{subsec:realizability}). We took the Hypothesis as the starting point and derived rigorous, general results from it (Sec.\,\ref{sec:consequences_hypothesis_physical_implications}). These results were born to be independent of which proposal is correct (or whether any proposal is correct at all, as long as the unknown complete classification satisfies the Hypothesis, which is plausible according to Sec.\,\ref{sec:justification_hypothesis}). They typically give relations between classifications of SPT phases in different dimensions and/or protected by different symmetry groups. They hold in arbitrarily high dimensions and apply equally to fermionic and bosonic SPT phases. Our formalism works not only for on-site symmetries but also, as we argued (Sec.\,\ref{subsec:spatiotemporal}), for discrete temporal translation (Sec.\,\ref{subsec:pumping_Floquet_eigenstates_classification_Floquet_SPT_phases}), discrete spatial translation (Secs.\,\ref{subsec:strong_weak_topological_indices_interacting_world} and \ref{subsec:hierarchy_strong_weak_topological_indices}), and other space group symmetries (Sec.\,\ref{subsec:applications_space_group_protected_SPT_phases} and Sec.\,\ref{sec:an_application}). In a sense, what we have accomplished was not a classification, but rather a meta-, or second-order classification of SPT phases, and the merit of this approach lies in the universality of our results.

More specifically, we have demonstrated the universal validity of the following statements (assuming unitary, spacetime orientation-preserving symmetries). (i) The classification of SPT phases in the ``invertible" sense \cite{FreedMoore2006, Kitaev_Stony_Brook_2011_SRE_2, Kitaev_Stony_Brook_2013_SRE, Kapustin_Boson, Freed_SRE_iTQFT, Freed_ReflectionPositivity, McGreevy_sSourcery}, as adopted in this paper, is precisely equal to the classification of SPT phases in the ``trivializable" sense \cite{Wen_Definition, Cirac} direct sum the classification of invertible topological orders (Sec.\,\ref{subsec:unification_old_new_definitions_SPT_phases}). (ii) The classification of SPT phases with translational symmetry in a given direction is precisely equal to the classification without translational symmetry in the same dimensions direct sum the classification without translational symmetry in one lower dimensions, which are the interacting analogues of strong and weak topological indices, respectively (Secs.\,\ref{subsec:strong_weak_topological_indices_interacting_world} and \ref{subsec:hierarchy_strong_weak_topological_indices}). (iii) The classification of Floquet SPT phases is precisely equal to the classification of stationary SPT phases in the same dimensions direct sum the classification of stationary SPT phases in one lower dimensions, and the latter captures to the phase pumped to the edge in one Floquet cycle (Sec.\,\ref{subsec:pumping_Floquet_eigenstates_classification_Floquet_SPT_phases}). (iv) Every point group-protected SPT phase can be canonically represented by a space group-protected SPT phase, if the space group is symmorphic (Sec.\,\ref{subsec:applications_space_group_protected_SPT_phases}). (v) More generally, a sufficient condition for being able to enlarge a symmetry group $G'$ to some larger $G$ is that $G = G'' \rtimes G'$ for some $G''$ (Sec.\,\ref{subsec:obstruction_free_enlargement_symmetry_group}).

In addition, as an application of the Hypothesis to concrete physical problems, we have made a prediction for the complete classification of 3D bosonic SPT phases with space group symmetry $G$: $H^4_{\rm Borel}(G;U(1)) \oplus H^1_{\rm group}(G;\ZZZ)$. The $H^4$ term represents crystalline analogues of internal SPT phases described by the Borel group cohomology proposal \cite{Wen_Boson}. The $H^1$ term was previously unknown and classifies crystalline SPT phases beyond group cohomology. This addressed the open question of what the complete classification of 3D bosonic SPT phases with space group symmetries might be. We have verified this prediction in the case of space group $Pmmm$ and will verify it in full generality in an upcoming work.

We believe the results presented in this paper are only the tip of an iceberg.
Generalized cohomology theories, and by extension infinite loop spaces and stable homotopy theory \cite{Adams1, Adams2}, are well-studied mathematical subjects with plenty of theorems one can draw from.
In fact, we have also mathematically derived the following results, but are yet to understand them thoroughly:
\begin{enumerate}[(i)]
\item Given coprime positive integers $m$ and $n$, we have $\tilde h^d\paren{B\ZZZ_{mn}} \isomorphic \tilde h^d\paren{B\paren{\ZZZ_m \times \ZZZ_n}} \isomorphic \tilde h^n\paren{B\ZZZ_m} \oplus \tilde h^n \paren{B\ZZZ_n}$ regardless of $\tilde h$.

\item There exist nontrivial discrete groups $G$ for which $\tilde h^d\paren{BG} = 0$ for all $d$ regardless of $\tilde h$.

\item There exist non-isomorphic finite groups $G_1$, $G_2$ for which $h^d\paren{BG_1} \isomorphic h^d\paren{BG_2}$ regardless of $h$, at least in low dimensions with an additional, well-founded physical input.
\end{enumerate}
We would also like to answer the following questions:
\begin{enumerate}[(i)]
\item How would our results generalize if effectively antiunitary symmetries were allowed, which would give rise to group actions on the $\Omega$-spectrum and necessitate twisted generalized cohomology theories?

\item Does the multiplicative structure of a multiplicative generalized cohomology theory have a physical meaning\footnote{We thank Ammar Husain for suggesting this.}?

\item Do generalized cohomology groups in negative dimensions have a physical meaning\footnote{We thank Ashvin Vishwanath for suggesting this.}?

\item Can the Hypothesis be derived from ``first principles"?

\item What is the counterpart of generalized cohomology theories for topological orders, or more generally $G$-protected topological phases?
\end{enumerate}
We will give a first answer to question (i) in the follow-up work Ref.\,\cite{Xiong_Alexandradinata}, which deals with the effectively antiunitary glide reflection symmetry. We will see that SPT phases with and without glide reflection symmetry fit into a short exact sequence, which will allow us to argue that the hourglass fermions \cite{Nonsymm_Shiozaki, Hourglass, Cohomological} -- a topological insulator whose hourglass-shaped surface bands structure has been recently observed using angle-resolved photoemission spectroscopy \cite{Ma_discoverhourglass} -- must be robust to interactions.


\newpage

\begin{appendices}

\renewcommand{\appendixname}{\empty}

\section*{Appendices}

The appendices are organized as follows.
In App.\,\ref{app:existing_proposals_generalized_cohomology_theories}, we explain in more detail how existing proposals for the classification of SPT phases are examples of generalized cohomology theories.
In App.\,\ref{app:homotopy_equivalence}, we complete the arguments in Sec.\,\ref{subsec:physical_interpretations_Omega_spectrum} for the pointed homotopy equivalences $F_d \homotopic \Omega F_{d+1}$. 
In App.\,\ref{app:field_theoretic_argument_weak_index_interpretation}, we propose a field-theoretic construction to corroborate the weak-index interpretation in Sec.\,\ref{subsec:strong_weak_topological_indices_interacting_world}.
In App.\,\ref{app:categorical_viewpoint}, we present an equivalent but more succinct version of the Hypothesis using the terminology of category theory.
In App.\,\ref{app:additivity_functoriality_group_cohomology_construction}, we explicitly show that the group cohomology construction \cite{Wen_Boson} is additive and functorial.
In App.\,\ref{app:proofs}, we collect the main mathematical theorems upon which the physical results in Sec.\,\ref{sec:consequences_hypothesis_physical_implications} are based.
App.\,\ref{app:mathematical_background} is a review of notions in algebraic topology, category theory, and generalized cohomology theories.

\section{Existing Classification Proposals as Generalized Cohomology Theories\label{app:existing_proposals_generalized_cohomology_theories}}

In this appendix, we explain how various proposals for the classification of SPT phases can be viewed as generalized cohomology theories. Below, we denote by $K(A, n)$ the $n$-th Eilenberg-Mac Lane space of $A$ (see App.\,\ref{subapp:generalized_cohomology_theories}).

\subsection{Borel group cohomology proposal\label{subapp:Borel_group_cohomology_proposal}}

Ref.\,\cite{Wen_Boson} proposed that $d$-dimensional $G$-protected bosonic SPT phases are classified by $H^{d+1}_{\rm group}\paren{G; U(1)}$ when $G$ is finite and acts in an on-site, unitary fashion. Here, $H^\bullet_{\rm group}$ denotes group cohomology. For infinite or continuous groups, Ref.\,\cite{Wen_Boson} conjectured a classification by a Borel group cohomology group $H_{\rm Borel}^{d+1}\paren{G;U(1)}$, which is naturally isomorphic to $H^{d+2}\paren{BG;\ZZZ}$ \cite{Borel_group_cohomology}. Here, $H^\bullet\paren{-;\ZZZ}$ is the ordinary (topological) cohomology theory with $\ZZZ$ coefficient \cite{Hatcher}. Ordinary cohomology theories are the most ordinary kind of generalized cohomology theories. We know from Table \ref{table:classic_generalized_cohomology_theories} that they are represented by Eilenberg-Mac Lane spectra. Taking into account the shift in dimension, we thus have
\begin{equation}
H_{\rm Borel}^{d+1}\paren{G;U(1)} \isomorphic H^{d+2}\paren{BG; \ZZZ} \isomorphic \brackets{BG, K(\ZZZ, d+2)}.
\end{equation}
It can be seen either at the level of $\Omega$-spectrum or by inspecting Definitions \ref{dfn:reduced_generalized_cohomology_theory_2} and \ref{dfn:unreduced_generalized_cohomology_theory_2} that a shift in dimension turns generalized cohomology theories into generalized cohomology theories. 

We will prove in App.\,\ref{app:additivity_functoriality_group_cohomology_construction} that the discrete abelian group and functorial structures of $H^{d+1}_{\rm group}\paren{G;U(1)}$ for finite $G$ correspond to stacking phases and replacing symmetry groups, respectively. This cannot be done for continuous or infinite discrete groups since no explicit construction was given in those cases.

It only remains to show that the $H^{d+2}\paren{BG;\ZZZ}$ reduces to $H^{d+1}_{\rm group}\paren{G;U(1)}$ in physical dimensions $d\geq 0$ when $G$ is finite. By comparing the definitions of group cohomology and cellular cohomology, one finds a natural isomorphism $H^{d+1}_{\rm group}\paren{G; U(1)} \isomorphic H^{d+1}\paren{BG; U(1)}$ for discrete groups and in particular finite groups. Since $H^{d+1}(-;A) = \widetilde H^{d+1}(-;A)$ for all $d\geq 0$ and coefficients $A$, the following lemma completes the proof.
\begin{lem}
For each $n\in \ZZZ$, there is a natural transformation,
\begin{equation}
\widetilde H^n\paren{X; U(1)} \fromto \widetilde H^{n+1}\paren{X; \ZZZ},
\end{equation}
that is an isomorphism when $X = BG$ and $G$ is a finite\footnote{This result was stated informally without proof in Ref.\,\cite{Kitaev_KITP}.}. \label{lem:Bockstein_homomorphism}
\end{lem}

\begin{pf}
The desired natural transformation is the Bockstein homomorphism associated with the short exact sequence
\begin{equation}
0 \fromto \ZZZ \fromto \RRR \fromto U(1) \fromto 0
\end{equation}
of abstract (i.e.\,without topology) abelian groups, which gives rise to a natural long exact sequence,
\begin{equation}
\cdots
\fromto \widetilde H^n\paren{X;\ZZZ} \fromto \widetilde H^n\paren{X;\RRR} \fromto \widetilde H^n\paren{X;U(1)}
\fromto \widetilde H^{n+1}\paren{X;\ZZZ} \fromto \widetilde H^{n+1}\paren{X;\RRR} \fromto \widetilde H^n\paren{X;U(1)}
\fromto \cdots
\end{equation}
The lemma will be established once we prove that
\begin{equation}
\widetilde H^n\paren{BG;\RRR} = 0
\end{equation}
for all $n\in \ZZZ$ and finite groups $G$. By the universal coefficient theorem, this amounts to showing that $\Ext^1\paren{ \widetilde H_n\paren{BG;\ZZZ}, \RRR } = \Hom\paren{ \widetilde H_n\paren{BG;\ZZZ}, \RRR } = 0$. The $\Ext$ group is trivial because $\RRR$ is a field. The $\Hom$ group is trivial because $\widetilde H_n\paren{BG;\ZZZ}$ is pure torsion, as per Remarks 3.6 and 3.7 and Corollary 5.4 in Chap.\,II of Ref.\,\cite{AdemMilgram}.
\qed\end{pf}


\subsection{Oriented cobordism proposal \label{subapp:oriented_cobordism_proposal}}

Ref.\,\cite{Kapustin_Boson} proposed that $d$-dimensional $G$-protected bosonic SPT phases are classified by
\begin{equation}
\Hom\paren{ MSO_{d+1}\paren{BG} , U(1) } \label{oriented_cobordism_proposal}
\end{equation}
when $G$ is finite and acts in an on-site, unitary fashion. 
Here, $MSO_\bullet\paren{X}$ denotes the $n$-th oriented bordism group, which is a discrete abelian group, of topological space $X$. Continuous symmetry groups were not dealt with in Ref.\,\cite{Kapustin_Boson}. In fact, the proposal was to further quotient out a subgroup of ``continuous theta-parameters," but we may as well do a classification with such parameters allowed and quotient them out at the end of the day. Ref.\,\cite{Kapustin_Boson} also assumed a ``vanishing thermal Hall response," but that is a matter of what the word ``system" means, which was put in a black box in Sec.\,\ref{subsubsec:new_definition_SPT_phases}.

To prove that the oriented cobordism proposal is a generalized cohomology theory, it is best to use the algebraic definitions \ref{dfn:reduced_generalized_cohomology_theory_2} and \ref{dfn:unreduced_generalized_cohomology_theory_2} of generalized cohomology theories, and the analogous algebraic definitions \cite{Hatcher} of generalized homology theories, rather than the topological definitions \ref{dfn:reduced_generalized_cohomology_theory} and \ref{dfn:unreduced_generalized_cohomology_theory}.
By inspecting these algebraic definitions, one can convince themselves that the functor $\Hom\paren{ - , U(1)}$ turns generalized homology theories into generalized cohomology theories. The only axiom that is perhaps nontrivial to check is the exactness axiom, for which one should invoke the fact that $U(1)$ is an injective $\ZZZ$-module. 
Knowing that oriented bordism $MSO_\bullet$ is a generalized homology theory \cite{Adams1, Adams2, Atiyah}, we conclude that the oriented cobordism proposal is a generalized cohomology theory.

It can only be partially verified that that the additive and functorial structures of the oriented cobordism proposal correspond to stacking phases and replacing symmetry groups, respectively, as no lattice model was given in Ref.\,\cite{Kapustin_Boson}.

Eq.\,(\ref{oriented_cobordism_proposal}) is different from the standard oriented cobordism group $MSO^{d+1}\paren{BG}$ \cite{Kapustin_Boson}, and hence is not represented by the Thom spectrum $MSO$ in the sense of Theorem \ref{thm:Brown_representability_theorem}. It is, however, still related to the Thom spectrum $MSO$ as oriented bordism groups $MSO_{d+1}\paren{BG}$ can be defined in terms of it \cite{Adams1, Adams2, Atiyah}.

\subsection{Kitaev's bosonic proposal \label{subapp:Kitaev_bosonic_proposal}}

Kitaev's proposal \cite{Kitaev_Stony_Brook_2011_SRE_1, Kitaev_Stony_Brook_2013_SRE} is unique among all existing classification proposals for bosonic SPT phases.
He took the Generalized Cohomology Hypothesis as a fundamental assumption and tried to construct an $\Omega$-spectrum from physical knowledge.
The key observation there was that $h^d\paren{\pt}$'s simultaneously classify invertible topological orders (see Physical Result \ref{rslt:invertible_topological_orders}) and determine homotopy groups of the $\Omega$-spectrum:
\begin{eqnarray}
h^d\paren{\pt} \isomorphic \pi_i \paren{F_{i+d}} \eqcolon \pi_{-d}\paren{F}, ~\forall i
\end{eqnarray}
Homotopy groups carry important information about a topological space. The additional information needed to determine the homotopy type of a space is given by so-called $k$-invariants \cite{Hatcher}, and they are sometimes unique for trivial reasons.
Given the homotopy groups and $k$-invariants of a space, the reconstruction proceeds by building a Postnikov tower from the bottom up \cite{Hatcher}.

Refs.\,\cite{Kitaev_Stony_Brook_2011_SRE_1, Kitaev_Stony_Brook_2013_SRE} assumed that
\begin{eqnarray}
F_0 \homeomorphic \CCC P^\infty, ~ h^1\paren{\pt} = 0, ~ h^2\paren{\pt} \isomorphic \ZZZ, ~ h^3\paren{\pt} = 0,
\end{eqnarray}
where $\CCC P^\infty$ is the space of rays of (the direct limit of) finite-dimensional Hilbert spaces (recall Sec.\,\ref{subsec:physical_interpretations_Omega_spectrum}), and $h^2\paren{\pt}$ is generated by the $E_8$-model \cite{Kitaev_honeycomb, 2dChiralBosonicSPT, Kitaev_KITP}. Physically, $\pi_2(\CCC P^\infty)\isomorphic \ZZZ$ can be identified with the integral of the Berry curvature, and $h^2\paren{\pt} \isomorphic \ZZZ$ can be identified with chiral central charge \cite{Kitaev_Stony_Brook_2011_SRE_1, Kitaev_Stony_Brook_2013_SRE}.
Accordingly, the homotopy groups of the $\Omega$-spectrum are
\begin{center}
\begin{tabular}{c|cccccccc}
$i$ & $<-3$ & $-3$ & $-2$ & $-1$ & $0$ & $1$ & $2$ & $>2$ \\
\hline
$\pi_i\paren{F}$ & $?$ & $0$ & $\ZZZ$ & $0$ & $0$ & $0$ & $\ZZZ$ & $0$ \\
\end{tabular}
\end{center}
Having a single nontrivial homotopy group, the homotopy type of $F_1$ can be trivially determined:
\begin{equation}
F_1 = K(\ZZZ, 3),
\end{equation}
since there is no $k$-invariant to worry about.
It turns out that the homotopy type of $F_2$ can also be determined:
\begin{equation}
F_2 = K(\ZZZ,4) \times \ZZZ,
\end{equation}
but for that one must utilize the fact that $F_2 \homotopic \Omega F_3$ is a loop space.
Though not mentioned in Refs.\,\cite{Kitaev_Stony_Brook_2011_SRE_1, Kitaev_Stony_Brook_2013_SRE}, we can go on to determine the homotopy type of $F_3$. It has two nontrivial homotopy groups in positive dimensions and one potentially nontrivial $k$-invariant, which takes value in $H^6\paren{K\paren{\ZZZ, 1}; \ZZZ}$. Incidentally, $H^6\paren{K\paren{\ZZZ, 1}; \ZZZ} = 0$, so this $k$-invariant must be trivial as well, and the homotopy type of $F_3$ can be determined:
\begin{equation}
F_3 = K(\ZZZ,5) \times K(\ZZZ, 1) \homotopic K(\ZZZ, 5) \times \SS^1.
\end{equation}
A similar argument ($H^7\paren{K\paren{\ZZZ,2};\ZZZ} = 0$ plus the fact that it is a loop space) shows that
\begin{equation}
F_4 = K(\ZZZ,6) \times K(\ZZZ,2) \times \pi_{-4}\paren{F} \homotopic K(\ZZZ,6) \times \CCC P^\infty \times h^4\paren{\pt},
\end{equation}
but $h^4\paren{\pt}$ is unknown. All higher dimensional $F_d$'s require further input.

It can only be partially verified that that the additive and functorial structures of Kitaev's bosonic proposal correspond to stacking phases and replacing symmetry groups, respectively, as the lattice model given in Ref.\,\cite{Kitaev_KITP} was schematic.

\subsection{Freed-Hopkins bosonic proposal \label{subapp:Freed_bosonic_proposal}}

We refer the reader to Refs.\,\cite{Freed_SRE_iTQFT, Freed_ReflectionPositivity} in view of the complexity of the proposal.

\subsection{Group supercohomology proposal \label{subapp:group_supercohomology_proposal}}

Ref.\,\cite{Wen_Fermion} proposed, when $G$ is finite and acts in an on-site, unitary fashion, that $d$-dimensional $G$-protected fermionic SPT phases are classified by a group supercohomology group whose cochains of are pairs\footnote{%
The cochains in Ref.\,\cite{Wen_Fermion} are actually triples
\begin{equation}
\paren{\nu_d, n_{d-1}, u_{d-1}} \in \Hom_\ZZZ \paren{\ZZZ\brackets{G^{d+1}}, U(1)} \times \Hom_{\ZZZ G} \paren{\ZZZ\brackets{G^d}, H^1\paren{\ZZZ_2^f, U(1)}} \times \Hom_{\ZZZ G}\paren{\ZZZ\brackets{G^d}, H^1_{\rm group}\paren{G_f, U(1)}},
\end{equation}
where $G_f$ is the full symmetry group including fermion parity, but at the level of equivalence classes, $u_{d-1}$ is irrelevant. See App.\,C of Ref.\,\cite{Wen_Fermion}.}
\begin{eqnarray}
\nu_d&:& G^{d+1} \fromto U(1), \\
n_{d-1}&:& G^d \fromto \ZZZ_2 \subset U(1).
\end{eqnarray}
According to Ref.\,\cite{Freed_SRE_iTQFT}, the proposal amounts to using the $\Omega$-spectrum with the homotopy groups
\begin{eqnarray}
\pi_{i-d}\paren{F} \coloneq \pi_i \paren{F_d} \isomorphic \begin{cases}
\ZZZ_2, & i=d, \\
\ZZZ, & i = d+2, \\
0, & \text{otherwise},
\end{cases}
\end{eqnarray}
and the $k$-invariants (see App.\,\ref{subapp:Kitaev_bosonic_proposal}) defined as follows. Having at most two nontrivial homotopy groups, each $F_d$ has at most one nontrivial $k$-invariant, $k_{d+1}$. If we denote by $\beta$ and $\beta'$ the Bockstein homomorphisms \cite{Hatcher} associated with the first and second rows of the commutative diagram
\begin{eqnarray}
\adjustbox{valign=m}{\includegraphics{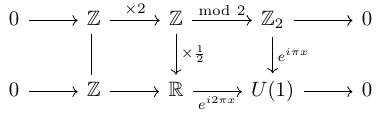}}
\end{eqnarray}
and by $\tau$ the map induced by the last vertical map,
then $k_{d+1}$ is defined to be the unique map making the following diagram commute:
\begin{eqnarray}
\adjustbox{valign=m}{\includegraphics{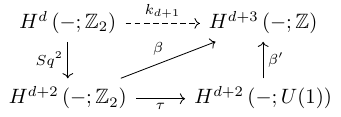}}
\end{eqnarray}
where $Sq^2$ is the Steenrod square \cite{Hatcher}, which Ref.\,\cite{Wen_Boson} mentioned in passing.
In other words,
\begin{eqnarray}
k_{d+1} = \beta \circ Sq^2.
\end{eqnarray}
One can think of the resulting theory as some sort of ``twisted product" between $H^{d+2}\paren{-;\ZZZ}$ and $H^d\paren{-;\ZZZ_2}$, which should correspond to $\nu_d$ and $n_{d-1}$, respectively (recall Lemma \ref{lem:Bockstein_homomorphism}). Indeed, if all $k_{d+1}$'s were trivial, then $F_d$ would simply be a product $K(\ZZZ,d+2) \times K(\ZZZ_2, d)$ and the generalized cohomology group would simply be $H^{d+2}\paren{-;\ZZZ} \oplus H^d\paren{-;\ZZZ_2}$. In reality, this is true in $d=0,1$ but not necessarily higher dimensions. Thus, we have
\begin{eqnarray}
F_0 &=& K(\ZZZ,2) \times \ZZZ_2 \homotopic \CCC P^\infty \times \ZZZ_2, \\
F_1 &=& K(\ZZZ,3) \times K(\ZZZ_2, 1) \homotopic K(\ZZZ,3) \times \RRR P^\infty,
\end{eqnarray}
while $F_d$ with $d\geq 2$ has to be obtained as a pull-back along $k_{d+1}$:
\begin{eqnarray}
\adjustbox{valign=m}{\includegraphics{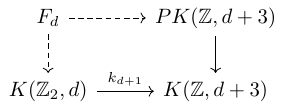}}
\end{eqnarray}
where the vertical arrow on the right is the path space fibration (see App.\,\ref{subapp:notions_algebraic_topology}).

\subsection{Spin cobordism proposal \label{subapp:spin_cobordism_proposal}}

We refer the reader to Ref.\,\cite{Kapustin_Fermion} in view of the complexity of the proposal.

\subsection{Kitaev's fermionic proposal \label{subapp:Kitaev_fermionic_proposal}}

Kitaev's proposal \cite{Kitaev_Stony_Brook_2013_SRE, Kitaev_IPAM} for the classification of fermionic SPT phases was in close analogy with the bosonic case discussed in App.\,\ref{subapp:Kitaev_bosonic_proposal}.
Again, he took the Generalized Cohomology Hypothesis as a fundamental assumption and tried to construct an $\Omega$-spectrum from physical knowledge.
This time, it was assumed that
\begin{eqnarray}
F_0 \homeomorphic \CCC P^\infty \times \ZZZ_2, ~ h^1\paren{\pt} \isomorphic \ZZZ_2, ~ h^2\paren{\pt} \isomorphic \ZZZ,
\end{eqnarray}
where $\CCC P^\infty$ is the space of rays of (the direct limit of) finite-dimensional Hilbert spaces (recall Sec.\,\ref{subsec:physical_interpretations_Omega_spectrum}), the $\ZZZ_2$ in $F_0$ is fermion parity, the $\ZZZ_2$ in $h^1\paren{\pt}$ is generated by the Majorana chain \cite{Majorana_chain}, and $\ZZZ$ is generated by $\paren{p+ip}$-superconductors \cite{Volovik_p+ip, Read_p+ip, Ivanov_p+ip}.
Physically, $\pi_2(\CCC P^\infty)\isomorphic \ZZZ$ can be identified with the integral of the Berry curvature.
Accordingly, the homotopy groups of the $\Omega$-spectrum are
\begin{center}
\begin{tabular}{c|cccccccc}
$i$ & $<-2$ & $-2$ & $-1$ & $0$ & $1$ & $2$ & $>2$ \\
\hline
$\pi_i\paren{F}$ & $?$ & $\ZZZ$ & $\ZZZ_2$ & $\ZZZ_2$ & $0$ & $\ZZZ$ & $0$ \\
\end{tabular}
\end{center}
Unfortunately, without further input, one can only determine the homotopy type of $F_d$ for $d\leq 0$. As for $F_1$, there are two path components, which are homotopy equivalent since $F_1$ is a loop space. The component containing the basepoint has two nontrivial homotopy groups and one potentially nontrivial $k$-invariant,
\begin{eqnarray}
k_2 \in H^4\paren{K\paren{\ZZZ_2, 1}; \ZZZ} \isomorphic \ZZZ_2.
\end{eqnarray}
Thus there are two possibilities:
\begin{equation}
F_1 = X_3 \times \ZZZ_2,
\end{equation}
where $X_3$ is either $K(\ZZZ,3) \times K(\ZZZ_2, 1)$ corresponding to $k_2=0$, or a more complicated space corresponding to $k_2\neq 0$. If one borrows $k_2$ from App.\,\ref{subapp:group_supercohomology_proposal}, then $k_2 = 0$, and $F_1 = K\paren{\ZZZ, 3} \times K\paren{\ZZZ_2,1} \times \ZZZ_2 \homotopic K(\ZZZ,3) \times \CCC P^\infty \times \ZZZ_2$.

It can only be partially verified that that the additive and functorial structures of Kitaev's fermionic proposal correspond to stacking phases and replacing symmetry groups, respectively, as the lattice model given in Ref.\,\cite{Kitaev_KITP} was schematic.

\subsection{Freed-Hopkins fermionic proposal \label{subapp:Freed_fermionic_proposal}}

We refer the reader to Refs.\,\cite{Freed_SRE_iTQFT, Freed_ReflectionPositivity} in view of the complexity of the proposal.

\section{Arguments for Homotopy Equivalence $F_d \homotopic \Omega F_{d+1}$\label{app:homotopy_equivalence}}

In this appendix, we complete the physical interpretation of $\Omega$-spectrum in Sec.\,\ref{subsec:physical_interpretations_Omega_spectrum} by arguing that the pair of maps
\begin{equation}
f: F_d \rightleftarrows \Omega F_{d+1} : g
\end{equation}
are pointed homotopy inverses of each other, that is, both compositions $f \circ g$ and $g \circ f$ are pointed homotopic to the identity. We do this for the pumping interpretation and domain wall interpretation in Apps.\,\ref{subapp:homotopy_equivalence_pumping} and \ref{subapp:homotopy_equivalence_domain_wall}, respectively. We will go on to show that the two interpretations are equivalent to each other and compatible with the additivity structure of SPT phases in Apps.\,\ref{subapp:homotopy_equivalence_equivalence} and \ref{subapp:homotopy_equivalence_compatibility}, respectively.

\subsection{Homotopy in pumping interpretation\label{subapp:homotopy_equivalence_pumping}}

\begin{figure}
\centering
\includegraphics[width=6in]{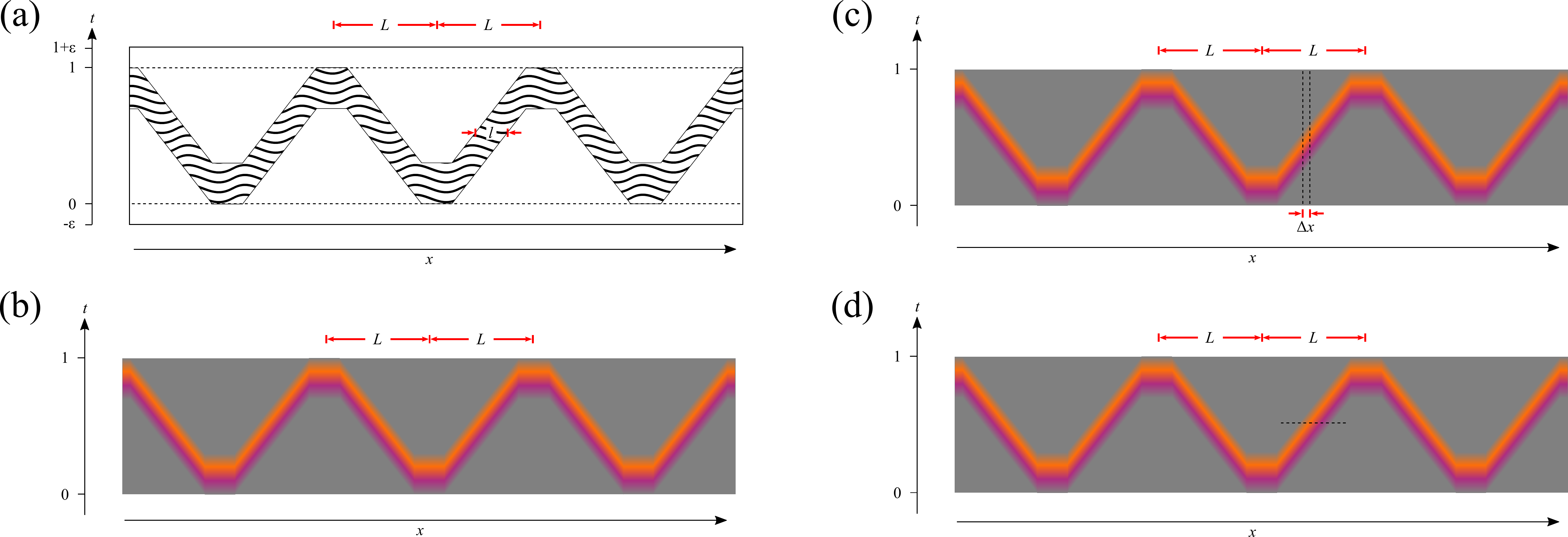}
\caption{(color online). Construction of a pointed homotopy $f \circ g \sim \identity$ in the pumping interpretation of $\Omega$-spectrum. (a) The subspace $A$ of $\RRR \times (-\epsilon, 1+\epsilon)$. (b) The final value $H_1(\mu)$ of the homotopy (\ref{homotopy_H}) evaluated at $\mu$. (c) A neighborhood $\Delta x$ of $x$ such that $l \gg \Delta x \gg \xi$. (d) A cut along the $x$-direction that intersects the trajectory in (b) exactly once.}
\label{fig:pumping_app}
\end{figure}

To see that $f$ and $g$ are pointed homotopy inverses of each other, we first note that
\begin{equation}
g \circ f = \identity
\end{equation}
by construction. As for
\begin{equation}
f \circ g \sim \identity,
\end{equation}
it is useful to regard $\mu$ as a function in two variables: $\mu(x,t) \in F_{d+1}$. The idea is $\mu(x,t)$ varies slowly with $x$ and that $\mu(-,t)$ describes an SRE state that is itself a spatially varying texture of $(d+1)$-dimensional SRE states; it looks like $\mu(x_1,t)$ near $x = x_1$, $\mu(x_2, t)$ near $x = x_2$, and so on. Now we will define a homotopy
\begin{eqnarray}
H: [0,1] \times \Omega F_{d+1} &\fromto& \Omega F_{d+1}, \nonumber\\
(s, \mu) &\mapsto& H_s(\mu), \label{homotopy_H}
\end{eqnarray}
such that
\begin{eqnarray}
H_0 &=& \identity, \\
H_1 &=& f\circ g, \\
H_s(\ast) &=& \ast, ~~ \forall s\in [0,1],
\end{eqnarray}
where $\ast \in \Omega F_{d+1}$ denotes the trivial loop. To do so we shall fix some positive number $\epsilon$ and a one-parameter family of homeomorphisms:
\begin{eqnarray}
\gamma: [0,1] \times \RRR \times (-\epsilon, 1+\epsilon) &\fromto& \RRR \times (-\epsilon, 1+\epsilon), \nonumber\\
(s, x, t) &\mapsto& \gamma_s(x,t),
\end{eqnarray}
such that
\begin{eqnarray}
&& \gamma_s \in {\rm Homeo}\paren{ \RRR \times (-\epsilon, 1+\epsilon) }, ~~ \forall s\in [0,1], \\
&& \gamma_s \paren{ \RRR \times (0,1) } \subset \RRR \times (0,1), ~~ \forall s\in [0,1], \\
&& \gamma_0 = \identity, \\
&& \gamma_1\paren{ \RRR \times [0,1] } = A.
\end{eqnarray}
Here, $A \subset \RRR \times (-\epsilon, 1+\epsilon)$ is the zigzag band shown in Fig.\,\ref{fig:pumping_app}(a), where the width $l$ is inessential except that we require $L \gg l \gg \xi$. It is easy to see that such a $\gamma$ exists. Now we define
\begin{eqnarray}
\brackets{H_s(\mu)}(x,t) \coloneq \mu \paren{ \gamma_s^{-1}(x,t) }, ~~ \forall (x,t) \in \RRR \times [0,1].
\end{eqnarray}

Clearly, $H_0 = \identity$ and $H_s(\ast) = \ast$ for all $s$. To see that $H_1 = f \circ g$, let us compare $H_1(\mu)$ and $f\paren{g\paren{\mu}}$ for an arbitrary $\mu$. If $\mu$ looks like Fig.\,\ref{fig:pumping}(c), then $H_1(\mu)$ will look like Fig.\,\ref{fig:pumping_app}(b), whereas $f\paren{g\paren{\mu}}$ will look like Fig.\,\ref{fig:pumping}(b) with $a = g(\mu)$. Note that both $\brackets{H_1(\mu)}(x,t)$ and $\brackets{f\paren{g\paren{\mu}}}(x,t)$ look trivial unless $(x,t)$ falls within the zigzag band $A$, so they describe transportation of $d$-dimensional SRE states along the same spacetime trajectory, $A$, which can be viewed as a trajectory due to its thinness. In fact, $\brackets{H_1(\mu)}(x,t)$ and $\brackets{f\paren{g\paren{\mu}}}(x,t)$ transport the same state, namely $a$. We can see this by fixing a neighborhood $\Delta x$ of $x$ such that $l \gg \Delta x \gg \xi$, as in Fig.\,\ref{fig:pumping_app}(c).
When restricted to $\Delta x$, $H_1(\mu)$ describes the same adiabatic evolution as $\mu$, except it is formally faster, that is, it begins at some $t > 0$ and ends at some $t<1$. However, being formally faster does not change the state that is pumped across $\Delta x$ in the adiabatic evolution. We thus conclude that $H_1 = f \circ g$. This completes the argument that there is a pointed homotopy
\begin{equation}
f \circ g \sim \identity.
\end{equation}


\subsection{Homotopy in domain wall interpretation\label{subapp:homotopy_equivalence_domain_wall}}

\begin{figure}
\centering
\includegraphics[width=5in]{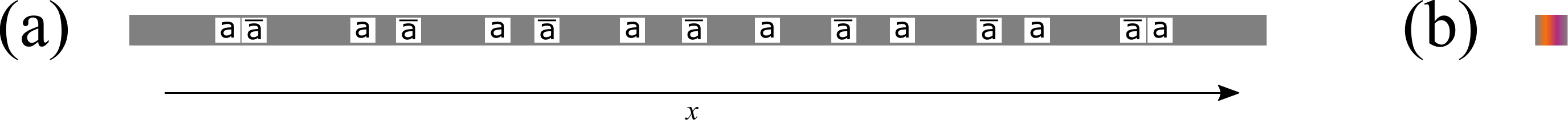}
\caption{(color online). The maps $f$ and $g$ in the domain wall interpretation of $\Omega$-spectrum. (a) The texture of $d+1$-dimensional SRE states represented by $f(a)$, where $a$ is a $d$-dimensional SRE state. (b) The $d$-dimensional SRE state represented by $g(\mu)$, where $\mu$ is the one-parameter family of $(d+1)$-dimensional SRE states shown in Fig.\,\ref{fig:pumping}(c).}
\label{fig:domain_wall_app}
\end{figure}

To show that $f$ and $g$ are homotopy inverses of each other, we first note that
\begin{equation}
g \circ f = \identity
\end{equation}
by construction. Indeed, given $a\in F_d$, $f(a)$ corresponds to the texture in Fig.\,\ref{fig:domain_wall_app}(a),
which becomes $a$ after squeezing. As for
\begin{equation}
f \circ g \sim \identity,
\end{equation}
we compare $f\paren{g\paren{\mu}}$ and $\mu$ for an arbitrary $\mu$. To that end, it is convenient to regard $\mu$ as a function of $t$ for $t\in [0,1]$, without interpreting it as a texture. Now, if $\mu \in \Omega F_{d+1}$ looks like Fig.\,\ref{fig:pumping}(c), then $f\paren{g\paren{\mu}} \in \Omega F_{d+1}$ will look like Fig.\,\ref{fig:pumping_app}(b).
The process of gradually squeezing $\mu$ in the $t$-direction defines a path from $\mu$ to $f\paren{g\paren{\mu}}$. In fact, it is a homotopy from $\identity$ to $f \circ g$, and can be achieved using the homotopy (\ref{homotopy_H}) defined in App.\,\ref{subapp:homotopy_equivalence_pumping}. We conclude that
\begin{equation}
f \circ g \sim \identity.
\end{equation}

\subsection{Equivalence between two interpretations\label{subapp:homotopy_equivalence_equivalence}}

Let us show that the definitions of the maps $f$ and $g$ are the same for the two interpretations. 

The map $f$ is the same for the two interpretations by definition. As for $g$, let $\mu \in \Omega F_{d+1}$ be a one-parameter family of $(d+1)$-dimensional SRE states that looks like Fig.\,\ref{fig:pumping}(c).
In the domain wall interpretation, we perform the change of variable $t = \frac{1 - \tanh x}{2}$ and treat $\mu$ as a $(d+1)$-dimensional SRE state that is itself a spatially varying texture of $(d+1)$-dimensional SRE states, as in Fig.\,\ref{fig:domain_wall}(b).
We then squeeze the pattern and define $g(\mu)$ to be the domain wall, as in Fig.\,\ref{fig:domain_wall}(c). Therefore, $g(\mu)$ looks like Fig.\,\ref{fig:domain_wall_app}(b).

In the pumping interpretation, we define $g(\mu)$ to be the state pumped to the boundary whose outward normal is in the $+x$-direction, as in Fig.\,\ref{fig:pumping}(d). As argued in App.\,\ref{subapp:homotopy_equivalence_pumping}, this is the same $d$-dimensional SRE state as the one transported along the following spacetime trajectory in Fig.\,\ref{fig:pumping_app}(b).
To determine this state, we fix a time $t\in (0,1)$ and consider a finite cut along the $x$-direction that intersects the trajectory exactly once, as in Fig.\,\ref{fig:pumping_app}(d).
We see that $g(\mu)$ is again the state in Fig.\,\ref{fig:domain_wall_app}(b), same as in the domain wall interpretation.

\subsection{Compatibility with additivity\label{subapp:homotopy_equivalence_compatibility}}

We now show that the abelian group structure (Sec.\,\ref{sec:generalized_cohomology_hypothesis}) of
\begin{equation}
h^d(BG) \coloneq \brackets{ BG, F_d } \isomorphic \brackets{BG, \Omega F_{d+1}}, \label{coh}
\end{equation}
which is defined by concatenating loops in $F_{d+1}$, matches the additive structure of SPT phases defined by stacking (Sec.\,\ref{subsubsec:additivity}).

It is easier to work at the level of representatives and show that the monoid structure of $\Omega F_{d+1}$ under concatenation matches the monoid structure of SRE states under stacking. Let us take two loops $\mu_1, \mu_2 \in \Omega F_{d+1}$, and write $\mu_1 + \mu_2 \in \Omega F_{d+1}$ for the concatenated loop (of, say, $\mu_1$ followed by $\mu_2$, for definiteness). In the pumping interpretation, if $g\paren{\mu_1}$ and $g\paren{\mu_2}$ are the states pumped to the boundary by $\mu_1$ and $\mu_2$, then obviously $g\paren{\mu_1} \otimes g\paren{\mu_2}$ will be the state pumped to the boundary by $\mu_1 + \mu_2$. In the domain wall interpretation, if $g\paren{\mu_1}$ and $g\paren{\mu_2}$ are the domain walls obtained by squeezing $\mu_1$ and $\mu_2$, then obviously $g\paren{\mu_1} \otimes g\paren{\mu_2}$ will be the domain wall obtained by squeezing $\mu_1 + \mu_2$. Therefore, the additive structure of (\ref{coh}) matches the additive structure of SPT phases.

\section{Field-Theoretic Argument for Weak-Index Interpretation\label{app:field_theoretic_argument_weak_index_interpretation}}

In this Appendix, we present a field-theoretic argument for Physical Result \ref{rslt:strong_weak_injection}. To do so, we must first stipulate how to associate physical phases to cohomology classes (Apps.\,\ref{subapp:Kitaev_construction} and \ref{subapp:generalization_translational_symmetry}). Then we can check if the map $\alpha$ in Physical Result \ref{rslt:strong_weak_injection} on the mathematical side corresponds to the layering construction on the physical side (App.\,\ref{subapp:weak_index_interpretation}).

The arguments below apply equally to the fermionic and the bosonic cases.

\subsection{Kitaev's construction \label{subapp:Kitaev_construction}}

We follow the prescription of Refs.\,\cite{Kitaev_Stony_Brook_2013_SRE, Kitaev_KITP} to associate $(d-1)$-dimensional SPT phases protected by on-site unitary symmetry $G$ to cohomology classes $\brackets{c} \in h^{d-1} \paren{BG}$. The construction is essentially a nonlinear sigma model with target space $BG$. There are some subtleties discussed in Refs.\,\cite{Kitaev_Stony_Brook_2013_SRE, Kitaev_KITP} that we will sweep under the rug here.

To wit, we first associate to each map $c: BG \fromto F_{d-1}$ and spatial slice $X$ the state
\begin{eqnarray}
\ket{\Psi\paren{c, X}} = \int_{\Map\paren{X, BG}} \ket{m} \otimes \ket{\psi\paren{c,m}} \D m,
\end{eqnarray}
where $m$ is a chiral field over $X$ with target space $BG$, and $\ket{\psi(c \circ m)}$ is a pattern of SRE states that looks like $c\paren{m(x)} \in F_{d-1}$ around $x \in X$. Then, to each cohomology class $\brackets{c} \in h^{d-1} \paren{BG}$, we associate the $(d-1)$-dimensional $G$-protected SPT phase represented by a system whose unique ground state on a spatial slice $X$ is $\ket{\Psi\paren{c,X}}$, where $c$ is any representative of $\brackets{c}$.

\subsection{A generalization to translational symmetry \label{subapp:generalization_translational_symmetry}}

We propose a generalization of the construction in Refs.\,\cite{Kitaev_Stony_Brook_2013_SRE, Kitaev_KITP} that will enable us to associate $d$-dimensional SPT phases protected by discrete spatial translational symmetry $\ZZZ$ and on-site unitary symmetry $G$ to cohomology classes $\brackets{c'} \in h^d\paren{B\paren{\ZZZ \times G}}$.

More specifically, over a spatial slice $Y = \RRR \times X$, where $\RRR$ is the direction along which discrete spatial translational symmetry is assumed, we let there be two fields: a chiral field $m'$ with target aspace $BG$ and a background field $e^{i\phi}$ with target space $\SS^1 \homeomorphic U(1)$\footnote{We thank Ryan Thorngren for suggesting the idea of a background field.}. The latter can be thought of as the vacuum expectation value of an order parameter characterizing the translational symmetry breaking. It should thus be constant over $X$ and wind around $\SS^1$ periodically along $\RRR$:
\begin{eqnarray}
\phi( x_0 + 1 ) = \phi(x_0) + 2\pi,
\end{eqnarray}
which guarantees that 
$e^{i\phi(x_0+1)} = e^{i\phi(x_0)}$.
Here, $x_0$ and $x$ are the coordinates for $\RRR$ and $X$, respectively. We have dropped $x$ from the arguments of $\phi$ for brevity.

\begin{figure}
\centering
\includegraphics[width=1.2in]{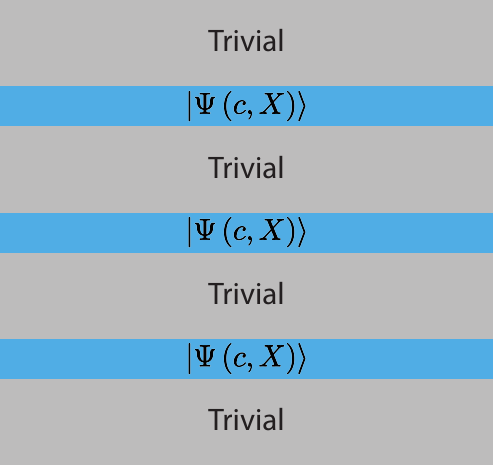}
\caption{(color online). A stack of identical copies of $\ket{\Psi\paren{c, X}}$ (blue) separated by trivial slabs (gray).}
\label{fig:layering_construction_alternation}
\end{figure}

\begin{figure}
\centering
\includegraphics[width=2in]{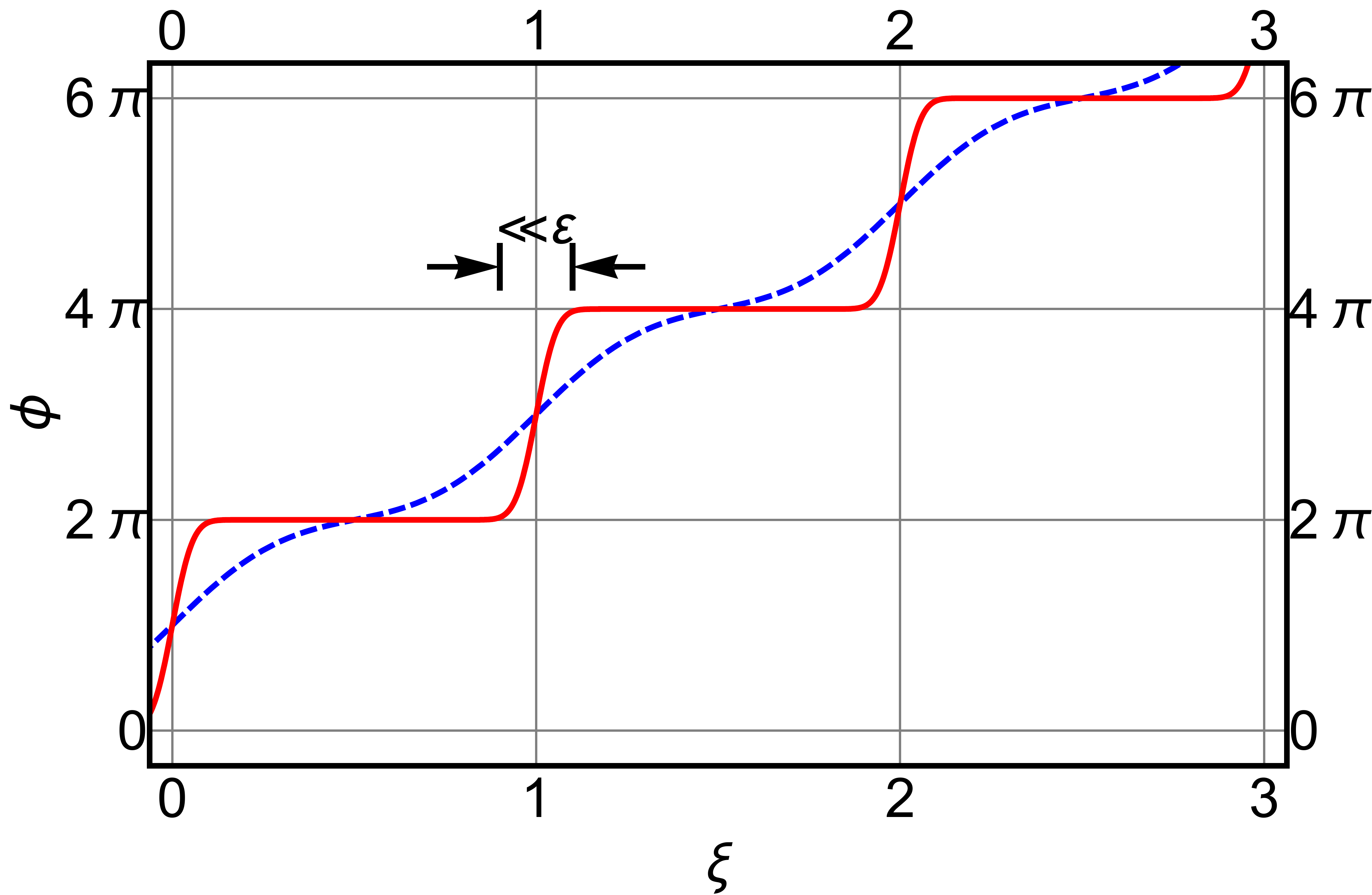}
\caption{(color online). We deform $\phi$ from the dashed blue curve to the solid red curve, so that transitions occur within intervals of size much less than the short-distance cutoff $\epsilon$ for $m'$.}
\label{fig:background_field_deformation}
\end{figure}

Now, we associate to each map $c': \SS^1 \times BG \fromto F_d$ and spatial slice $Y = \RRR \times X$ the state
\begin{eqnarray}
\ket{\Psi\paren{c',\phi, X}}
= \int_{\Map\paren{Y, F_d}} \ket{m'} \otimes \ket{\psi\paren{c',\phi, m'}} \D m',
\end{eqnarray}
where $\ket{\psi\paren{c',\phi, m'}}$ is the pattern of SRE states that looks like $c'\paren{e^{i\phi(x_0)}, m'(x_0, x)}$ around $(x_0, x)\in \RRR \times X$. Then, to each cohomology class $\brackets{c'} \in h^d\paren{B \paren{\ZZZ \times G}} \isomorphic h^d\paren{\SS^1 \times BG}$, we associate the $d$-dimensional $\paren{\ZZZ\times G}$-protected SPT phase represented by a system whose unique ground state on a spatial slice $Y = \RRR \times X$ is $\ket{\Psi\paren{c',\phi, X}}$, where $c'$ is any representative of $\brackets{c'}$.

\subsection{Weak-index interpretation \label{subapp:weak_index_interpretation}}

Take any $\brackets{c} \in h^{d-1} \paren{BG}$ and let $\brackets{c'} \in h^d\paren{B\paren{\ZZZ \times G}}$ be its image under $\alpha$. Since $F_{d-1} \homotopic \Omega F_d$, the cohomology class $\brackets{c}$ can be represented by a map
\begin{equation}
c: BG \fromto \Omega F_d,
\end{equation}
which sends each point of $BG$ to a loop in $F_d$, or equivalently a map
\begin{equation}
c: \SS^1 \times BG \fromto F_d
\end{equation}
subject to the constraint that it sends all of $\braces{s_0} \times BG$ to the basepoint of $F_d$, where $s_0$ denotes the basepoint of $\SS^1$. On the other hand, since $B\paren{\ZZZ \times G} \homotopic \SS^1 \times BG$, the cohomology class $\brackets{c'}$ can also be represented by a map
\begin{equation}
c': \SS^1 \times BG \fromto F_d,
\end{equation}
but without any constraint. One can show that $\alpha$ can be defined by setting
\begin{equation}
c' = c.
\end{equation}

We now argue, by tinkering with the background field, that $\ket{\Psi\paren{c,\phi, X}}$ can be obtained by stacking identical copies of $\ket{\Psi\paren{c, X}}$ separated by trivial slabs (see Fig.\,\ref{fig:layering_construction_alternation}). To that end, let us assume, in the spirit of Ref.\,\cite{Kitaev_IPAM}, that there is a short distance cutoff $\epsilon$ for the chiral field $m'$. We deform $\phi$ according to Fig.\,\ref{fig:background_field_deformation}: we create a series of plateaus and squeeze transitions between them to within a distance much less than $\epsilon$ from integral values of $x_0$. Symmetry is preserved during the deformation, presumably so is the gap. Since the constant loop in $F_d$ corresponds to a trivial $(d-1)$-dimensional state, the $\ket{\Psi\paren{c,\phi, X}}$ must now look trivial away from integral values of $x_0$. This effectively decouples layers corresponding to different transitions between plateaus, each of which is nothing but a copy of $\ket{\Psi\paren{c, X}}$. We have achieved the factorization
\begin{equation}
\ket{\Psi\paren{c,\phi, X}} = \cdots \otimes \ket{\Psi\paren{c, X}} \otimes \ket{\rm trivial} \otimes \ket{\Psi\paren{c, X}} \otimes \ket{\rm trivial} \otimes \cdots.
\end{equation}

\section{Categorical Viewpoint\label{app:categorical_viewpoint}}

In this appendix, we revisit the Generalized Cohomology Hypothesis from a categorical perspective. As we will see, the Hypothesis can be stated more succinctly in categorical language (see App.\,\ref{subapp:categories_functors_natural_transformations} for background).

\subsection{Paraphrase of the Generalized Cohomology Hypothesis\label{subapp:paraphrase_generalized_cohomology_hypothesis}}

The classification of SPT phases can be viewed as a sequence of contravariant functors
\begin{eqnarray}
\SPT^d: \Grp \fromto \Ab^\delta
\end{eqnarray}
indexed by nonnegative integers $d\in \NNN$. Given a group $G$, $\SPT^d\paren{G}$ is the discrete abelian group of $d$-dimensional $G$-protected SPT phases. Given a group homomorphism $\varphi$, $\SPT^d(\varphi)$ is the map defined by pulling back representations, as in Sec.\,\ref{subsubsec:functoriality}.
We can paraphrase the Generalized Cohomology Hypothesis as follows:

\begin{nameddef}[Generalized Cohomology Hypothesis (Categorical Version)]
There exists a generalized cohomology theory $h$ such that there are natural isomorphisms
\begin{equation}
\SPT^d(G) \isomorphic h^d(BG), ~ \forall d\in \NNN.
\end{equation}
\end{nameddef}

Note the left-hand side is defined physically while the right-hand side is purely mathematical. The Hypothesis bridges physics and mathematics.

But life is not always as good as natural isomorphisms. In practice, what one can do is to propose a \emph{construction}, which can be viewed as a family of maps
\begin{equation}
h^d\paren{BG} \fromto \SPT^d\paren{G}.
\end{equation}
Such maps may or may not be bijective, but they had better be homomorphisms between discrete abelian groups and respect the functorial structure. In other words, they had better form a natural transformation for each $d$. Under certain conditions, this can be achieved through a redefinition of the additive or functorial structures of $h^d$ if it is not already the case. Alternatively, one can propose a \emph{topological invariant}, which can be viewed as a family of maps
\begin{equation}
\SPT^d\paren{G} \fromto h^d\paren{BG}.
\end{equation}
The target $h^d\paren{BG}$ is the abelian group that the topological invariant takes values in.\footnote{In practice, one often uses multiple topological invariants at once, but we can as well group them together and treat the collection of invariants as one invariant, which is valued in some more complicated abelian group.} A legitimate topological invariant should stay constant as one deforms a system within the same phase, which corresponds to the well-definedness of the maps. We generally want these maps to form a natural transformation for each $d$ as well.

\subsection{Further examples\label{subapp:further_examples}}

Let us exemplify how this categorical lingo can be used.

We can say that Ref.\,\cite{Wen_Boson} proposed a construction (at least for finite groups)
\begin{equation}
H^{d+2}\paren{BG;\ZZZ} \fromto \SPT^d\paren{G}, \label{further_examples_Borel_group_cohomology_construction}
\end{equation}
and proved that the maps were well-defined. They actually form a natural transformation for each $d$ as per App.\,\ref{app:additivity_functoriality_group_cohomology_construction}, though the original paper did not set out to prove this.

Ref.\,\cite{Kapustin_Boson}, on the other hand, proposed a topological invariant (at least for finite groups)
\begin{equation}
\SPT^d(G) \fromto \Hom\paren{MSO_{d+1}(BG), U(1)},
\end{equation}
which was in principle defined by first gauging a given system and then taking the long-distance limit. Since the expotentiated action (which takes values in $U(1)$) multiplies under stacking, this map should be a homomorphism. The functorial structure was presumably also respected, so the map should form a natural transformation. Unfortunately, the limiting process was not made explicit in Ref.\,\cite{Kapustin_Boson}, so these statements cannot be verified in detail.

Composing the two maps above, we obtain a commutative diagram:
\begin{eqnarray}
\adjustbox{valign=m}{\includegraphics{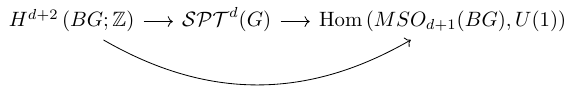}}
\end{eqnarray}
Ref.\,\cite{Kapustin_Boson} conjectured that the composition -- the curved arrow -- was induced by the canonical map between the bordism group and homology group of $BG$. All three arrows in the diagram are natural transformations, but none of them is a priori a natural isomorphism. In fact, as Ref.\,\cite{Kapustin_Boson} pointed out, the curved arrow is in general neither surjective, reflecting the fact that the cobordism proposal may predict phases that the group cohomology proposal does not capture (which occurs in $d=3$ dimensions), nor injective, reflecting the fact that systems that are predicted to be nontrivial in the group cohomology proposal may be seen as being trivial (which occurs in $d=6$ dimensions).

Finally, we can say that what we did in App.\,\ref{app:field_theoretic_argument_weak_index_interpretation} was to specify the horizontal arrows in the diagram below and argue that the diagram commutes:
\begin{eqnarray}
\adjustbox{valign=m}{\includegraphics{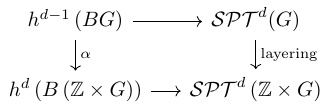}}
\end{eqnarray}
We used Kitaev's construction \cite{Kitaev_IPAM} for the upper horizontal arrow and proposed a generalized construction for the lower horizontal arrow.

\section{Additivity and Functoriality of the Group Cohomology Construction\label{app:additivity_functoriality_group_cohomology_construction}}

In this subsection we will show, within the group cohomology construction \cite{Wen_Boson} of bosonic SPT phases (for finite groups), that adding cohomology classes corresponds to stacking SPT phases (see Sec.\,\ref{subsubsec:additivity}), and that the map induced by a homomorphism between symmetry groups corresponds to replacing the symmetry group (see Sec.\,\ref{subsubsec:functoriality}). We will begin with the 1-dimensional case.

\subsection{1-dimensional case\label{subapp:1_dimensional_case}}

Let us review the construction in Ref.\,\cite{Wen_Boson}, specializing to 1 dimension. Take a finite symmetry group $G$. Consider a ring with $N$ sites and associate to each site the $\bars{G}$-dimensional Hilbert space $\CCC G$, which has orthonormal basis $\braces{\ket{g} | g\in G}$ and on which $G$ acts according to $\rho_{g} \ket{g_i} = \ket{gg_i}$. We define $\ket{\phi} \coloneq \frac{1}{\sqrt{\bars G}} \sum_{g\in G} \ket{g}$ and $\hat P_i \coloneq \mathbb I^{\otimes (i-1)} \otimes \ket{\phi}\bra{\phi} \otimes \mathbb I^{\otimes (N-i)} $. Then the Hamiltonian
\begin{equation}
\hat H(0) \coloneq - \sum_{i =1}^N \hat P_i
\end{equation}
is local, preserves the symmetry, and has a unique, gapped ground state,
\begin{equation}
\ket{\Psi(0)} = \ket{\phi}^{\otimes N}.
\end{equation}
Given a 2-cocycle $\nu \in \Hom_{G}\paren{G^3, U(1)}$, we define a diagonal, local unitary operator,
\begin{equation}
\hat U(\nu) \coloneq \sum_{g_1, \ldots, g_N \in G} 
\brackets{ \nu\paren{1, g_1, g_N}^{-1} \prod_{i=1}^{N-1} \nu\paren{1, g_i, g_{i+1} } } \ketbra{ \braces{g_i} }{ \braces{g_i} }.\label{1_dimensional_case_U(nu)}
\end{equation}
Then the Hamiltonian corresponding to $\nu$ is given by
\begin{equation}
\hat H(\nu) \coloneq \hat U(\nu) \hat H(0) \hat U(\nu)^\dag,
\end{equation}
which is local and symmetry-preserving because $\hat H(0)$ and $\hat U(\nu)$ are. It has a unique, gapped ground state,
\begin{eqnarray}
\ket{ \Psi(\nu) } = \frac{1}{\sqrt{\bars{G}^N}} \sum_{ g_1, \ldots, g_N \in G }
\brackets{ \nu\paren{1, g_1, g_N}^{-1} \prod_{i=1}^{N-1} \nu\paren{1, g_i, g_{i+1} } }
\ket{g_1, \ldots, g_N }.\label{1_dimensional_case_Psi(nu)}
\end{eqnarray}

\subsubsection{Adding cohomology classes = stacking SPT phases\label{subsubapp:adding_cohomology_classes_stacking_SPT_phases}}

Envision two rings as in App.\,\ref{subapp:1_dimensional_case}, corresponding to 2-cocycles $\nu$ and $\nu'$, respectively. Stacking one ring on top of the other produces another 1-dimensional system. With an augmented Hilbert space $\CCC G \otimes \CCC G$ associated to each (composite) site, this composite system is no longer given by the group cohomology construction per se. It is, nevertheless, in the same phase as a system constructed as such, namely the one corresponding to the sum $\nu \nu'$ of $\nu$ and $\nu'$, as we show below\footnote{%
Recall that there is an additive structure on the set of 2-cocycles, defined by $\paren{\nu\nu'}(g_0, g_1, g_2) \coloneq \nu(g_0,g_1,g_2) \nu(g_0,g_1,g_2)$. Addition of cocycles is written multiplicatively because, in physics, the composition law of $U(1)$ is usually considered multiplicative rather than additive.}.
Thus, the mathematical addition of cocycles, and hence cohomology classes, corresponds precisely to the physical stacking of SPT phases.

\begin{figure}[t]
\centering
\includegraphics[width=3.3in]{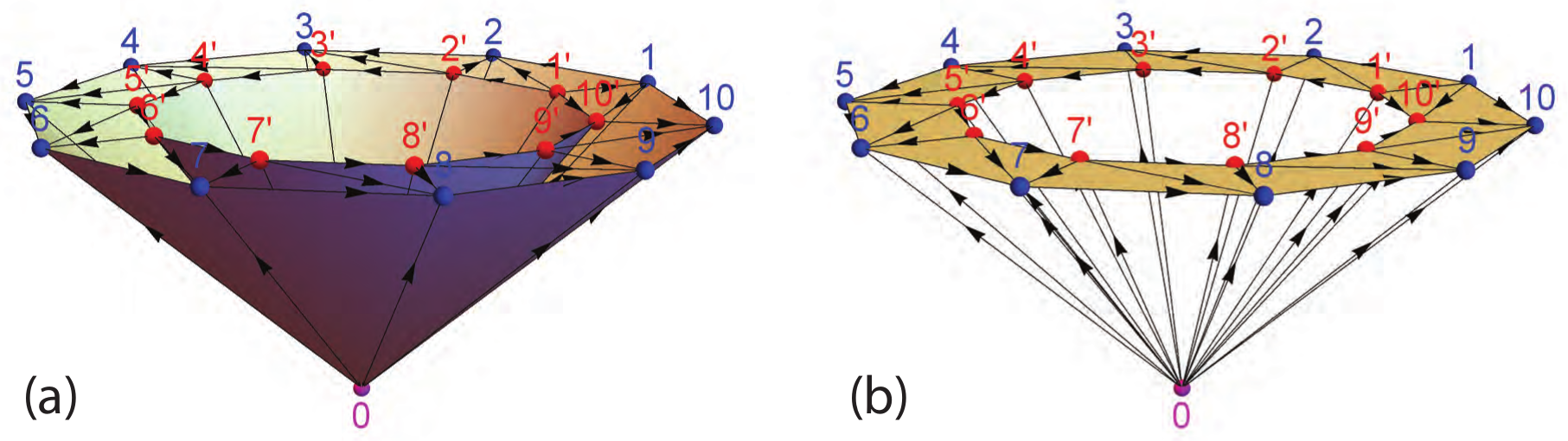}
\caption{(color online). Two 1-dimensional systems, which consist of vertices labeled 1 through 10 (blue) and $1'$ through $10'$ (red), respectively, are stacked together to form a new 1-dimensional system. With the introduction of an auxiliary vertex $0$ (magenta), a cone is formed for each system. The ground states $\ket{\Psi(\nu)}$ and $\ket{\Psi(\nu')}$ are then given by ``integrating" $\nu$ and $\nu'$ over the two cones, respectively -- this is a standard procedure in topological quantum field theories \cite{Atiyah_TQFT}. 
The coefficients in Eq.\,(\ref{adding_cohomology_classes_stacking_SPT_phases_U1}) and Eq.\,(\ref{conjugation_unitary_2}) are the ``integrals" of $\nu'$ over the shaded ``surfaces" (i.e.\,chains) in (a) and (b), respectively.
The two are equal because the chains in (a) and (b) are homologous.
}
\label{fig:homologous_chains}
\end{figure}

To show that the composite system with the Hamiltonian
\begin{equation}
\hat H(\nu) \otimes \hat H(\nu') = \hat U(\nu) \hat H(0) \hat U(\nu)^\dag \otimes \hat U(\nu') \hat H(0) \hat U(\nu')^\dag
\end{equation}
is in the same phase as the system with the Hamiltonian
\begin{equation}
\hat H(\nu \nu') = \hat U(\nu \nu') \hat H(0) \hat U(\nu \nu')^\dag,
\end{equation}
we first tensor the latter with a trivial ancillary ring, yielding $\hat H(\nu\nu') \otimes \hat H(0)$. Since $\hat H(\nu\nu') \otimes \hat H(0)$ is related to $\hat H(\nu) \otimes \hat H(\nu')$ by conjugation by the unitary operator
\begin{eqnarray}
\hat{\widetilde U}_1 &\coloneq& U(\nu \nu') U(\nu)^\dag \otimes U(\nu')^\dag \nonumber\\
&=& \sum_{\braces{g_i}, \braces{g_i'}} 
\brackets{
\frac{ \nu'\paren{1,g_1',g_N'} }{ \nu'\paren{1,g_1,g_N} }
\prod_{i=1}^{N-1} \frac{ \nu'\paren{ 1, g_i, g_{i+1} } }{ \nu'\paren{ 1, g_i', g_{i+1}' } } 
}
\ketbra{ \begin{matrix} \braces{g_i} \\ \braces{g_i'} \end{matrix} }{ \begin{matrix} \braces{g_i} \\ \braces{g_i'} \end{matrix} },\label{adding_cohomology_classes_stacking_SPT_phases_U1}
\end{eqnarray}
it suffices to find a path from $\mathbb I$ to $\hat{\widetilde U}_1$ via local unitary operators that preserve the symmetry. Here, $\braces{g_i}$ and $\braces{g_i'}$ are variables on the first and the second rings, respectively. By the cocycle condition $d\nu' = 0$, we have
\begin{equation}
\frac{ \nu'\paren{1, g_i, g_j} }{ \nu'\paren{1, g_i', g_j' } } = \frac{ \nu'\paren{g_i' , g_i, g_j} }{ \nu\paren{ g_i' , g_j', g_j } } \frac{ \nu'\paren{ 1, g_j', g_j } }{ \nu'\paren{ 1, g_i', g_i } }\label{adding_cohomology_classes_stacking_SPT_phases_nu}
\end{equation}
for all $i$ and $j$,
which enables us to rewrite
\begin{eqnarray}
\hat{\widetilde U}_1 =
\sum_{\braces{g_i}, \braces{g_i'}} 
\brackets{
\frac{ \nu'\paren{g_1,g_N'
,g_N} }{ \nu'\paren{g_1',g_1,g_N'} }
\prod_{i=1}^{N-1} \frac{ \nu'\paren{ g_i', g_i, g_{i+1} } }{ \nu'\paren{ g_i', g_{i+1}', g_{i+1} } } }
\ketbra{ \begin{matrix} \braces{g_i} \\\braces{g_i'} \end{matrix} }{ \begin{matrix} \braces{g_i} \\\braces{g_i'} \end{matrix} }.\label{conjugation_unitary_2}
\end{eqnarray}
Geometrically, this amounts to replacing the chain shown in Fig.\,\ref{fig:homologous_chains}(a) by the chain shown in Fig.\,\ref{fig:homologous_chains}(b). In this new form, $\hat{\widetilde U}_1$ would preserve the symmetry even if $\nu'$ failed to satisfy the cocycle condition. Take a path $\nu'_t$ in the space of 2-cochains that begins at the trivial 2-cochain and ends at $\nu'$. Then
\begin{eqnarray}
\hat{\widetilde U}_t \coloneq
\sum_{\braces{g_i}, \braces{g_i'}} 
\brackets{
\frac{ \nu_t'\paren{g_1,g_N',g_N} }{ \nu_t'\paren{g_1',g_1,g_N'} }
\prod_{i=1}^{N-1} \frac{ \nu'_t\paren{ g_i', g_i, g_{i+1} } }{ \nu'_t\paren{ g_i', g_{i+1}', g_{i+1} } } 
} 
\ketbra{ \begin{matrix} \braces{g_i} \\\braces{g_i'} \end{matrix} }{ \begin{matrix} \braces{g_i} \\\braces{g_i'} \end{matrix} },
\end{eqnarray}
for $0 \leq t \leq 1$, is a path from $\mathbb I$ to $\hat{\widetilde U}_1$ via local unitary operators that preserve the symmetry, as desired.

\subsubsection{Induced cohomology class = replaced symmetry group \label{subsubapp:induced_cohomology_class_replaced_symmetry_group}}

Consider two possible symmetry groups $G'$ and $G$ and a homomorphism $\varphi: G' \fromto G$ between them. 
A 2-cocycle $\nu$ of $G$ determines a 1-dimensional system representing a $G$-protected SPT phase via the construction in App.\,\ref{subapp:1_dimensional_case}. It has the Hilbert space $\CCC G$ associated to each site, the $G$-action $\rho_g \ket{g_i} = \ket{gg_i}$, and the Hamiltonian $\hat H(\nu)$. We denote this system by $\paren{\CCC G, \rho, \hat H(\nu)}$.

Precomposing $\rho$ with $\varphi$, we obtain a $G'$-action on $\CCC G$:
\begin{equation}
\paren{\rho \circ \varphi}_{g'} \ket{g_i} = \rho_{\varphi(g')} \ket{g_i} = \ket{ \varphi(g') g_i }.
\end{equation}
The Hamiltonian $\hat H(\nu)$ commutes with $\rho\circ \varphi$ since it does with $\rho$. Thus the same physical system can also be viewed as a representative of a $G'$-protected SPT phase. Physically, this amounts to forgetting those symmetry operations in $G$ that are not in the image of $\varphi$, and relabelling those in the image of $\varphi$ by elements of $G'$ in a possibly redundant manner. We denote this system by $\paren{\CCC G, \rho\circ\varphi, \hat H(\nu)}$.

On the other hand, the mathematical structure of group cohomology is such that every homomorphism $\varphi:G'\fromto G$ gives rise to an induced homomorphism $\varphi^*$ from the discrete abelian group of 2-cocycles of $G$ to the discrete abelian group of 2-cocycles of $G'$. More explicitly, this $\varphi^*$ sends a 2-cocycle $\nu$ of $G$ to the 2-cocycle
\begin{eqnarray}
\varphi^* \nu: G' \times G' \times G' &\fromto& U(1) \nonumber\\
(g_0', g_1', g_2') &\mapsto& \nu\paren{\varphi(g_0'), \varphi(g_1'), \varphi(g_2')}
\end{eqnarray}
of $G'$. For the given $\varphi$ and $\nu$, the 2-cocycle $\varphi^* \nu$ determines, via the construction in App.\,\ref{subapp:1_dimensional_case}, a system that represents a $G'$-protected SPT phase. We denote this system by $\paren{\CCC G', \rho', \hat H(\varphi^* \nu)}$.

A good construction of SPT phases should have functoriality built into its mathematical structure. It would therefore be ideal if the systems $\paren{\CCC G', \rho', \hat H(\varphi^* \nu)}$ and $\paren{\CCC G, \rho\circ\varphi, \hat H(\nu)}$ were actually the same, which is unfortunately false unless $\varphi$ is an isomorphism. They are, however, in the same $G'$-protected SPT phase, as we now show.

To that end, let us recall that every group homomorphism can be factored as the composition of a surjective homomorphism and an inclusion. Thus it suffices to consider these two special cases.

First, suppose $\varphi: G' \fromto G$ is an inclusion. We will deform the system $\paren{\CCC G, \rho\circ\varphi, \hat H(\nu)}$ into the system $\paren{\CCC G', \rho', \hat H(\varphi^* \nu)}$ step by step. To begin, let $S$ be a set of representatives of the right cosets of $G'$ in $G$. That is,
\begin{eqnarray}
&&G' s_1 \cap G' s_2 = \emptyset,~\forall s_1 \neq s_2\in S,\\
&& \cup_{s\in S} G' s = G.
\end{eqnarray}
We can assume that the identity $1\in G$ is contained in $S$. Given any $g\in G$, there is a unique pair $(g',s)\in G' \times S$ for which $g's = g$. We can thus rewrite every basis state $\ket g$ in the form $\ket{g'} \otimes \ket{s}$ and pretend that the Hilbert space $\CCC G$ is the tensor product of $\CCC G'$ and $\CCC S$. The $G'$-action $\rho \circ \varphi$ on $\CCC G$ then goes over into
\begin{equation}
\paren{\rho\circ \varphi}_{g'} \paren{\ket{g'_i} \otimes \ket{s} } = \ket{g' g'_i} \otimes \ket{s}.
\end{equation}
Next, we choose a path $\hat{\widetilde W}_t$ of unitary operators on $\CCC G' \otimes \CCC S$ that acts trivially on $\CCC G'$ for all $t\in [0,1]$, equals $\mathbb I$ at $t=0$, and sends $\ket{g_i'} \otimes \frac{1}{\sqrt{\bars S}} \sum_{s\in S} \ket{s} $ to $\ket{g_i'} \otimes \ket{1}$ at $t=1$. Since $\hat{\widetilde W}_t$ commutes with the $G'$-action for all $t$, so does the family of local unitary operators
\begin{eqnarray}
\hat{\widetilde U}_t \coloneq \hat U(\nu) \hat{\widetilde W}\,_t^{\otimes N} \hat U(\nu)^\dag.
\end{eqnarray}
The path $\hat{\widetilde U}_t \ket{\Psi(\nu)}$ establishes an equivalence between $\ket{\Psi(\nu)} = \hat{\widetilde U}_0 \ket{\Psi(\nu)}$ and
\begin{eqnarray}
\hat{\widetilde U}_1 \ket{\Psi(\nu)} &=& \hat U(\nu) \hat{\widetilde W}\,_1^{\otimes N} \hat U(\nu)^\dag \hat U(\nu) \ket{\Psi(0)} \nonumber\\
&=& \hat U(\nu) \hat{\widetilde W}\,_1^{\otimes N} \ket{\Psi(0)} \nonumber\\
&=& \hat U(\nu) \paren{ \hat{\widetilde W}_1 \frac{1}{\sqrt{\bars {G}}} \sum_{g'\in G'} \ket{g'} \otimes \sum_{s\in S} \ket{s} }^{\otimes N} \nonumber\\
&=& \hat U(\nu) \paren{ \frac{1}{\sqrt{\bars{G'}}} \sum_{g'\in G'} \ket{g'} \otimes \ket{1} }^{\otimes N}.
\end{eqnarray}
Restoring the old notation, the last expression reads
\begin{eqnarray}
\hat U(\nu) \sum_{ g_1, \ldots, g_N \in G } \frac{ \eta\paren{\braces{g_i}} }{ \sqrt{\bars{G'}^N} } \ket{ \braces{g_i} } 
= \sum_{ g_1, \ldots, g_N \in G } \frac{ \eta\paren{\braces{g_i}} }{ \sqrt{\bars{G'}^N} }
\brackets{
\nu\paren{1,g_1,g_N}^{-1} \prod_{i=1}^{N-1} \nu\paren{1, g_i, g_{i+1} }
}
\ket{ \braces{g_i} },
\end{eqnarray}
where $\eta\paren{\braces{g_i}} = 1$ if $g_i\in G'$ for all $i$ and $0$ otherwise. But this is related to the ground state
\begin{equation}
\ket{\Psi\paren{\varphi^*\nu}} = \frac{1}{ \sqrt{\bars{G'}^N} }
\sum_{ g_1', \ldots, g_N' \in G' } \brackets{ \nu\paren{1,g_1',g_N'}^{-1} \prod_{i=1}^{N-1} \nu\paren{ 1, g_i', g_{i+1}' } } \ket{ \braces{g_i'} }
\end{equation}
of $\paren{\CCC G', \rho', \hat H(\varphi^* \nu)}$ by a symmetry-preserving isometry (the one induced by the inclusion $\CCC G' \subset \CCC G$), and hence equivalent to it.

Next, suppose $\varphi: G' \fromto G$ is a surjective homomorphism. We will deform the system $\paren{\CCC G', \rho', \hat H(\varphi^* \nu)}$ into the system $\paren{\CCC G, \rho\circ\varphi, \hat H(\nu)}$ step by step. To begin, let $R = \kernel(\varphi)$, and $T$ be a set of representatives of the left cosets of $R$ in $G'$. That is,
\begin{eqnarray}
&&t_1 R \cap t_2 R = \emptyset,~\forall t_1\neq t_2\in T, \\
&& \cup_{t\in T} t R = G'.
\end{eqnarray}
Given any $g'\in G'$, there is a unique pair $(t,r)\in T \times R$ for which $tr = g'$. We can thus rewrite every basis state $\ket {g'}$ in the form $\ket{t} \otimes \ket{r}$ and pretend that the Hilbert space $\CCC G'$ is the tensor product of $\CCC T$ and $\CCC R$. In this new form, the $G'$-action satisfies
\begin{equation}
\rho'_{g'} \paren{ \ket t \otimes \ket{\phi_R} } = \ket {g'.t} \otimes \ket{\phi_R},
\end{equation}
where $\ket{\phi_R} = \frac{1}{\sqrt{\bars R}} \sum_{r\in R} \ket r$ and $g'.t$ is the unique element of $T$ for which $\varphi(g'.t) = \varphi(g') \varphi(t)$. The ground state $\ket{\Psi(\varphi^* \nu)}$ of $\paren{ \CCC G', \rho', \hat H(\varphi^* \nu) }$ goes over into
\begin{eqnarray}
&& \frac{1}{ \sqrt{\bars{G'}^N} }
\sum_{ \stackrel{ t_1, \ldots, t_N\in T }{ r_1, \ldots, r_N \in R }}
\brackets{
\paren{\varphi^*\nu} \paren{ 1, t_1r_1, t_Nr_N }^{-1}
\prod_{i=1}^{N-1} \paren{\varphi^*\nu} \paren{ 1, t_ir_i, t_{i+1}r_{i+1} } 
}
\ket{ \braces{t_i} } \otimes \ket{ \braces{r_i} } \nonumber\\
&=& \frac{1}{ \sqrt{\bars{G'}^N} }
\sum_{ \stackrel{ t_1, \ldots, t_N\in T }{ r_1, \ldots, r_N \in R }}
\brackets{
\nu \paren{ 1, \varphi(t_1), \varphi(t_N) }^{-1}
\prod_{i=1}^{N-1} \nu \paren{ 1, \varphi(t_i), \varphi(t_{i+1}) } }
\ket{ \braces{t_i} } \otimes \ket{ \braces{r_i} } \nonumber\\
&=& \braces{
\frac{1}{ \sqrt{\bars{T}^N} }
\sum_{ t_1, \ldots, t_N\in T }
\brackets{
\nu \paren{ 1, \varphi(t_1), \varphi(t_N) }^{-1}
\prod_{i=1}^{N-1} \nu \paren{ 1, \varphi(t_i), \varphi(t_{i+1}) } 
}
\ket{ \braces{t_i} } } 
\otimes \ket{\phi_R}^{\otimes N}.
\end{eqnarray}
Removing the trivial ancilla $\ket{\phi_R}^{\otimes N}$, we obtain the equivalent state
\begin{equation}
\frac{1}{ \sqrt{\bars{T}^N} } \sum_{ t_1, \ldots, t_N\in T } \brackets{ \nu \paren{ 1, \varphi(t_1), \varphi(t_N) }^{-1} \prod_{i=1}^{N-1} \nu \paren{ 1, \varphi(t_i), \varphi(t_{i+1}) } } \ket{ \braces{t_i} }.
\end{equation}
Since $\varphi$ gives a bijection between $T$ and $G$, we can relabel the states $\ket{t_i}$ by elements of $G$, yielding
\begin{eqnarray}
&& \frac{1}{ \sqrt{\bars{T}^N} }
\sum_{ t_1, \ldots, t_N\in T }
\brackets{ \nu \paren{1, \varphi(t_1), \varphi(t_N) }^{-1} \prod_{i=1}^{N-1} \nu \paren{ 1, \varphi(t_i), \varphi(t_{i+1}) } }
\ket{ \braces{\varphi(t_i)} }\nonumber\\
&=& \frac{1}{ \sqrt{\bars{G}^N} }
\sum_{ g_1, \ldots, g_N\in G }
\brackets{ \nu \paren{ 1, g_1, g_N }^{-1} \prod_{i=1}^{N-1} \nu \paren{ 1, g_i, g_{i+1} } }
\ket{ \braces{g_i} }.
\end{eqnarray}
This is nothing but the ground state $\ket{\Psi(\nu)}$ of $\paren{\CCC G, \rho\circ\varphi, \hat H(\nu)}$.

\subsection{Higher-dimensional case\label{subapp:higher_dimensional_case}}

Take a finite symmetry group $G$. Consider a triangulated $d$-dimensional oriented closed manifold $M$ together with a total ordering of the vertices\footnote{%
Ref.\,\cite{Wen_Boson} considered ``branching structures" instead of total orderings of vertices, but this distinction is inconsequential.}, which we accordingly label by 1, 2, $\ldots$, $N$.
We denote by $\Delta_0, \ldots, \Delta_d$ the vertices of a $d$-simplex $\Delta$, with $\Delta_0 < \cdots < \Delta_d$. The ordering $\Delta_0 < \cdots < \Delta_d$ determines an orientation of $\Delta$, which may or may not agree with that of $M$. We set $\mathcal O(\Delta)=1$ if it does and $\mathcal O(\Delta)=-1$ otherwise.
Given a $(d+1)$-cocycle $\nu$, the construction in Ref.\,\cite{Wen_Boson} of $\hat H(\nu)$ and $\ket{\Psi(\nu)}$ is the same as in App.\,\ref{subapp:1_dimensional_case} except that the unitary operator (\ref{1_dimensional_case_U(nu)}) should be replaced by
\begin{equation}
\hat U(\nu) \coloneq \sum_{\braces{g_i}} \prod_{\Delta} \nu \paren{1, g_{\Delta_0}, \ldots, g_{\Delta_d} }^{ \mathcal O(\Delta) } 
\ketbra{ \braces{g_i} }{ \braces{g_i} },
\end{equation}
where $\Delta$ runs over the $d$-simplices of $M$.

\subsubsection{Adding cohomology classes = stacking SPT phases\label{subsubapp:adding_cohomology_classes_stacking_SPT_phases_2}}

Take any $d$-cocycle $\nu'$ of $G$. Since $d\nu'=0$, we have
\begin{eqnarray}
\prod_{k=0}^d d\nu' \paren{ 1, g_{\Delta_0}', \ldots, g_{\Delta_k}', g_{\Delta_k}, \ldots, g_{\Delta_d} }^{(-1)^k} = 1
\end{eqnarray}
for all $g_{\Delta_0}', \ldots, g_{\Delta_d}', g_{\Delta_1}, \ldots, g_{\Delta_d}\in G$. Expanding the left-hand side, one can show that
\begin{eqnarray}
\prod_\Delta \brackets{ \frac{ \nu' \paren{1, g_{\Delta_0}, \ldots, g_{\Delta_d} } }{ \nu' \paren{1, g_{\Delta_0}', \ldots, g_{\Delta_d}' } } }^{ \mathcal O(\Delta) }
= \prod_\Delta \brackets{ \prod_{k=0}^d \nu'\paren{g_0', \ldots, g_k', g_k, \ldots, g_d }^{(-1)^k} }^{\mathcal O(\Delta)}.\label{adding_cohomology_classes_stacking_SPT_phases_nu_2}
\end{eqnarray}
The proof in App.\,\ref{subsubapp:adding_cohomology_classes_stacking_SPT_phases} can be immediately generalized to $d$ dimensions by substituting Eq.\,(\ref{adding_cohomology_classes_stacking_SPT_phases_nu_2}) for Eq.\,(\ref{adding_cohomology_classes_stacking_SPT_phases_nu}), where the vertices of the composite system may be ordered either so that $1'<1<2'<2<\cdots<N'<N$ or so that $1'<\cdots<N'<1<\cdots<N$.

\subsubsection{Induced cohomology class = replaced symmetry group\label{subsubapp:induced_cohomology_class_replaced_symmetry_group_2}}

To generalize the proof in App.\,\ref{subsubapp:induced_cohomology_class_replaced_symmetry_group} to $d$ dimensions, one simply replaces all expressions of the form
\begin{equation}
\nu(1, g_1, g_N)^{-1} \prod_{i=1}^{N-1} \nu(1, g_i, g_{i+1}),
\end{equation}
where $\nu$ is some 2-cocycle and $\braces{g_i}$ is some indexed family of elements of either $G'$ or $G$, by corresponding expressions of the form
\begin{equation}
\prod_{\Delta} \nu\paren{1, g_{\Delta_0}, \ldots, g_{\Delta_d}}^{\mathcal O(\Delta)},
\end{equation}
where $\Delta$ runs over the $d$-simplices of $M$.


\section{Main Mathematical Theorems\label{app:proofs}}

In this appendix, we collect some major theorems upon which the physical results in Sec.\,\ref{sec:consequences_hypothesis_physical_implications} are based. These theorems hold for all generalized cohomology theories. We will denote by $\paren{F_d}_{d\in \ZZZ}$ an arbitrary $\Omega$-spectrum, and by $h$ and $\tilde h$ the unreduced and reduced generalized cohomology theories it defines, respectively. We will begin with notation and conventions, proceed to some lemmas, and then prove the main theorems.



\subsection{Notation and conventions\label{subsec:mathematical_notation_conventions}}

We denote bijections and homeomorphisms by $\homeomorphic$, isomorphisms of algebraic structures by $\isomorphic$, homotopy or pointed homotopy by $\sim$, and homotopy equivalences or pointed homotopy equivalences by $\homotopic$.
We denote the one-point set, the unit interval (i.e.\,$[0,1]$), the boundary of the unit interval (i.e.\,$\braces{0,1}$), the $n$-sphere, the $n$-disk, and the boundary of the $n$-disk by $\pt$, $I$, $\partial I$, $\SS^n$, $\DD^n$, and $\partial \DD^n$, respectively.

Unless stated otherwise, ``map" always means continuous map, ``group" always means topological group, and ``homomorphism" between groups always means continuous homomorphism. The technical conventions in App.\,\ref{subapp:technical_conventions} are observed throughout the paper except in Apps.\,\ref{subapp:notions_algebraic_topology}-\ref{subapp:technical_conventions}.

\subsection{Some lemmas\label{subapp:some_lemmas}}

\begin{lem}\label{lem:reduced_unreduced}
Let $(F_n)$ be an $\Omega$-spectrum and $\paren{X,x_0}$ be a pointed CW-complex. There is a natural split short exact sequence,
\begin{eqnarray}
\adjustbox{valign=m}{\includegraphics{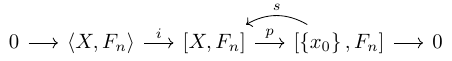}}
\end{eqnarray}
with $s$ induced by the projection $X \onto \braces{x_0}$, $p$ induced by the inclusion $\braces{x_0} \oneone X$, and $i$ given by forgetting basepoints. 
\end{lem}

\begin{pf}
The long exact sequence of reduced cohomology groups of the pair
\begin{equation}
\paren{ \paren{X \times I}/\paren{X \times \partial I}, \paren{\braces{x_0} \times I}/\paren{\braces{x_0} \times \partial I} }
\end{equation}
breaks into short exact sequences, since there is an obvious retraction
\begin{equation}
\paren{X \times I}/\paren{X \times \partial I} \fromto \paren{\braces{x_0} \times I}/\paren{\braces{x_0} \times \partial I}.
\end{equation}
Now apply the suspension-loop adjunction and use the fact that $F_n \homotopic \Omega F_{n+1}$.
\qed\end{pf}

\begin{lem}\label{lem:retraction}
Let $(F_n)$ be an $\Omega$-spectrum and $\paren{X, A}$ be a CW-pair with basepoint $x_0$ together with a retraction $\rho: X \onto A$. There is a natural commutative diagram,
\begin{eqnarray}
\adjustbox{valign=m}{\includegraphics{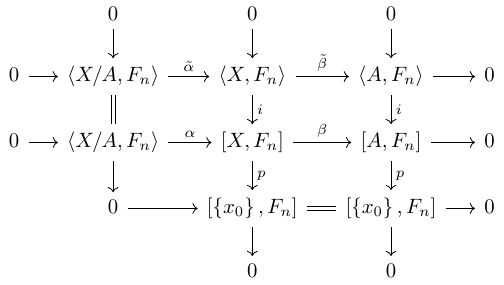}}
\end{eqnarray}
consisting of exact rows and columns,
with $\tilde \alpha$ and $\alpha$ induced by the quotient map $X \onto X/A$, $\tilde \beta$ and $\beta$ induced by the inclusion $A \oneone X$, and $i$ and $p$ as in Lemma \ref{lem:reduced_unreduced}. Furthermore, $\rho$ induces splittings $\tilde \sigma$ and $\sigma$ of the first and second rows, which fit into the commutative diagram
\begin{eqnarray}
\adjustbox{valign=m}{\includegraphics{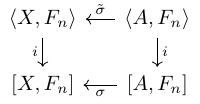}}
\end{eqnarray}
\end{lem}

\begin{pf}
The exactness of the columns follows from Lemma \ref{lem:reduced_unreduced}. The split exactness of the first row follows from the fact that the long exact sequence of reduced cohomology groups of $(X,A)$ breaks into short exact sequences due to the existence of a retraction. The split exactness of the second row follows from diagram chasing. Commutativity and naturality are trivial to check.
\qed\end{pf}

\begin{lem}\label{lem:stable_splitting}
Let $\paren{F_n}$ be an $\Omega$-spectrum and $X$, $Y$ be pointed CW-complexes.
There exists an isomorphism,
\begin{equation}
\angles{X \times Y, F_n} \isomorphic \angles{X\vee Y \vee (X\wedge Y), F_n},
\end{equation}
whose composition,
\begin{equation}
\tilde \lambda: \angles{X \times Y, F_n} \xfromto\isomorphic \angles{X, F_n} \oplus \angles{ X \wedge Y, F_n } \oplus \angles{Y, F_n},
\end{equation}
with the obvious isomorphism
\begin{equation}
\angles{ X \vee \paren{X \wedge Y} \vee Y , F_n } \isomorphic \angles{X, F_n} \oplus \angles{ X \wedge Y, F_n } \oplus \angles{Y, F_n}
\end{equation}
is such
that the canonical inclusions
\begin{eqnarray}
\angles{X, F_n} &\oneone& \angles{X\times Y, F_n}, \\
\angles{X \wedge Y, F_n} &\oneone& \angles{X \times Y, F_n}, \\
\angles{Y, F_n} &\oneone& \angles{X\times Y, F_n}
\end{eqnarray}
are induced by the canonical projections $X\times Y \onto X$, $X\times Y \onto X\wedge Y$, and $X\times Y \onto Y$, respectively,
and that the canonical projections
\begin{eqnarray}
\angles{X \times Y, F_n} &\onto& \angles{X, F_n}, \\
\angles{X \times Y, F_n} &\onto& \angles{Y, F_n}
\end{eqnarray}
are induced by the canonical inclusions $X \oneone X\times Y$ and $Y \oneone X\times Y$, respectively.
\end{lem}

\begin{pf}
Recall there is a stable splitting (Proposition 4I.1 of \cite{Hatcher}),
\begin{equation}
\Sigma (X \times Y) \homotopic \Sigma \paren{X \vee (X \wedge Y) \vee Y}.
\end{equation}
Now apply the suspension-loop adjunction and use the fact that $F_n \homotopic \Omega F_{n+1}$. The rest can be verified straightforwardly.
\qed\end{pf}

\begin{lem}\label{lem:stable_splitting_2}
Let $\paren{F_n}$ be an $\Omega$-spectrum and $X$, $Y$ be pointed CW-complexes.
There exists an isomorphism $\lambda$ fitting into a natural commutative diagram
\begin{eqnarray}
\adjustbox{valign=m}{\includegraphics{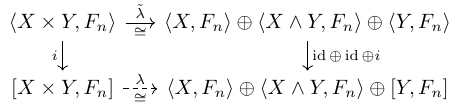}}
\end{eqnarray}
where $i$ is as in Lemma \ref{lem:relationship_reduced_unreduced_generalized_cohomology_theories}, $\tilde \lambda$ is as in Lemma \ref{lem:stable_splitting}, and the canonical injection and projection
\begin{eqnarray}
&& \brackets{Y, F_n} \oneone \brackets{X \times Y, F_n}, \\
&& \brackets{X \times Y, F_n} \onto \brackets{Y, F_n} 
\end{eqnarray}
are induced by the canonical projection $X \times Y \onto Y$ and injection $Y \oneone X \times Y$, respectively.
\end{lem}

\begin{pf}
Extend the columns into short exact sequences according to Lemma \ref{lem:reduced_unreduced}. Then apply the Five Lemma to $\tilde \lambda^{-1}$ and the putative $\lambda^{-1}$.
\qed\end{pf}




\begin{lem}
Let $G$ be any group and $0$ be the trivial group. There is a natural split short exact sequence,
\begin{eqnarray}
\adjustbox{valign=M}{\includegraphics{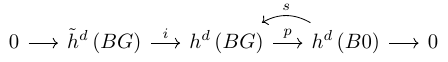}}
\end{eqnarray}
with $s$ induced by the epimorphism $G \onto 0$, $p$ induced by the monomorphism $0 \oneone G$, and $i$ given by forgetting basepoints. \label{lem:relationship_reduced_unreduced_generalized_cohomology_theories}
\end{lem}

\begin{pf}
Set $X=BG$ in Lemma \ref{lem:reduced_unreduced}.
\qed\end{pf}



\begin{cor}
Let $G$ be any group and $0$ be the trivial group. There is a natural isomorphism,
\begin{equation}
h^d(BG) \isomorphic \tilde h^d(BG) \oplus h^d(B 0).
\end{equation}\qed\label{cor:relationship_reduced_unreduced_generalized_cohomology_theories}
\end{cor}

\subsection{Main theorems\label{subapp:main_proofs}}


\begin{prp}\label{prp:generalized_Kunneth_formula}
Let $G$ be any group. There is a natural commutative diagram,
\begin{eqnarray}
\adjustbox{valign=M}{\includegraphics{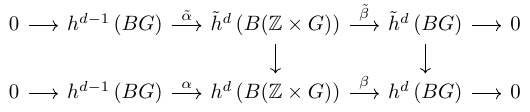}}
\label{generalized_Kunneth_formula_diagram}
\end{eqnarray}
where $\tilde \beta$ and $\beta$ are induced by the monomorphism $G \oneone \ZZZ\times G$, $g\mapsto (0,g)$, the two vertical maps are obtained by forgetting basepoints, $\tilde \alpha$ is the composition of the obvious maps
\begin{eqnarray}
\adjustbox{valign=M}{\includegraphics{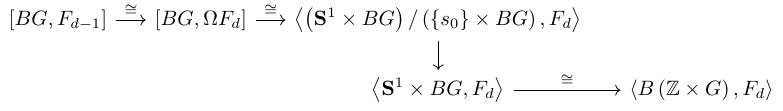}}
\end{eqnarray}
and $\alpha$ is the unique map making the diagram commute. Here, $s_0$ is the basepoint of $\SS^1$. In diagram (\ref{generalized_Kunneth_formula_diagram}), each row is a naturally split short exact sequence, with splitting induced by the epimorphism $\ZZZ \times G \onto G$, $(i,g) \mapsto g$.
\end{prp}

\begin{pf}
This is a special case of Proposition \ref{prp:generalization_arbitrary_semidirect_product}.
\qed\end{pf}



\begin{cor}
Let $G$ be any group. There are natural isomorphisms,
\begin{eqnarray}
\tilde h^d\paren{ B(\ZZZ\times G) } &\isomorphic& h^{d-1}\paren{BG} \oplus \tilde h^d\paren{BG}, \label{generalized_Kunneth_formula_isomorphism} \\
h^d\paren{ B(\ZZZ\times G) } &\isomorphic& h^{d-1}\paren{BG} \oplus h^d\paren{BG}. \label{generalized_Kunneth_formula_isomorphism_2}
\end{eqnarray}\qed\label{cor:generalized_Kunneth_formula}
\end{cor}



Recall, given any semidirect product $\ZZZ \rtimes G$, that the composition of the canonical monomorphism $G \oneone \ZZZ \rtimes G$ and the canonical epimorphism $\ZZZ \rtimes G \onto G$ is the identity on $G$. It follows that the induced map $BG \fromto B\paren{\ZZZ \rtimes G}$ is an embedding.


\begin{prp}\label{prp:generalization_semidirect_product}
Let $G$ be any group and $\ZZZ \rtimes G$ be any semidirect product. There is a natural commutative diagram,
\begin{eqnarray}
\adjustbox{valign=m}{\includegraphics{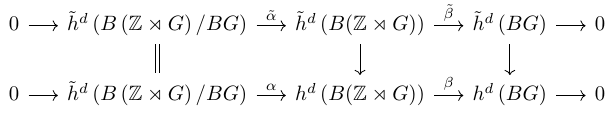}}
\label{generalization_semidirect_product_diagram}
\end{eqnarray}
where $\tilde \beta$ and $\beta$ are induced by the monomorphism $G \oneone \ZZZ\rtimes G$, $g\mapsto (0,g)$, the two vertical maps are obtained by forgetting basepoints, $\tilde \alpha$ is induced by the quotient map $B\paren{\ZZZ\rtimes G} \fromto B\paren{\ZZZ \rtimes G}/ BG$, and $\alpha$ is the unique map making the diagram commute. Here, $BG$ denotes its homeomorphic image in $B\paren{\ZZZ \rtimes G}$ under the induced map $BG \fromto B\paren{\ZZZ\rtimes G}$. In diagram (\ref{generalization_semidirect_product_diagram}), each row is a naturally split short exact sequence, with splitting induced by the epimorphism $\ZZZ \rtimes G \onto G$, $(i,g) \mapsto g$.
\end{prp}

\begin{pf}
This is a special case of Proposition \ref{prp:generalization_arbitrary_semidirect_product}.
\qed\end{pf}


\begin{cor}\label{cor:generalization_semidirect_product}
Let $G$ be any group and $\ZZZ \rtimes G$ be any semidirect product. There are natural isomorphisms,
\begin{eqnarray}
\tilde h^d\paren{ B(\ZZZ\rtimes G) } &\isomorphic& \tilde h^{d}\paren{ B\paren{\ZZZ \rtimes G} / BG } \oplus \tilde h^d\paren{BG}, \\
h^d\paren{ B(\ZZZ\rtimes G) } &\isomorphic& \tilde h^{d}\paren{ B\paren{\ZZZ \rtimes G} / BG } \oplus h^d\paren{BG}.
\end{eqnarray}\qed
\end{cor}



\begin{prp}\label{prp:generalization_arbitrary_product}
Let $G_1$ and $G_2$ be any groups. There is a natural commutative diagram,
\begin{eqnarray}
\adjustbox{valign=m}{\includegraphics{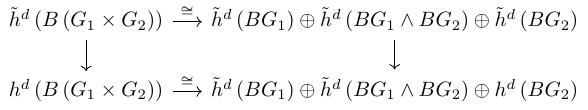}}
\label{generalization_arbitrary_product_diagram}
\end{eqnarray}
with the vertical maps obtained by forgetting basepoints,
such that the canonical inclusions
\begin{eqnarray}
\tilde h^d\paren{BG_1} &\oneone& \tilde h^d\paren{B\paren{G_1 \times G_2}}, \\
\tilde h^d\paren{BG_2} &\oneone& \tilde h^d\paren{B\paren{G_1 \times G_2}}, \\
h^d\paren{BG_2} &\oneone& h^d\paren{B\paren{G_1 \times G_2}}
\end{eqnarray}
are induced by the canonical epimorphisms $G_1\times G_2 \onto G_1$, $G_1\times G_2 \onto G_2$, and $G_1\times G_2 \onto G_2$, respectively, and that the canonical projections
\begin{eqnarray}
\tilde h^d\paren{B\paren{G_1 \times G_2}} &\onto& \tilde h^d\paren{BG_1}, \\
\tilde h^d\paren{B\paren{G_1 \times G_2}} &\onto& \tilde h^d\paren{BG_2} , \\
h^d\paren{B\paren{G_1 \times G_2}} &\onto& h^d\paren{BG_2} 
\end{eqnarray}
are induced by the canonical monomorphisms $G_1 \oneone G_1\times G_2$, $G_2 \oneone G_1 \times G_2$, and $G_2 \oneone G_1 \times G_2$, respectively.
\end{prp}

\begin{pf}
Set $X = BG_1$ and $Y=BG_2$ in Lemma \ref{lem:stable_splitting_2}.
\qed\end{pf}


\begin{cor}\label{cor:generalization_arbitrary_product}
Let $G_1$ and $G_2$ be any groups and $0$ be the trivial group. There are natural isomorphisms,
\begin{eqnarray}
\tilde h^d\paren{B\paren{G_1\times G_2}} &\isomorphic& \tilde h^d\paren{BG_1} \oplus \tilde h^d\paren{BG_1 \wedge BG_2} \oplus \tilde h^d\paren{BG_2},
\end{eqnarray}
and
\begin{eqnarray}
h^d\paren{B\paren{G_1 \times G_2}} &\isomorphic& \tilde h^d\paren{BG_1} \oplus \tilde h^d\paren{BG_1 \wedge BG_2} \oplus h^d\paren{BG_2} \\
&\isomorphic& h^d\paren{BG_1} \oplus \tilde h^d\paren{BG_1 \wedge BG_2} \oplus \tilde h^d\paren{BG_2} \\
&\isomorphic& \tilde h^d\paren{BG_1} \oplus \tilde h^d\paren{BG_1 \wedge BG_2} \oplus \tilde h^d\paren{BG_2} \oplus h^d\paren{B0}.
\end{eqnarray}\qed
\end{cor}


Recall, given any semidirect product $G_1 \rtimes G_2$, that the composition of the canonical monomorphism $G_2 \oneone G_1 \rtimes G_2$ and the canonical epimorphism $G_1 \rtimes G_2 \onto G_2$ is the identity on $G_2$. It follows that the induced map $BG_1 \fromto B\paren{G_1 \rtimes G_2}$ is an embedding.

\begin{prp}\label{prp:generalization_arbitrary_semidirect_product}
Let $G_1 \rtimes G_2$ be any semidirect product of any groups $G_1$ and $G_2$. There is a natural commutative diagram,
\begin{eqnarray}
\adjustbox{valign=m}{\includegraphics{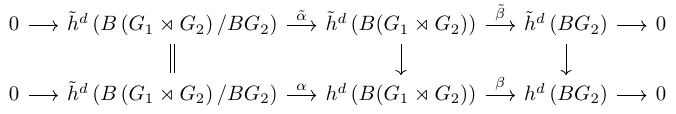}}
\label{generalization_arbitrary_semidirect_product_diagram}
\end{eqnarray}
where $\tilde \beta$ and $\beta$ are induced by the canonical monomorphism $G_2 \oneone G_1 \rtimes G_2$, the two vertical maps are obtained by forgetting basepoints, $\tilde \alpha$ is induced by the quotient map $B\paren{G_1\rtimes G_2} \fromto B\paren{G_1\rtimes G_2}/BG_2$, and $\alpha$ is the unique map making the diagram commute. Here, $BG_2$ denotes its homeomorphic image in $B\paren{G_1 \rtimes G_2}$ under the induced map $BG_2 \fromto B\paren{G_1 \rtimes G_2}$. In diagram (\ref{generalization_arbitrary_semidirect_product_diagram}), each row is a naturally split short exact sequence, with splitting induced by the canonical epimorphism $G_1 \rtimes G_2 \onto G_2$.
\end{prp}

\begin{pf}
In Lemma \ref{lem:retraction}, set $X = B\paren{G_1 \rtimes G_2}$, $A = BG_2$, and $\rho$ to be induced by the canonical epimorphism $G_1 \rtimes G_2 \onto G_2$.
\qed\end{pf}



\begin{cor}\label{cor:generalization_arbitrary_semidirect_product}
Let $G_1 \rtimes G_2$ be any semidirect product of any groups $G_1$ and $G_2$.
There are natural isomorphisms,
\begin{eqnarray}
\tilde h^d\paren{ B\paren{G_1 \rtimes G_2} } &\isomorphic& \tilde h^{d}\paren{ B\paren{G_1 \rtimes G_2}/ BG_2 } \oplus \tilde h^d\paren{BG_2}, \\
h^d\paren{ B\paren{G_1 \rtimes G_2} } &\isomorphic& \tilde h^{d}\paren{ B\paren{G_1 \rtimes G_2}/ BG_2 } \oplus h^d\paren{BG_2}.
\end{eqnarray}\qed
\end{cor}


\section{Mathematical Background\label{app:mathematical_background}}

\subsection{Notions in algebraic topology\label{subapp:notions_algebraic_topology}}

The definitions and constructions below are standard in algebraic topology. See e.g.\,Ref.\,\cite{Hatcher} for detail.

\begin{dfn}[pointed topological space]
A pointed topological space $(X,x_0)$ is a nonempty topological space $X$ together with a privileged point $x_0\in X$ called the basepoint. When the choice of $x_0$ is clear from the context, one may simply write $X$ instead of $(X,x_0)$.
\end{dfn}

Recall from Sec.\,\ref{subsec:mathematical_notation_conventions} that ``map" always means continuous map.

\begin{dfn}[pointed map]
A pointed map between pointed topological spaces is a map that preserves basepoint.
\end{dfn}

\begin{dfn}[topological group]
A topological group is a topological space with a group structure such that both the multiplication and the inversion maps are continuous.
\end{dfn}

As in the main text (see Sec.\,\ref{subsec:mathematical_notation_conventions}), we will abbreviate ``topological group" to simply ``group" and assume that homomorphisms between topological groups are continuous.

\begin{cnstr}
Given a topological space $X$, we can form the quotient space $X/A$ from $X$ by collapsing a subspace $A\subset X$. The image of $A$ is the default basepoint of $X/A$.
\end{cnstr}

\begin{cnstr}
Given two pointed topological spaces $(X,x_0)$ and $(Y, y_0)$, we define the wedge sum $X\vee Y$ to be $\paren{X \sqcup Y}/ \braces{x_0, y_0} $. That is, it is formed from the disjoint union $X \sqcup Y$ by identifying $x_0$ and $y_0$.
\end{cnstr}

\begin{cnstr}
Given two pointed topological spaces $(X,x_0)$ and $(Y, y_0)$, we define the smash product $X\wedge Y$ to be $\paren{X \times Y}/ \paren{ \paren{ X \times \braces{y_0} } \cup \paren{ \braces{x_0} \times Y } }$. It can be viewed as $\paren{X \times Y }/\paren{X \vee Y}$.
\end{cnstr}

\begin{cnstr}
Given a topological space $X$, we form the suspension $SX$ from $X \times I$ by collapsing $X \times \braces{0}$ to a point and $X \times \braces{1}$ to another point.
\end{cnstr}

\begin{cnstr}
Given a pointed topological space $(X,x_0)$, we define the reduced suspension $\Sigma X$ to be $\paren{X \times I}/\paren{\paren{X\times \partial I} \cup \paren{\braces{x_0} \times I}}$. Equivalently, it can be formed from $SX$ by further collapsing $\braces{x_0} \times I$. It can also be viewed as $\SS^1 \wedge X$.
\end{cnstr}

\begin{figure}[t]
\centering
\includegraphics[width=6in]{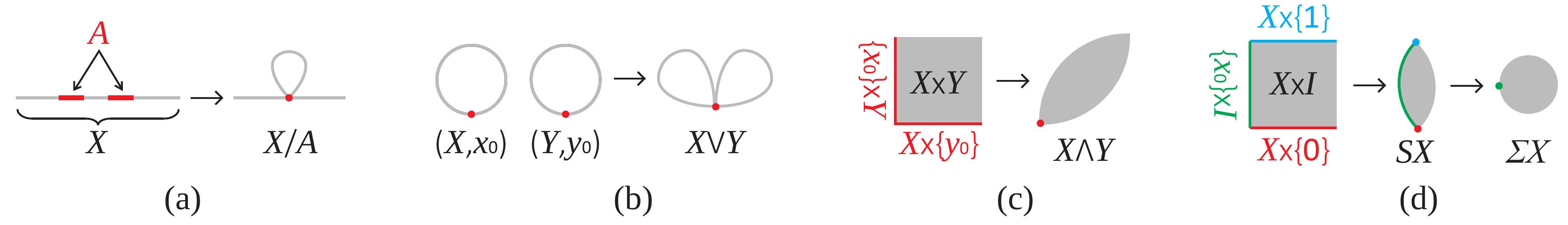}
\caption{(color online). Illustration of the (a) quotient, (b) wedge sum, (c) smash product, (d) suspension, and reduced suspension constructions.\label{fig:topological_construction}}
\end{figure}

These constructions are illustrated in Fig.\,\ref{fig:topological_construction}.

\begin{dfn}[homotopy]
A homotopy between two maps $f_0, f_1: X \fromto Y$ is a map $f: X \times I \fromto Y$ such that
\begin{eqnarray}
f(x,0) = f_0(x), ~ f(x,1)=f_1(x),~ \forall x.
\end{eqnarray}
When such a map exists, $f_0$ and $f_1$ are said to be homotopic, and we write $f_0 \sim f_1$. This defines an equivalence relation, an equivalence class with respect to which is called a homotopy class. The set of homotopy classes of maps from $X$ to $Y$ is denoted by $\brackets{X, Y}$.
\end{dfn}

\begin{dfn}[pointed homotopy]
A pointed homotopy between two pointed maps $f_0, f_1: (X,x_0) \fromto (Y,y_0)$ is a map $f: X \times I \fromto Y$ such that
\begin{eqnarray}
&&f(x,0) = f_0(x), ~ f(x,1)=f_1(x),~ \forall x, \\
&&f(x_0, t) = y_0, ~\forall t.
\end{eqnarray}
When such a map exists, $f_0$ and $f_1$ are said to be homotopic in the pointed sense, and we write $f_0 \sim f_1$. This defines an equivalence relation, an equivalence class with respect to which is called a pointed homotopy class. The set of pointed homotopy classes of maps from $(X,x_0)$ to $(Y,y_0)$ is denoted by $\angles{X, Y}$.
\end{dfn}

\begin{exm}
The $n$-th homotopy group of a pointed topological space $(Y,y_0)$ is $\pi_n(Y) \coloneq \angles{\SS^n, Y}$. In particular, the fundamental group is $\pi_1(Y) \coloneq \angles{\SS^1, Y}$, while the set of path components is $\pi_0(Y) \coloneq \brackets{\pt, Y} \approx \angles{\SS^0, Y}$.
\end{exm}

\begin{dfn}[homotopy equivalence]
A homotopy equivalence between topological spaces $X$ and $Y$ is a pair of maps $f: X \leftrightarrows Y: g$ such that both $g\circ f$ and $f \circ g$ are homotopic to the identities. When such maps exist, $X$ and $Y$ are said to be homotopy equivalent, and we write $X \homotopic Y$. This defines an equivalence relation, an equivalence class with respect to which is called a homotopy type.
\end{dfn}

\begin{dfn}[pointed homotopy equivalence]
A pointed homotopy equivalence between pointed topological spaces $(X,x_0)$ and $(Y,y_0)$ is a pair of pointed maps $f: (X,x_0) \leftrightarrows (Y,y_0): g$ such that both $g\circ f$ and $f \circ g$ are homotopic to the identities in the pointed sense. When such maps exist, $(X,x_0)$ and $(Y,y_0)$ are said to be homotopy equivalent in the pointed sense, and we write $(X,x_0) \homotopic (Y, y_0)$. This defines an equivalence relation, an equivalence class with respect to which is called a pointed homotopy type.
\end{dfn}

A single map $f: X \fromto Y$ or pointed map $f: (X,x_0) \fromto (Y, y_0)$ is sometimes said to be a homotopy equivalence or pointed homotopy equivalence, respectively, if a $g$ with the above properties exists. Thus $f$ is a homotopy equivalence or pointed homotopy equivalence if and only if it represents an invertible map in $\brackets{X, Y}$ or $\angles{X, Y}$, respectively. A homotopy equivalence or pointed homotopy equivalence is precisely an isomorphism in the homotopy category (see App.\,\ref{subapp:categories_functors_natural_transformations}).

\begin{cnstr}
Given topological spaces $X$ and $Y$, we can form the space $\Map(X, Y)$ of maps from $X$ to $Y$, endowed with the compact-open topology \cite{Hatcher}.
\end{cnstr}

\begin{cnstr}
Given pointed topological spaces $(X,x_0)$ and $(Y,y_0)$, we can form the space $\Map_\star(X, Y)$ of pointed maps from $(X,x_0)$ to $(Y,y_0)$, endowed with the compact-open topology.
\end{cnstr}

\begin{exm}
Provided that $X$ is sufficiently well-behaved (e.g.\,locally compact; see Proposition A.14 of Ref.\,\cite{Hatcher}), a homotopy or pointed homotopy can alternatively be defined to be a path in the space $\Map(X, Y)$ or $\Map_\star(X, Y)$, respectively. In this case, $\brackets{X,Y}$ and $\angles{X, Y}$ can be viewed as the sets of path components of $\Map(X, Y)$ and $\Map_\star(X, Y)$, respectively.
\end{exm}

\begin{exm}[path space]\label{exm:path_space}
The path space $PY$ of a pointed topological space $(Y, y_0)$ is defined to be the space $\Map_\star\paren{(I,0), (Y,y_0)}$. Intuitively, it is the space of paths in $Y$ with $y_0$ as the initial point. There is a canonical map $PY \fromto Y$ sending a path $p$ to its endpoint $p(1)$. The default basepoint of $P Y$ is the constant path.
\end{exm}

\begin{exm}[loop space]
The loop space $\Omega Y$ of a pointed topological space $(Y,y_0)$ is defined to be the space $\Map_\star\paren{(\SS^1,s_0), (Y,y_0)}$. It can be viewed as the preimage of $y_0$ with respect to the map $PY \fromto Y$. Intuitively, it is the space of loops in $Y$ based at $y_0$. The default basepoint of $\Omega Y$ is the constant loop.
\end{exm}

\begin{thm}
The sequence $\Omega Y \fromto PY \fromto Y$, where the first map is the inclusion and the second map is as in Example \ref{exm:path_space}, is a fibration. It is called the path space fibration.
\qed\end{thm}

The definition of topological space is general enough to harbor wild examples. It is common in algebraic topology to work with better-behaved spaces, such as CW-complexes.

\begin{cnstr}
Let us construct a topological space $X$ inductively, as follows. Begin with a discrete topological space $X^0$, called the 0-skeleton. For each $n \geq 1$, we form the $n$-skeleton $X^n$ by ``gluing" the boundaries of a family of $n$-disks to $X^{n-1}$ along some maps $\varphi_\alpha: \partial \DD^n \fromto X^{n-1}$. That is, we form the disjoint union $X^{n-1} \sqcup \paren{\sqcup_\alpha \DD^n_\alpha}$ and then identify $x\in \partial \DD^n_\alpha$ with $\varphi_\alpha(x) \in X^{n-1}$ for all $x$ and $\alpha$. Finally, define $X = \cup_n X^n$ and declare a set in $X$ to be open if and only if its intersections with all $X^n$'s are open.\label{cnstr:CW-complex}
\end{cnstr}

The homeomorphic image $e^n_\alpha$ of the interior of a $\DD^n_\alpha$ is called an $n$-cell. A point in $X^0$ is called a 0-cell. Note that the $\varphi_\alpha$'s need not be injective.

\begin{dfn}[CW-complex]
A CW-complex is a topological space constructed as in Construction \ref{cnstr:CW-complex}, with the partition into cells retained as part of the data.
\end{dfn}

\begin{figure}[t]
\centering
\includegraphics[width=3in]{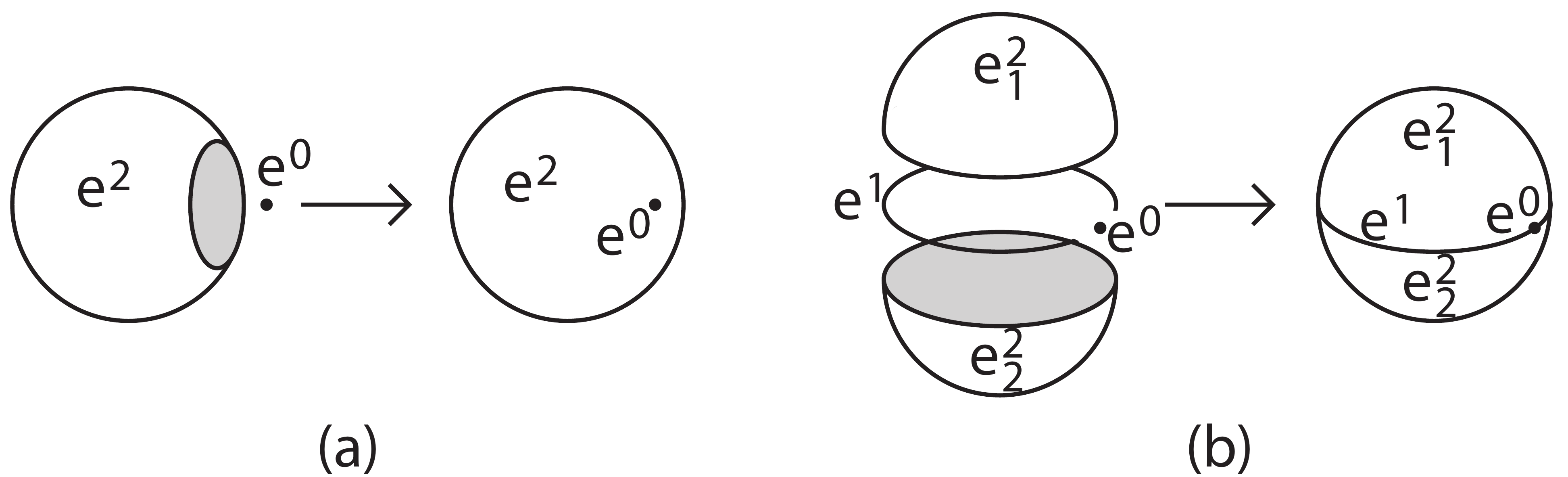}
\caption{$\SS^2$ can be constructed either (a) by attaching a single 2-cell $e^2$ to a single 0-cell $e^0$, or (b) by attaching a single 1-cell $e^1$ (equator) to a single 0-cell $e^0$ and then attaching two 2-cells $e^2_1$ (northern hemisphere) and $e^2_2$ (southern hemisphere).}
\label{fig:sphere_CW_structures}
\end{figure}

\begin{exm}
There are two common CW structures on $\SS^2$ as illustrated in Fig.\,\ref{fig:sphere_CW_structures}.
\end{exm}%

\begin{exm}
All closed manifolds of dimension $\neq 4$ can be given CW structures (the 4-dimensional case is an open question) \cite{KirbySiebenmann, FreedmanQuinn}.
\end{exm}

\begin{dfn}[CW-group]
A CW-group $G$ is a CW-complex together with a topological group structure with the following properties \cite{AdemMilgram, Milnor1956_I, Milnor1956_II}:
\begin{enumerate}[(i)]
\item the inversion map sends $n$-cells to $n$-cells;

\item $\forall g_1, g_2\in G$ contained in some $n_1$- and $n_2$-cells respectively, $g_1g_2$ is contained in a cell of dimension $\leq n_1+n_2$.
\end{enumerate}
These properties imply that the identity is a 0-cell.
\end{dfn}

\begin{exm}
All discrete groups can be viewed as CW-groups with each group element viewed as a 0-cell.
\end{exm}

\begin{exm}
$O(n)$, $U(n)$, $Sp(n)$, and $SO(n)$ can all be given CW-group structures \cite{SteenrodEpstein}.
\end{exm}


\subsection{Categories, functors, and natural transformations\label{subapp:categories_functors_natural_transformations}}

The definitions below are standard in category theory. See e.g.\,Ref.\,\cite{MacLane} for detail.

\begin{dfn}[category]
A category $\C$ consists of
\begin{enumerate}[(i)]
\item a class $\Obj(\C)$ of objects;

\item a class $\Mor(\C)$ of morphisms (or arrows);

\item a function $\dom: \Mor(\C) \fromto \Obj(\C)$ called domain (or source) and a function $\cod: \Mor(\C) \fromto \Obj(\C)$ called codomain (or target)
--
we denote by $\Hom_{\C}\paren{a,b}$ or simply $\Hom\paren{a,b}$, called the hom-class, the class of morphisms with domain $a$ and codomain $b$, and use $f: a \fromto b$ to indicate that $\dom(f) = a$ and $\cod(f) = b$ --

\item a function
\begin{eqnarray}
\identity: \Obj(\C) &\fromto& \Mor(\C) \nonumber \\
a &\mapsto& \identity_a
\end{eqnarray}
called identity;

\item for each triple $(a,b,c)$ of objects, a map
\begin{eqnarray}
\Hom\paren{a,b} \times \Hom\paren{b,c} &\fromto& \Hom\paren{a,c} \nonumber \\
(f,g) &\mapsto& g \circ f \text{~or~} gf
\end{eqnarray}
called composition -- we say two morphisms $f,g$ are composable if $g \circ f$ is defined --
\end{enumerate}
such that the following axioms are satisfied:
\begin{enumerate}
\item associativity: $\paren{h\circ g} \circ f = h \circ \paren{g \circ f}$ for all composable morphisms $f, g, h$;

\item identity: $\identity_a \in \Hom\paren{a,a}$ and $\identity_b \circ f = f \circ \identity_a = f$ for all objects $a,b$ and morphisms $f\in \Hom\paren{a,b}$.
\end{enumerate}
\end{dfn}

\begin{exm}
The category $\Set$ of sets has as objects the class of all sets, and as morphisms the class of all functions between sets. That is, $\Obj(\Set)$ consists of all sets, and given sets $a,b$, $\Hom(a,b)$ consists of all functions from $a$ to $b$. The composition is the usual composition of functions. Given $a$, $\identity_a$ is the constant function on $a$.
\end{exm}

\begin{exm}
The category $\Top$ of topological spaces has as objects all topological spaces, and as morphisms all maps between them. 
\end{exm}

\begin{exm}
The category $\Top_\star$ of pointed topological spaces has as objects all pointed topological spaces, and as morphisms all pointed maps between them. 
\end{exm}

\begin{exm}
The category $\Top^2$ of topological pairs has as objects all pairs $(X,A)$ of topological spaces with $A \subset X$, and as $\Hom_{\Top^2}\paren{\paren{X,A}, \paren{Y,B}}$ all maps $f: X \fromto Y$ such that $f(A) \subset B$. 
\end{exm}

\begin{exm}
The category $\Grp$ of groups has as objects all groups, and as morphisms all homomorphisms between them. 
\end{exm}

\begin{exm}
The category $\Ab^\delta$ of discrete abelian groups has as objects all discrete abelian groups, and as morphisms all homomorphisms between them. 
\end{exm}

\begin{exm}
The homotopy category $\Toph$ of topological spaces has as objects all topological spaces, and $\Hom_{\Toph}(X, Y) \coloneq \brackets{X, Y}$. 
\end{exm}

\begin{exm}
The homotopy category $\Toph_\star$ of pointed topological spaces has as objects all pointed topological spaces, and $\Hom_{\Toph}(X, Y) \coloneq \angles{X, Y}$. 
\end{exm}

\begin{dfn}
A monomorphism, epimorphism, or isomorphism is a morphism that is left-cancellative, right-cancellative, or invertible (in the two-sided sense), respectively. Recall that $f$ is called left- or right-cancellative if $f\circ g_1 = f\circ g_2 \Rightarrow g_1 = g_2$ or $g_1 \circ f = g_2 \circ f \Rightarrow g_1 = g_2$, respectively.
\end{dfn}

\begin{exm}
A monomorphism, epimorphism, or isomorphism in $\Set$ is an injective, surjective, or bijective function, respectively.
\end{exm}

\begin{exm}
A monomorphism, epimorphism, or isomorphism in $\Top$ is an injective, surjective, or bijective map, respectively.
\end{exm}

\begin{exm}
A monomorphism, epimorphism, or isomorphism in $\Grp$ is an injective, surjective, or bijective homomorphism, respectively.
\end{exm}

\begin{exm}
An isomorphism in $\Toph$ or $\Toph_\star$ is a homotopy equivalence or pointed homotopy equivalence, respectively.
\end{exm}

\begin{dfn}[covariant functor]
A covariant functor (or functor) $\F$ from category $\C$ to category $\D$, often written $\F: \C \fromto \D$, consists of
\begin{enumerate}[(i)]
\item a function $\F: \Obj(\C) \fromto \Obj(\D)$;

\item a function $\F: \Mor(\C) \fromto \Mor(\D)$;
\end{enumerate}
such that the following axioms are satisfied:
\begin{enumerate}[(i)]
\item $\F$ maps $\Hom_{\C}(a,b)$ into $\Hom_{\D}\paren{\F(a),\F(b)}$ for all $a, b\in \Obj(\C)$;

\item $\F(\identity_a) = \identity_{\F(\identity_a)}$ for all $a \in \Obj(\C)$;

\item $\F(g \circ f) = \F(g) \circ \F(f)$ for all composable $f,g\in \Mor(\C)$.
\end{enumerate}
When $\F$ is clear from the context, one often writes $f_\ast$ instead of $\F(f)$.
\end{dfn}

\begin{dfn}[contravariant functor]
A contravariant functor (or cofunctor) $\F$ from category $\C$ to category $\D$, often written $\F: \C \fromto \D$ (or $\F: \C^{\rm op} \fromto \D^{\rm op}$), consists of
\begin{enumerate}[(i)]
\item a function $\F: \Obj(\C) \fromto \Obj(\D)$;

\item a function $\F: \Mor(\C) \fromto \Mor(\D)$;
\end{enumerate}
such that the following axioms are satisfied:
\begin{enumerate}[(i)]
\item $\F$ maps $\Hom_{\C}(a,b)$ into $\Hom_{\D}\paren{\F(b),\F(a)}$ for all $a, b\in \Obj(\C)$;

\item $\F(\identity_a) = \identity_{\F(\identity_a)}$ for all $a \in \Obj(\C)$;

\item $\F(g \circ f) = \F(f) \circ \F(g)$ for all composable $f,g\in \Mor(\C)$.
\end{enumerate}
When $\F$ is clear from the context, one often writes $f^\ast$ instead of $\F(f)$.
\end{dfn}

\begin{exm}
The forgetful functor $\For: \Top_\star \fromto \Top$ is a covariant functor that assigns to each pointed topological space $(X,x_0)$ the topological space $X$ with the basepoint forgotten, and to each pointed map $f: (X,x_0) \fromto (Y,y_0)$ the same $f$ viewed as a map between unpointed topological spaces.
\end{exm}

\begin{exm}
The loop space functor $\Omega: \Top_\star \fromto \Top_\star$ is a covariant functor that assigns to each $(X,x_0) \in \Top_\star$ the loopspace $\Omega X$, and to each pointed map $f: (X,x_0) \fromto (Y,y_0)$ the map $\Omega f: \Omega X \fromto \Omega Y$ given by composition with $f$. That is, it sends a loop $l: (\SS^1, s_0) \fromto (X,x_0)$ in $(X,x_0)$ to the loop $f \circ l: (\SS^1, s_0) \fromto (Y, y_0)$ in $(Y,y_0)$. 
\end{exm}

\begin{exm}
The classifying space functor $B: \Grp \fromto \Top_\star$ is a covariant functor that assigns to each topological group $G$ its classifying space $BG$, and to each homomorphism $\varphi: G' \fromto G$ a pointed map $\varphi_\ast: BG' \fromto BG$ (see App.\,\ref{subapp:generalized_cohomology_theories}).
\end{exm}

\begin{dfn}[natural transformation]
Let $\F, \G: \C \fromto \D$ be covariant functors. A natural transformation $\T$ from $\F$ to $\G$, often written $\T : \F \fromto \G$, is an assignment of a morphism $\T(a) : \F(a) \fromto \G(a)$ to each $a\in \Obj(\C)$ such that the following diagram commutes for all $a, b\in \Obj(\C)$ and all $f\in \Hom_{\C}(a,b)$:
\begin{eqnarray}
\adjustbox{valign=m}{\includegraphics{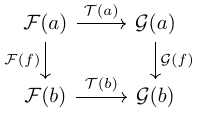}}
\end{eqnarray}
A natural transformation between contravariant functors is defined the same way but with the vertical arrows in the diagram reversed.
\end{dfn}

\begin{dfn}[natural isomorphism]
A natural isomorphism $\T$ is a natural transformation with all $\T(a)$ being isomorphisms.
\end{dfn}

\subsection{Technical conventions\label{subapp:technical_conventions}}

It is not only mathematically customary, but also physically justifiable, to work with ``nice" categories of topological spaces and groups, because after all, pathological spaces and groups may be unphysical. Throughout the paper, apart from Apps.\,\ref{subapp:notions_algebraic_topology}-\ref{subapp:technical_conventions}, the following conventions shall be observed (adapted from Ref.\,\cite{AdemMilgram}):
\begin{enumerate}[(i)]
\item Unless a topological construction makes it impossible\footnote{For instance, the path or loop space of a pointed CW-complex may or may not be a pointed CW-complex. It is, however, always pointed homotopy equivalent to one \cite{Milnor1959}.}, all topological spaces shall be CW-complexes, and the basepoints of all pointed topological spaces shall be $0$-cells.

\item All subspaces of CW-complexes shall be subcomplexes.

\item All groups shall be CW-groups.

\item All subgroups shall be subcomplexes.

\item All binary products of topological spaces shall be compactly generated products.

\item All objects in $\Top$, $\Top_\star$, $\Top^2$, $\Toph$, $\Toph_\star$, and $\Grp$ shall be unpointed or pointed CW-complexes or CW-groups, as appropriate.
\end{enumerate}

The CW approximation theorem implies that every topological space is weakly homotopy equivalent to a CW complex \cite{Hatcher}. The following theorem (a generalization of Proposition 4.22 of Ref.\,\cite{Hatcher}) then indicates that restricting to CW-complexes is hardly a loss of generality. It was also the reason why we were able to freely switch between homotopy equivalent spaces on numerous occasions in the main text.

\begin{thm}
Let $f: Y \fromto Z$ be a homotopy equivalence, or more generally weak homotopy equivalence, between topological spaces $Y$ and $Z$. Then the induced maps
\begin{eqnarray}
f_\ast: \brackets{X, Y} &\fromto& \brackets{X, Z}, \\
f_\ast: \angles{X, Y} &\fromto& \angles{X, Z}, \\
f^\ast: \brackets{Z, X} &\fromto& \brackets{Y,X}, \\
f^\ast: \angles{Z, X} &\fromto& \angles{Y, X}
\end{eqnarray}
are bijections for all CW-complexes $X$.
\qed\end{thm}

\subsection{Generalized cohomology theories\label{subapp:generalized_cohomology_theories}}


\begin{dfn}[Eilenberg-Mac Lane space]
Let $G$ be a discrete group and $n$ be a non-negative integer. If $n>1$, we further require $G$ to be abelian. A space $X$ is called an Eilenberg-Mac Lane space $K(G, n)$ if
\begin{eqnarray}
\pi_i(X) \isomorphic \begin{cases}
G, & i=n, \\
0, & i \neq n,
\end{cases}
\end{eqnarray}
for non-negative integers $i$.
$K(G,n)$ exists and is unique up to homotopy equivalence. This allows us to abuse the terminology and speak of \emph{the} Eilenberg-Mac Lane space $K(G,n)$.
\end{dfn}

\begin{exm}
$\RRR P^\infty$, $\ZZZ$, $\SS^1$, and $\CCC P^\infty$ are $K(\ZZZ_2, 1)$, $K(\ZZZ,0)$, $K(\ZZZ, 1)$, and $K(\ZZZ, 2)$, respectively.\label{exm:Eilenberg_Mac_Lane_spaces}
\end{exm}

\begin{dfn}[classifying space]
Let $G$ be a group. A space $BG$ is called a classifying space of $G$ if there exists a principal $G$-bundle $\xi_G: EG \fromto BG$ satisfying either of the following equivalent conditions \cite{AdemMilgram}:
\begin{enumerate}[(i)]
\item Given any $X$, every principal $G$-bundle over $X$ is isomorphic to the pull-back of $\xi_G$ along a unique homotopy class of maps $f: X \fromto BG$.

\item The map
\begin{eqnarray}
\brackets{X, BG} &\fromto& \braces{\parbox{1.7in}{isomorphism classes of principal $G$-bundles over $X$}} \nonumber\\
\brackets{f} &\mapsto& \brackets{f^\ast(\xi_G)}
\end{eqnarray}
is a bijection.
\end{enumerate}
$BG$ exists and is unique up to homotopy equivalence.
\end{dfn}


Some simple examples of classifying spaces are given in Table\,\ref{table:classifying_spaces}. 
It turns out \cite{Hatcher, AdemMilgram} that
\begin{eqnarray}
\pi_i\paren{BG} = \pi_{i-1}(G).
\end{eqnarray}
Thus if $G$ is a discrete group, then $BG$ is a $K(G,1)$. More generally, if a group $G$ is a $K(G',n)$ as a topological space for some discrete $G'$, then $BG$ is a $K(G', n+1)$. This is consistent with Example \ref{exm:Eilenberg_Mac_Lane_spaces} and Table \ref{table:classifying_spaces}.

\begin{table}[t]
\caption{Examples of classifying spaces. Recall that $BG$ is unique only up to homotopy equivalence. Given here are the most widely used models for $\xi_G: EG \fromto BG$.}
\begin{tabular}{llll}
\hline
\hline
$G$ & $EG$ & $BG$ & $\xi_G: EG \fromto BG$ \\
\hline
$\ZZZ$ & $\RRR$ & $\SS^1 \homeomorphic U(1)$ & $x \mapsto e^{i 2\pi x}$ \\
$U(1)$ & $\SS^\infty = \cup_{n=1}^\infty \SS^{2n-1} \subset \cup_{n=1}^\infty \CCC^{n}$ & $\CCC P^\infty = \cup_{n=0}^\infty \CCC P^n$ & Identify $\paren{z_1, \ldots, z_n} \sim \paren{z_1 e^{i\theta}, \ldots, z_n e^{i\theta}}$ \\
$\ZZZ_2$ & $\SS^\infty = \cup_{n=0}^\infty \SS^n$ & $\RRR P^\infty = \cup_{n=0}^\infty \RRR P^n$ & Identify antipodes \\
\hline
\hline
\end{tabular}
\label{table:classifying_spaces}
\end{table}

\begin{cnstr}[explicit construction of classifying spaces]
There is an explicit construction of $\xi_G: EG \fromto BG$ based on the usual geometric realization \cite{AdemMilgram, Segal1968}. It has the following properties:
\begin{enumerate}[(i)]
\item Each $EG$ is a CW-complex and each $BG$ is a pointed CW-complex.

\item $B: \Grp \fromto \Top_\star$ is a covariant functor.

\item $B(G_1\times G_2)$ is homeomorphic to $BG_1 \times BG_2$.

\item $B\paren{G_1 \rtimes G_2}$ homotopy equivalent to $BG_1 \times_{G_2} EG_2$.

\item $BG$ can be given an abelian group structure if $G$ is abelian.
\end{enumerate}
This will be our default model for $BG$.
\end{cnstr}

The last property enables us to iterate the construction to produce $B^2G$, $B^3G$, $\ldots$ when $G$ is an abelian group. If $G$ is in addition discrete, then $B^nG$ is a $K(G,n)$.

\begin{dfn}[$\Omega$-spectrum]
An $\Omega$-spectrum \cite{Hatcher, Adams1, Adams2} is a family of pointed topological spaces indexed by integers,
\begin{equation}
\ldots, F_{-2}, F_{-1}, F_0, F_1, F_2, \ldots
\end{equation}
together with pointed homotopy equivalences
\begin{equation}
F_n \xfromto{\homotopic} \Omega F_{n+1}
\end{equation}
for all $n$.
\end{dfn}

One can show that $F_n$ determines all $F_m$'s with $m < n$ up to pointed homotopy equivalence. Moreover, shifting the index $n$ turns an $\Omega$-spectrum into another $\Omega$-spectrum.

\begin{exm}[Eilenberg-Mac Lane spectrum]\label{exm:Eilenberg_MacLane_spectrum}
Given any discrete abelian group $A$, the Eilenberg-Mac Lane spaces $K(A,n)$ form an $\Omega$-spectrum, called the Eilenberg-Mac Lane spectrum of $A$ \cite{Hatcher, Adams1, Adams2}. More precisely, the Eilenberg-Mac Lane spectrum of $A$ consists of
\begin{equation}
F_n = \begin{cases}
K(A,n), & n \geq 0, \\
\pt, & n < 0.
\end{cases}
\end{equation}
\end{exm}


A generalized cohomology theory \cite{Adams1, Adams2} is a theory that satisfies the first six of the seven Eilenberg-Steenrod axioms \cite{EilenbergSteenrod1, EilenbergSteenrod2} plus Milnor's additivity axiom \cite{Milnor1962}. Inclusion of the seventh, dimension axiom of Eilenberg and Steenrod's would force the theory to be an ordinary cohomology theory. Here we define generalized cohomology theories in an equivalent but more compact way \cite{Hatcher}.

\begin{dfn}[reduced generalilzed cohomology theory]
\label{dfn:reduced_generalized_cohomology_theory_2}
A reduced generalized cohomology theory consists of
\begin{enumerate}[(i)]
\item a family of contravariant functors
\begin{equation}
\tilde h^n: \Top_\star \fromto \Ab^\delta
\end{equation}
indexed by integers $n$;

\item a natural transformation, called the coboundary map,
\begin{equation}
\delta: \tilde h^n(A) \fromto \tilde h^{n+1}(X/A)
\end{equation}
for topological pairs $(X,A)$, for each $n$;
\end{enumerate}
such that the following axioms are satisfied:
\begin{enumerate}[(i)]
\item homotopy: pointed homotopic maps in $\Top_\star$ induce identical homomorphisms in $\Ab^\delta$;

\item exactness: given any pair $(X,A)$, there is a long exact sequence
\begin{eqnarray}
\cdots && \xfromto{\delta} \tilde h^n\paren{X/A} \xfromto{q^*} \tilde h^n(X) \xfromto{i^*} \tilde h^n(A) \nonumber\\
&&\xfromto{\delta} \tilde h^{n+1}\paren{X/A} \xfromto{q^*} \tilde h^{n+1}(X) \xfromto{i^*} \tilde h^{n+1}(A) \nonumber\\
&&\xfromto{\delta} \cdots
\end{eqnarray}
where $i: A \oneone X$ is the inclusion map and $q: X \onto X/A$ is the quotient map;

\item wedge: given any family of pointed spaces, $\paren{X_\alpha}$, the inclusion maps $X_\alpha \oneone \vee_\alpha X_\alpha$ induce an isomorphism
\begin{equation}
\tilde h^n\paren{ \vee_\alpha X_\alpha } \xfromto{\isomorphic} \prod_\alpha \tilde h^n\paren{X_\alpha}.
\end{equation}
\end{enumerate}
\end{dfn}


\begin{dfn}[unreduced generalized cohomology theory]
\label{dfn:unreduced_generalized_cohomology_theory_2}
An (unreduced) generalized cohomology theory consists of
\begin{enumerate}[(i)]
\item a family of contravariant functors
\begin{equation}
h^n: \Top^2 \fromto \Ab^\delta
\end{equation}
indexed by integers $n$;

\item a natural transformation, called the coboundary map,
\begin{equation}
\delta: h^n\paren{A,\emptyset} \fromto h^{n+1}\paren{X,A}
\end{equation}
for topological pairs $(X,A)$, for each $n$;
\end{enumerate}
such that the following axioms are satisfied:
\begin{enumerate}[(i)]
\item homotopy: homotopic maps in $\Top^2$ induce identical homomorphisms in $\Ab^\delta$;

\item exactness: given any pair $(X,A)$, there is a long exact sequence
\begin{eqnarray}
\cdots && \xfromto{\delta} h^n\paren{X,A} \xfromto{j^*} h^n\paren{X,\emptyset} \xfromto{i^*} h^n\paren{A,\emptyset} \nonumber\\
&& \xfromto{\delta} h^{n+1}\paren{X,A} \xfromto{j^*} h^{n+1}\paren{X,\emptyset} \xfromto{i^*} h^{n+1}\paren{A,\emptyset} \nonumber\\
&& \xfromto{\delta} \cdots \label{LES_reduced}
\end{eqnarray}
where $i: \paren{A,\emptyset} \fromto \paren{X,\emptyset}$ and $j: \paren{X,\emptyset} \fromto \paren{X,A}$ are the inclusion maps.

\item excision: given a triple $(X,A,B)$ with $B\subset A \subset X$, the quotient map $\paren{X,A} \fromto \paren{X/B, A/B}$ induces an isomorphism
\begin{equation}
h^n\paren{X/B, A/B} \xfromto{\isomorphic} h^n\paren{X,A};
\end{equation}

\item additivity: given any family of pairs, $(X_\alpha, A_\alpha)$, the inclusion maps $\paren{X_\alpha, A_\alpha} \fromto \paren{\sqcup_\alpha X_\alpha, \sqcup_\alpha A_\alpha}$ induce an isomorphism
\begin{equation}
h^n\paren{ \sqcup_\alpha X_\alpha, \sqcup_\alpha A_\alpha } \xfromto{\isomorphic} \prod_\alpha h^n\paren{X_\alpha, A_\alpha}.
\end{equation}
\end{enumerate}
\end{dfn}


Every reduced generalized cohomology theory canonically determines an unreduced generalized cohomology theory, and vice versa, as follows. Given a reduced theory $\tilde h$, we define an unreduced theory $h$ according to
\begin{equation}
h^n\paren{X, A} \coloneq \tilde h^n\paren{X/A},
\end{equation}
with the convention $X/\emptyset \coloneq X \sqcup \pt$. Given an unreduced theory $h$, we define a reduced theory $\tilde h$ according to
\begin{equation}
\tilde h^n(X) \coloneq h^n(X, \pt).
\end{equation}
To make contact with Definitions\,\ref{dfn:unreduced_generalized_cohomology_theory} and \ref{dfn:reduced_generalized_cohomology_theory}, we need the pivotal Brown representability theorem (see e.g.\,Ref.\,\cite{DavisKirk} or Theorems 4.58 and 4E.1 of Ref.\,\cite{Hatcher}).

\begin{thm}[Brown representability theorem]
Every $\Omega$-spectrum $\paren{F_n}_{n\in \ZZZ}$ defines a reduced generalized cohomology theory $\tilde h$ according to
\begin{equation}
\tilde h^n\paren{X} \coloneq \angles{X, F_n}.
\end{equation}
Conversely, every reduced generalized cohomolog theory can be represented by an $\Omega$-spectrum this way.\label{thm:Brown_representability_theorem}
\qed\end{thm}

Definitions \ref{dfn:unreduced_generalized_cohomology_theory} and \ref{dfn:reduced_generalized_cohomology_theory} differ from Definitions \ref{dfn:unreduced_generalized_cohomology_theory_2} and \ref{dfn:reduced_generalized_cohomology_theory_2} in two subtle ways, even when the Brown representability theorem is assumed. First, Definitions \ref{dfn:unreduced_generalized_cohomology_theory} and \ref{dfn:reduced_generalized_cohomology_theory} treated $\Omega$-spectrum as part of the data of a generalized cohomology theory, but in reality different $\Omega$-spectra can represent the same theory (although, in the category of spectra, a representing spectrum is determined by the theory up to isomorphism, in view of the Yoneda lemma). It was because of the physical interpretations of $\Omega$-spectrum that we decided to treat it as part of the data. Second, in Definition \ref{dfn:unreduced_generalized_cohomology_theory}, an unreduced generalized cohomology theory was only evaluated on individual spaces not pairs. The connection is given by
\begin{equation}
h^n (X) \coloneq h^n (X, \emptyset).
\end{equation}
It is then easy to show that
\begin{equation}
h^n (X) \isomorphic \brackets{X, F_n }
\end{equation}
for any $\Omega$-spectrum $\paren{F_n}$ that represents the corresponding reduced theory $\tilde h$, in accord with Definition \ref{dfn:unreduced_generalized_cohomology_theory}.

%

Table\,\ref{table:classic_generalized_cohomology_theories} contains some classic generalized cohomology theories alongside with $\Omega$-spectra that represent them.

\begin{table}[t]
\caption{Classic examples of generalized cohomology theories and spectra that represent them \cite{Adams1, Adams2}. Here, $K(A,n)$ denotes the $n$-th Eilenberg-Mac Lane space of $A$ (see App. \ref{subapp:generalized_cohomology_theories}), and $U$ denotes the infinite unitary group $U(\infty) = \bigcup_{i=1}^\infty U(i)$.}
\scriptsize
\resizebox{\columnwidth}{!}{
\begin{tabular}{p{3.5cm}p{3.2cm}p{2.4cm}p{3.1cm}}
\hline
\hline
Theory & Spectrum & Standard notation & Explicit expression \\
\hline
Ordinary cohomology theory with coefficient group $A$ & Eilenberg-Mac Lane \par spectrum of $A$ & $HA$ or $H^\bullet\paren{-;A}$ & $F_n = \begin{cases} K(A,n), & n \geq 0 \\\pt, & n<0 \end{cases}$\\
Real $K$-theory & Real $K$-theory spectrum & $KO$ & Periodic: $F_n \homotopic F_{n+8}$ \\
Complex $K$-theory & Complex $K$-theory \par spectrum & $KU$ & $F_n = \begin{cases} \ZZZ \times BU, & n\text{~even} \\U, & n \text{~odd} \end{cases}$ \\
Stable cohomotopy & Sphere spectrum & $S$ & $F_n = \colim \Omega^m \SS^{n+m}$\\
Oriented cobordism & Thom spectrum of $SO$ & $MSO$ & $F_n = \colim \Omega^m MSO_{n+m}$\\
Unoriented cobordism & Thom spectrum of $O$ & $MO$ & $F_n = \colim \Omega^m MO_{n+m}$\\
Spin cobordism & Thom spectrum of $Spin$ & $MSpin$ & $F_n = \colim \Omega^m MSpin_{n+m}$\\
Pin$^\pm$ cobordism & Thom spectrum of $Pin^\pm$ & $MPin^\pm$ & $F_n = \colim \Omega^m MPin^\pm_{n+m}$\\
\hline
\hline
\end{tabular}
}
\label{table:classic_generalized_cohomology_theories}
\end{table}

\end{appendices}


\bibliographystyle{elsarticle-num}

\renewcommand{\bibfont}{\footnotesize}


\begin{thebibliography}{100}
\bibitem{MarkTwain}
Quote by Mark Twain, which, however, seems to originate from a poem by Thomas
  Campbell.
\end{thebibliography}

\end{document}